\documentclass[a4paper,11pt]{article}
\usepackage[english]{babel}
\usepackage{amsmath, amssymb,graphicx}
\usepackage{yfonts}
\usepackage{caption}
\usepackage{subcaption}
\pdfoutput=1
\usepackage{jheppub}
\hypersetup{colorlinks=true,linkcolor=magenta,linktoc=page,citecolor=blue}
\usepackage{tikz}
\newcommand{\be}{\begin{equation}}
\newcommand{\ee}{\end{equation}}
\newcommand{\bea}{\begin{eqnarray}}
\newcommand{\eea}{\end{eqnarray}}
\newcommand{\nn}{\nonumber}
\usepackage{xcolor}
\usepackage{cancel}
\usepackage{pdflscape}
\usepackage{rotating, graphicx}
\usepackage{lipsum}
\usepackage{graphicx}
\usepackage{transparent}

\title{Holographic Photosynthesis}
\author{Irina  Aref'eva and Igor  Volovich}
\affiliation{Steklov Mathematical Institute, Russian Academy of Sciences,\\Gubkina str. 8, 119991, Moscow, Russia}

\emailAdd{arefeva@mi.ras.ru, volovich@mi.ras.ru}

\abstract{There are successful applications of the holographic AdS/CFT
correspondence to high energy  and condensed matter physics. We apply the holographic approach to photosynthesis that is an important example of nontrivial quantum phenomena relevant for life which is being studied in the emerging field of quantum biology.
Light harvesting complexes of photosynthetic organisms are  many-body quantum
systems, in which quantum coherence has recently been experimentally shown to survive for relatively long
time scales even at the physiological temperature despite the decohering effects of their environments.

We use
the holographic  approach to evaluate the time dependence of entanglement entropy and
quantum mutual information in the Fenna-Matthews-Olson (FMO) protein-pigment complex in green sulfur bacteria during
the transfer of an excitation from a chlorosome antenna to a reaction center. It is demonstrated that the time evolution of the mutual information
simulating the Lindblad master equation in some cases can be obtained by means
of a dual gravity describing black hole formation in the AdS-Vaidya spacetime.
The wake up and scrambling times for various partitions of the FMO complex are discussed.
}


\begin{document}

\maketitle
\newpage

\section{Introduction}
 The  anti-de Sitter - conformal field theory (AdS/CFT) correspondence \cite{Aharony:1999ti}
 and more general holographic gravity/gauge duality play an important role in  modern theoretical physics. They have been used
 for description of strong interacting equilibrium
and non-equilibrium systems in high energy physics, in particular, to describe heavy-ion collisions
and formation of quark-gluon plasma  \cite{Solana,IA,DeWolf}, as well as in  condensed matter physics \cite{Hartnoll08kx,Sachdev:2010ch}.

According to the holographic  correspondence
the quantum gravity (string theory) on  anti-de Sitter
spacetime (AdS) is equivalent to a certain quantum field theory on the AdS boundary. The holographic approach provides a powerful method for studying
strongly coupled quantum field theories
by means of the dual classical gravitational theory in  the AdS space which is more tractable.

In last years, there has been a growing  interest to study the entanglement entropy
and quantum mutual information for various quantum systems by using
the holographic approach \cite{Ryu:2006bv}-\cite{Mirabi:2016elb} and refs therein.
The entanglement entropy of a boundary region is determined by the area of the minimal bulk surface that coincides with the entangling surface at the boundary \cite{Ryu:2006bv,Ryu:2006ef,Hubeny:2007xt}.

In this paper
we apply the holographic approach to photosynthesis. Light harvesting complexes (LHC) in bacteria and plants are important examples of nontrivial quantum phenomena relevant for life which are being studied in the emerging field of quantum biology.
Recent investigations of quantum effects in biology
include also the process of vision, the olfactory sense, the magnetic orientation of migrant birds as well as photon antibunching in proteins, the quantum delocalization of biodyes in matter-wave interferometry
and quantum tunneling in biomolecules,  see for instance \cite{Markus Arndt,
 Graham R. Fleming, 
 OhyaVol, QBIC, 
 %Neill Lambert,   
 Quantum Effects in Biology, IV}.

Photosynthesis is vital for life on Earth \cite{Blanck}. Photosynthesis changes the energy from
the sun into chemical energy and splits water to liberate oxygen and convert carbon
dioxide into organic compounds, especially sugars. Nearly all life either depends on it directly as a source of energy, or indirectly as the ultimate source of the energy
in their food.

Light-harvesting complexes in plants and photosynthetic bacteria include
protein scaffolds into which pigment molecules are embedded, e.g. chlorophyll or
bacteriochlorophyll molecules. The pigment molecules absorb light and the resulting electronic excitation, exciton,
is transported between the pigment molecules until it reaches a reaction center
complex, where its energy is converted into separated charges.
The process whereby the light energy is transported
through the cell is extremely efficient\, -- higher than any artificial energy transport process.

One models many photosynthetic light harvesting
complexes by a general
three-part structure comprising the antenna, the transfer network, and the reaction center. The antenna
captures photons from sunlight and subsequently excites the electrons of the pigment  from their ground state. The excited electrons, which combine with
holes to make excitons, travel from the antenna through an intermediate protein exciton transfer
complex to the reaction center where they
participate in the chemical reaction that generates oxygen.

Quantum coherences have  been observed in two-dimensional spectroscopic studies of energy transfer within
several different light harvesting complexes.
The simplest light harvesting
complex is  the Fenna-Matthews-Olson (FMO) protein-pigment complex in green sulfur
bacteria. We demonstrate that some numerical results on the time evolution of the mutual information for the FMO complex \cite{FM}
can be obtained by using the holographic approach.

 Experiments with the FMO complex have shown the presence of quantum beats between excitonic levels at both cryogenic
 $(77^{\circ }K)$ and ambient $(300^{\circ }K)$ temperatures \cite{Engel, Collini, Panit, SFOG}. Quantum coherence has  also been seen in light
harvesting antenna complexes of green plants \cite{Calh} and marine algae \cite{Coll}. There are also  studies of quantum coherence within the reaction center \cite{Lee}.

Theoretical studies of excitation dynamics in the FMO complex in the single excitation subspace have demonstrated the presence of long-lived, multipartite entanglement.
A  study of the temporal duration  of
entanglement in the FMO complex using a simulation of excitation energy transfer  dynamics under conditions
that approximate the real environment was performed in \cite{MRL}-\cite{TAD}.
Entanglement  within the Markovian description of the FMO complex has been considered in \cite{Car,Car-2}. The time evolution of the mutual information in the FMO complex was studied in \cite{0912.5112}
by using simulation of the Gorini-Kossakowski-Sudarshan-Lindblad master equation. Quantum nonlocality as a function of time is studied in
\cite{Charlotta Bengtson}.

The paper is organized as follows.

In Sect.\ref{Sect:Setup}  has an introductory character. Here we remind the main objects and tools of our  study of
the FMO complex.
In Sect.\ref{Sect:Meq} has an introductory character. Here we briefly describe  the FMO complex and write down the  corresponding master equation.
In  Sect.\ref{EE-MI-FMO}  the standard consideration of LHC  as  quantum system and
entropy of entanglement and mutual information for the FMO complex are sketched.
In Sect.\ref{Sect:reductions} we  list different reductions of the FMO complex used in  modern theoretical studies.
In Sect.\ref{HEE} definitions of holographic entanglement entropy and mutual information, as well as the basic formula that we use in the main text are presented.
In Sect.\ref{Decom} two iterative procedures to calculate the holographic entanglement entropy ${\cal S}(A_1\cup A_2\cup ...\cup A_n)$
are presented. The fist one takes into account contributions only of primitive diagrams and the second one incorporates also  Boltzmann rainbow  diagrams.
In Sect.\ref{HMI-phases} we remind the phase structure of the holographic  mutual information $I(A,B)$
for two belts in the static backgrounds, empty AdS$_{d+1}$ and  AdS$_{d+1}$ black brane.

Sect.\ref{HMI-simple-FMO} is devoted to the study of the time evolution of
holographic mutual information for the simplest reduction of the FMO complex. We start, Sect.\ref{HMI-simple-FMO-m0}, by calculation of
holographic entanglement entropy for two site system during a quench at nonzero temperature. Then in Sect.\ref{Sect:HMI}
we study holographic mutual information for two site system at nonzero temperature.
In Sect.\ref{Comparison}  we compare the results of our calculations with the mutual information
 $I\left(  A,B\right) $ calculated for the reduced FMO system in \cite{0912.5112}.
In Sect.\ref{Sect:wakeup} we discuss the dependencies of wake up time   and scrambling times on the geometrical parameters and the initial temperature.

 In Sect.\ref{Sect:HMI-comp} we consider the holographic mutual information for the  reduction of the FMO complex with one composite part.
 For this purpose in Sect.\ref{Sect:HEE-comp} we study  the phase structure of  holographic entanglement entropy  $S(A\cup B\cup C)$ for the Vaidya shell in the AdS$_d$ black brane background.
In Sect.\ref{MI-3segm} the mutual information for 3 segments and scrambling time for this background is calculated
and in Sect.\ref{Sect-Comparision-comp} we make a fit of time dependence of
the  mutual information at the physiological temperature ($300^\circ K$) calculated in \cite{0912.5112} for  one mixed state
by the time dependence  of the holographic mutual information $I(A_1\cup A_6,A_3)$ under the global quench by the Vaidya shell in AdS$_4$.

\section{Setup}\label{Sect:Setup}

\subsection{ FMO complex}\label{Sect:Meq}

\subsubsection {Seven bacteriochlorophills}

\begin{figure}[h!]
$$\,$$
\centering
 \includegraphics[scale=0.25]{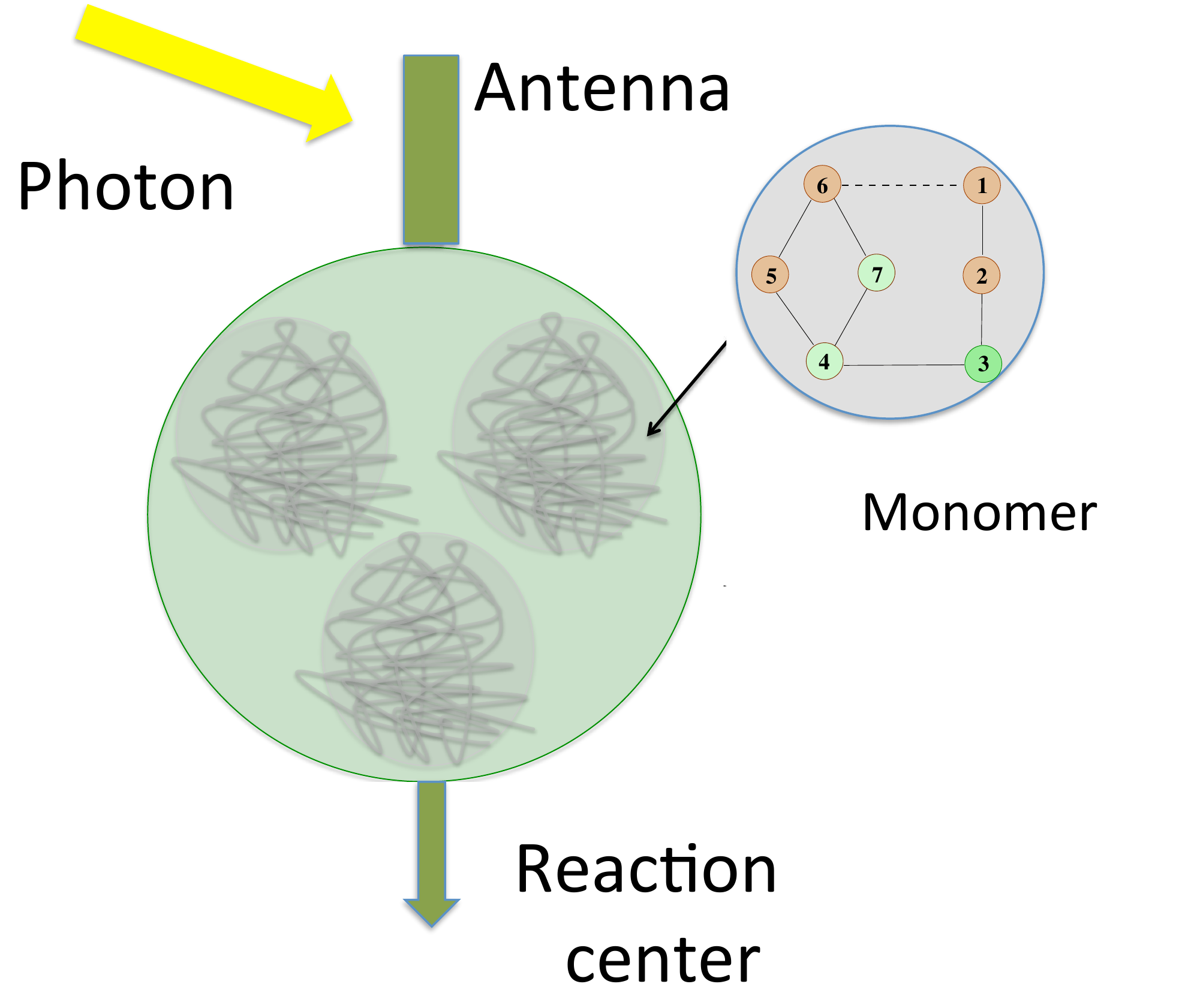}A$\,\,\,\,\,\,\,\,$$\,\,\,\,\,\,\,\,$$\,\,\,\,\,\,\,\,$
 \includegraphics[scale=0.7]{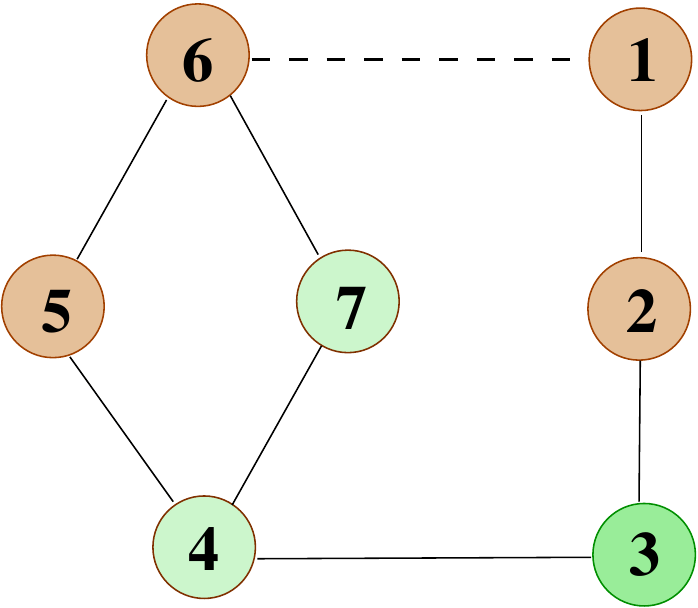}$\,\,\,\,\,\,\,$B
\caption{The schematic picture  of the FMO complex. A. The FMO complex trimer.  B.  The schematic picture for one monomer.}\label{Fig:FMO}
\end{figure}

The FMO protein complex \cite{FM} is the main light-harvesting component of the green sulfur bacteria {\it Prosthecochloris
aestuarii}. It is a trimer,
consisting of three identical molecular sub-units, Fig.\ref{Fig:FMO}A.
Each of the sub-units is a network
of seven\footnote{Recently, an additional bacteriochlorophyll (the eighth) pigment was discovered in each subunit of this trimer\cite{Schmidt}.} interconnected bacteriochlorophylls
$\{1,2,...,7\}$ arranged in two weakly connected
branches  that are separately connected to the antenna (bacteriochlorophyll sites one and six) and jointly connected to the reaction center via site three, Fig.\ref{Fig:FMO}B.

The theoretical
models in the literature \cite{Car-2, Sar, 0912.5112, Car}   study the
dynamics of one sub-unit of the trimer. The total Hamiltonian of
the system includes the non-relativistic QED Hamiltonian in the dipole
approximation, phonon Hamiltonian and other environmental fields, for a derivation of the master equation in the weak coupling stochastic limit
see \cite{ALV} and also \cite{AKV, KV}.

We consider one sub-unit of the trimer which consists of seven bacteriochlorophyll sites $\{1,2,...,7\}$ transferring energetic excitations from a photon-receiving antenna to a reaction center,
see Fig.\ref{Fig:FMO}.

\subsubsection{Master equation}

We consider the  nine-state
model \cite{Sar, 0912.5112}  for excitation transfer
in the single excitation approximation which is described by a 9-dimensional subspace in the Hilbert space $(\mathbb{C}^2)^{\otimes 8}$. The possible
states for the exciton will be expressed in the site basis
$\{|m\rangle\}^{7}_{m=1}$
where the state $|m\rangle$ indicates that the excitation
is present at site $m$. There are  also a ground state $|G\rangle$ corresponding to the loss
or recombination of the excitation and a sink state $|S\rangle$ corresponding to the trapping of the exciton at the reaction
center (there is  a discussion of the "local" and "global" bases in theory of  open quantum systems in \cite{TV}). The density
operator  for this quantum system has the
following representation in the site basis:
\be
\rho=\sum_{m,n\in\{G,1,...,7,S\}}\rho_{m,n}|m\rangle\langle n|
\ee
It is assumed that the density matrix satisfies the GKS-Lindblad master equation of the following form
\be
\frac{d}{dt}\rho=-i[H,\rho]+{\cal L}(\rho)
\ee
Here the Hamiltonian $H$ describes the coupling between the seven sites states $|1\rangle,..., |7\rangle  $:
\be
H=\sum_m E_m|m\rangle\langle m|+\sum_{m<n}V_{mn}(|m\rangle\langle n|+
|n\rangle\langle m|),
\ee
where $E_m$ is the energy of the site $m$ and $V_{mn}$ describes the coupling between sites $m$ an $n$. A Lindblad superoperator ${\cal L}(\rho)$ has the general form
\be
{\cal L}(\rho)=\sum_m\gamma_m(2L_m\rho L_{m}^{\dag}-\{L_{m}^{\dag}L_m,
\rho \}),
\ee
where $L_m$ are arbitrary operators and $\gamma_m$ are positive constants, see for example \cite{OhyaVol}.
In our case for the FMO complex the Lindblad superoperator is taken in the form \cite{MRL,Car-2,Sar,0912.5112}
\be
{\cal L}(\rho)={\cal L}_{\text{diss}}(\rho)
+{\cal L}_{\text{deph}}(\rho)+{\cal L}_{\text{sink}}(\rho)
\ee
Here the first term ${\cal L}_{\text{diss}}(\rho)$
describes the dissipative recombination  of the exciton,
\be
{\cal L}_{\text{diss}}(\rho)=\sum_m\Gamma_m(2|G\rangle\langle m|\rho|m\rangle\langle G|-\{|m\rangle\langle m|,\rho\})
\ee
where $\Gamma_m$ is the rate   of the recombination at site $m$.

The second Lindblad superoperator  ${\cal L}_{\text{deph}}$
accounts for the dephasing interaction with the environment,
\be
{\cal L}_{\text{deph}}(\rho)=\sum_m\lambda_m(2|m\rangle\langle m|\rho|m\rangle\langle m|-\{|m\rangle\langle m|,\rho\})
\ee
where $\lambda_m$ is the rate of dephasing at site $m$.

 The final term ${\cal L}_{\text{sink}}(\rho)$ accounts for the trapping of the exciton in the reaction center:
\be
{\cal L}_{\text{sink}}(\rho)=\Gamma_{\text{sink}}(2|S\rangle\langle 3|\rho|3\rangle\langle S|-\{|3\rangle\langle 3|,\rho\})
\ee
 It is supposed that the initial state of the FMO complex is a pure excitation at
site one or site six, or a mixture of these two states.

\subsubsection{Entropy of entanglement and mutual information }\label{EE-MI-FMO}

Let two parties $A$ and $B$ share a quantum state $\rho_{AB}$
in a Hilbert space ${\cal H}_A\otimes{\cal H}_B$.
The von Neumann entropy of this state is
\be
 S(AB)=-\text{tr}(\rho_{AB}\log\rho_{AB})
\ee
Similarly, one can define the entanglement entropies
\be
 S(A)=-\text{tr}(\rho_{A}\log\rho_{A}),\,\,\,
S(B)=-\text{tr}(\rho_{B}\log\rho_{B})\,,
\ee
where the reduced density matrices $\rho_{A}=\text{tr}_B\rho_{AB}$ and
$\rho_{B}=\text{tr}_A\rho_{AB}$.

The quantum mutual information $I(A;B)$ measures the correlations
shared between the two parties,
\be\label{MI-general}
I(A;B)=S(A)+S(B)-S(AB).
\ee
This can be written as a relative entropy and is therefore non-negative:
\be
I(A;B)=S(\rho_{AB}||\rho_A\otimes\rho_B)\geq 0
\ee
where
$$
S(\rho||\sigma)=\text{Tr}(\rho\log\rho)-\text{Tr}(\rho\log\sigma),
$$
see for example  \cite{OhyaVol}.

Several simulations of the quantum mutual information as a function of time at cryogenic $(77^{\circ }K)$ and physiological  $(300^{\circ }K)$ temperatures were conducted in \cite{0912.5112}. In particular the simulations calculate
the quantum mutual information with respect to several "bipartite cuts" of the sites in the FMO complex. The different cases are as follows,
(Figs.1-6 in \cite{0912.5112}):

1. The first cut was picked up where system $A$ consists of site three and system $B$ consists of sites one and six, (Figs.1 and 2 in \cite{0912.5112}). Remind that the initial state of the complex is at site one, six, or the mixture, and
the FMO complex  transfers the excitation from these initial sites to site three.

2. The next bipartite cut is with the A system consisting of sites one and two and the B system
consisting of site three, (Figs.3 and 4 in \cite{0912.5112}).

3. Finally, the third cut was  the cut where system
A consists of site three and system B consists of all other sites,
 (Figs.5 and 6 in \cite{0912.5112}). The quantum mutual information for this case
should be larger than for the case of the other cuts.

\subsubsection{Reductions of the FMO complex}\label{Sect:reductions}

\begin{figure}[h!]
\centering
$$\,$$\\
 \includegraphics[scale=0.8]{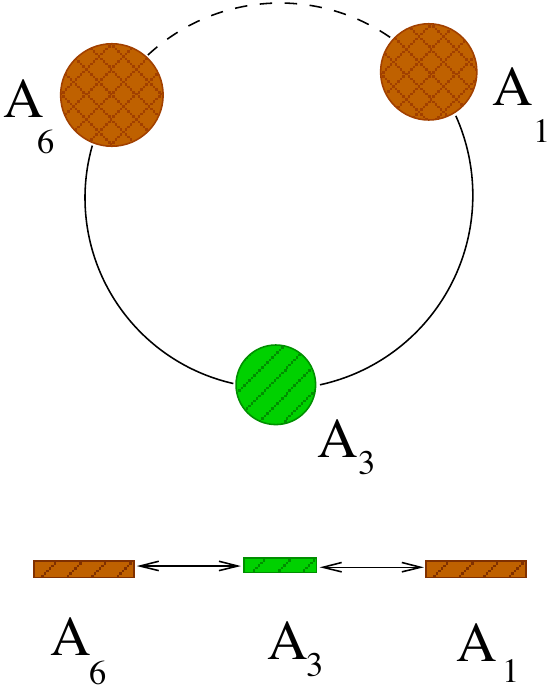}
 \caption{Schematic picture of  $(1,6|3)$ reduced FMO system. }
 \label{Fig:fmo-1-3-6}
\end{figure}

\begin{figure}[h!]
$$\,$$
\centering
 \includegraphics[scale=1]{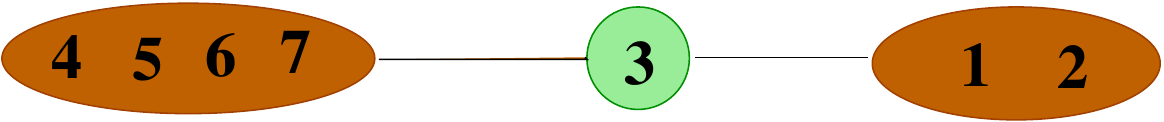}
\caption{$(1,2,4,5,6,7|3)$  reduction of the FMO-complex}\label{Fig:fmo-4567-3-12}
\end{figure}

\begin{figure}[h!]
$$\,$$
\centering
 \includegraphics[scale=1]{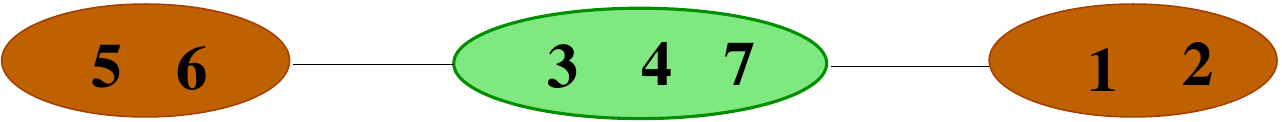}
\caption{$(1,2,5,6|3,4,7)$  reduction of the FMO-complex}\label{Fig:fmo-12-347-56}
\end{figure}

We start from Fig.\ref{Fig:FMO}.  Several reductions of this system to more simple ones are considered. Some of them are schematically presented in  Fig.\ref{Fig:fmo-1-3-6}--Fig.\ref{Fig:fmo-12-347-56}.

\newpage
\subsection{Holographic entanglement entropy}\label{HEE}
\subsubsection{Static AdS background}

Consider a quantum field theory on a $d$-dimensional
manifold $\mathbb{R}\times \mathbb{R}^{d-1}$, where $\mathbb{R}$ and
$\mathbb{R}^{d-1}$
denote the time axis and the $(d-1)$-dimensional space-like manifold, respectively. Let be given a $(d-1)$-dimensional submanifold $A\subset \mathbb{R}^{d-1}$ at fixed time $t=t_0$ and let $\partial A$ be its boundary. Then the formula for the holographic entanglement entropy $S_A$ in a CFT on
$\mathbb{R}\times \mathbb{R}^{d-1}$ reads
 \cite{Ryu:2006bv}
\begin{equation} S_{A}=\frac{{\cal A}(\gamma_{A})}{4G^{(d+1)}_N},
\label{arealaw}
\end{equation}
where ${\cal A}(\gamma_{A})$ is the area of $\gamma_{A}$, that is  the $d-1$ dimensional static minimal surface in
AdS$_{d+1}$ with metric
\be
ds^2=\frac{R^2}{
z^{2}} \left(dz^2-dt^2+\sum_{i=1}^{d-1} dx_i^2\right), \label{Poincare}
\ee
   whose boundary is given by $\partial A$, and $G^{(d+1)}_{N}$
is the $d+1$ dimensional Newton constant, $R$ is the radius of AdS$_{d+1}$.

\begin{figure}[h!]
\centering
 \includegraphics[scale=1]{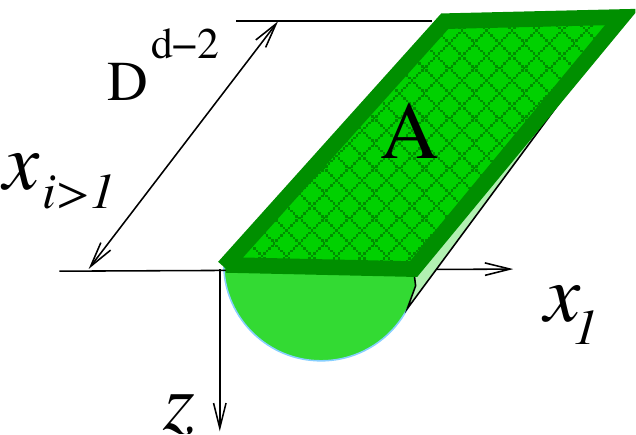}
 \caption{The standard holographic picture of a strip configuration in three spacetime dimensions at a constant time slice.
 $\ell$ is the width of the belt and $D>>\ell$. The bulk surface extended in the direction $z$ ends on the entanglement surface $A$.}\label{Fig:1}
\end{figure}

The simplest   example for the shape of $A$ is a
straight belt at the boundary $z=0$,  see Fig.\ref{Fig:1},
\be
A=\{x_i|x_1\in [-l,l\,], x_{2,3,...,d-1}\in
(-\infty,\infty)\}.\ee

 In this
case the area of the minimal
surface, divided on the suitable constants, is \cite{Ryu:2006bv} \be
\frac{{\cal A}_{vac,reg}}{R^{d-1}D^{d-2}}=
\left(\frac{1}{(d-2)\epsilon^{d-2}}-\frac{c_0}{\ell^{d-2}}\right)
,\;\;d> 2,
\label{Avacreg}
\ee
where $D$ is the length of $A$ in the traversal
$x_{2,3,\cdots,d-1}$-direction, $\epsilon>0$ is the UV regularization,
 $c_0$ %$=\frac{\cancel{2^{d-2}}}{d-2}\left(\sqrt{\pi}\Gamma(\frac{d}{2(d-1)})/\Gamma(\frac{1}{2(d-1)})\right)^{d-1}$.
 is a positive constant depending on $d$.
  The first term is divergent when $\epsilon\to 0$, but for our purpose it will not play a role, since we will consider the differences of entanglement entropies.

 For an excited state whose gravitational dual is provided by the black brane solution
 with mass $m$ and the Hawking temperature $T_H=dm^{1/d}/4\pi$ (here we assume $R=1$)
\be \label{BB}
ds^2=\frac{1}{
z^{2}} \left(-f(z)dt^2+\frac{dz^2}{f(z)}+\sum_{i=1}^{d-1} dx_i^2\right), \,\,\,\,\,f(z)=1-mz^d,
\ee in general, it is not possible to find an explicit expression for
the  entanglement entropy for $d>3$. It can be found from the integral equation relating the entropy to an auxiliary parameter $z_*$
   \bea\label{BHren}
\frac{\mathcal{A}_{exc,ren}} {2R^{d-1}D^{d-2} } &=&    \frac{1}{z_*^{d-2}} \int _{0}^{1}\frac{dw}{w^{d-1 }}\left[\frac{1}{
\sqrt{f(z_* w)\left(1-w^{2(d-1)}\right) } }-1\right]-\frac{ 1}{(d-2)z_*^{d-2}},\nn\\
\label{BHrend}
\eea
 meanwhile the parameter $z_*$ is related to the width of the belt
\bea\nn
\frac{\ell}{2} &=& z_*\int^{1}_{0}\frac{w^{2(d-1)}dw}{\sqrt{f(z_* w)(1-w^{2(d-1)}) }}.\eea

%%%%%%%%%%%%%%%%%%%%%%%

\subsubsection{Time dependent AdS Vaidya background}
The proposal (\ref{arealaw}) has been generalized to time dependent geometries in \cite{Hubeny:2007re}.
In the Vaidya AdS$_{d+1}$ spacetime the corresponding minimal surface describes the thermalization process in the d-dimensional boundary theory \cite{Balasubramanian:2011ur}, see also \cite{AbajoArrastia:2010yt,Albash:2010mv,Allais:2011ys}. The Vaidya AdS$_{d+1}$ metric has the form
\be \label{Vaidya}
ds^2=\frac{R^2}{z^2} \left(-f(v,z)dv^2-dvdz+\sum_{i=1}^{d-1} dx_i^2\right),
\ee
where the function
\be\label{fvz}
f(v,z)=1-m(v)z^d.
\ee
 The form of $m(v)$ is usually chosen to be
\be\label{m-v}
m(v)=\frac{m}{2}\left(1+\tanh \frac{v}{\alpha}\right)
\ee
The metric (\ref{Vaidya}) with this function describes a spacetime model
which evolves from pure AdS at early times to the Schwarzschild black brane at late times because of the  shell of null dust infalling along
$v=0$. The parameter $\alpha$ determines the thickness of the shell. The case $\alpha\to 0$
corresponds to a step function and one deals with a shock wave.

We will consider also the shell in the   AdS$_{d+1}$ black brane background (the Vaidya AdS black brane metric), in this case
\be\label{m-v-BB}
m(v)=m_0+\frac{m}{2}\left(1+\tanh \frac{v}{\alpha}\right)
\ee

The entanglement entropy is given by the extremum of the  functional
\cite{Balasubramanian:2011ur,Allais:2011ys}
\bea
\label{action} \frac{\mathcal{A} }{2R^{d-1}D^{d-2} }
=
\int_{0}^{\ell}
\frac{dx}{z^{d-1}} \sqrt{1-\big[1-m(v)z^d\big](v')^2-2v' z'},
\eea
where  $z=z(x), v=v(x)$ and $v^{'} = dv/dx$.

The Euler-Lagrange equations corresponding to the action  (\ref{action}) read
\bea
\label{EOM-v}
\big[1-m(v)z^d\big]v''+z''  -\frac{\partial_v m(v)}{2}\,z^d (v')^2 -d\, m(v) z^{d-1} z' v'
&=& 0 \,,
\\
\label{EOM-z}
z\, v''-\frac{d-2}{2}\,m(v) z^d (v')^2+(d-1)\big[\,(v')^2+2v' z'-1\,\big]
&=& 0\;.
\eea

For the system of equations  (\ref{EOM-v}), (\ref{EOM-z}) we will solve numerically the Cauchy problem with    initial data (\ref{ID-0}),
(\ref{ID-0})
\be\label{IDstar}
z(0)=z_*,\,\,\,\,\,\,v(0)=v_*,
\ee
\be\label{ID-0}
z'(0)=v'(0)=0,
\ee
We are interested in finding solutions that reach the boundary at some point, that we identify\footnote{In our calculation the width of the belt is $2\ell$, although on some pictures we omit the factor $2$. } with   $\ell$ in (\ref{action}) at the boundary time $t$:
\be
z(\ell)=0,\,\,\,\,\,v(\ell)=t.
\ee

Equations  (\ref{EOM-v}), (\ref{EOM-z}) have an  integral of motion
 \bea\label{zstar}
\left(\frac{z_*}{z}\right)^{2(d-1)} &=&
1-\big[\,1-m(v)z^d\,\big](v')^2-2v' z'\,.
\eea
 Due to this identity the integral in (\ref{action})
can be substantially simplified to give
\bea
\label{reg-action}
\frac{\mathcal{A}_{reg}}{2R^{d-1}D^{d-2}} &=&
\int_{0}^{\ell-\epsilon}
\frac{z_\ast^{d-1}}{z^{2(d-1)}}\,dx\;,\eea
where we introduce  regularization $\epsilon >0$. This integral contains the UV divergence when $\epsilon\to 0$. The   renormalized version of (\ref{reg-action}) can be
written as \cite{Balasubramanian:2011ur,Allais:2011ys}
\bea
\label{ren-action}
\frac{\mathcal{A}_{ren}}{2R^{d-1}D^{d-2}} &=&
\int_{0}^{\ell-\epsilon}
\frac{z_*^{d-1}}{z^{2(d-1)}}\,dx
-\frac{1}{(d-2)\, (z(\ell-\epsilon))^{d-2}},\,\,\,\,\,\,d>2,
\eea
and regularization can be removed. Note, that in the case of $d=2$
\cite{Balasubramanian:2011ur,Allais:2011ys}
\bea
\label{ren-action-d2}
\frac{\mathcal{A}_{ren}}{2R} &=&
\int_{0}^{\ell-\epsilon}
\frac{z_*}{z^2}\,dx
+ z_*\log (z(\ell-\epsilon)).
\eea

It is quite straightforward to find the dependence of the renormalized entanglement entropy on $z_*$
and $v_*$. To find the dependence of the  entanglement entropy on $\ell$
and $t$ one has to parametrize the solutions  to (\ref{ID-0})-(\ref{EOM-z}), not by  $(z_*,v_*)$,
but by $\ell$ and $t$.

It happens  that
the values of $\ell$ and $t$ are very sensitive to  the initial data (\ref{IDstar}),
and one needs a special numerical procedure to find  the pair $(z_*,v_*)$ corresponding to the given  pair $\ell,t$
with large values of $\ell,t$,
see \cite{Allais:2011ys,1512.05666,1601.06046} and refs therein for more details. It is interesting to note that  the case when $m_0 > 0$ in
 (\ref{m-v-BB}) is more stable than the case  $m_0=0$.
%%%%%%%%%%%%

In Fig.\ref{Fig:Ent3V}  the dependence of the renormalized holographic entanglement entropy on $t$ and $\ell$
  for the propagating Vaidya shell in the four dimensional black brane background with $f=f(z,v)$
 given by \eqref{fvz} and \eqref{m-v-BB} is presented.  As compare with the dependence of the holographic entanglement entropy
for the Vaidya AdS$_4$ metric the entropy does not go to fixed value for large  $\ell$, but as both go to constant values for large times.
The same is shown in Fig.\ref{Fig:Ent2V} for the three dimensional black brane background.

\begin{figure}[h!]
$$\,$$
\centering
\begin{picture}(250,100)
\put(-80,0){\includegraphics[scale=0.28]{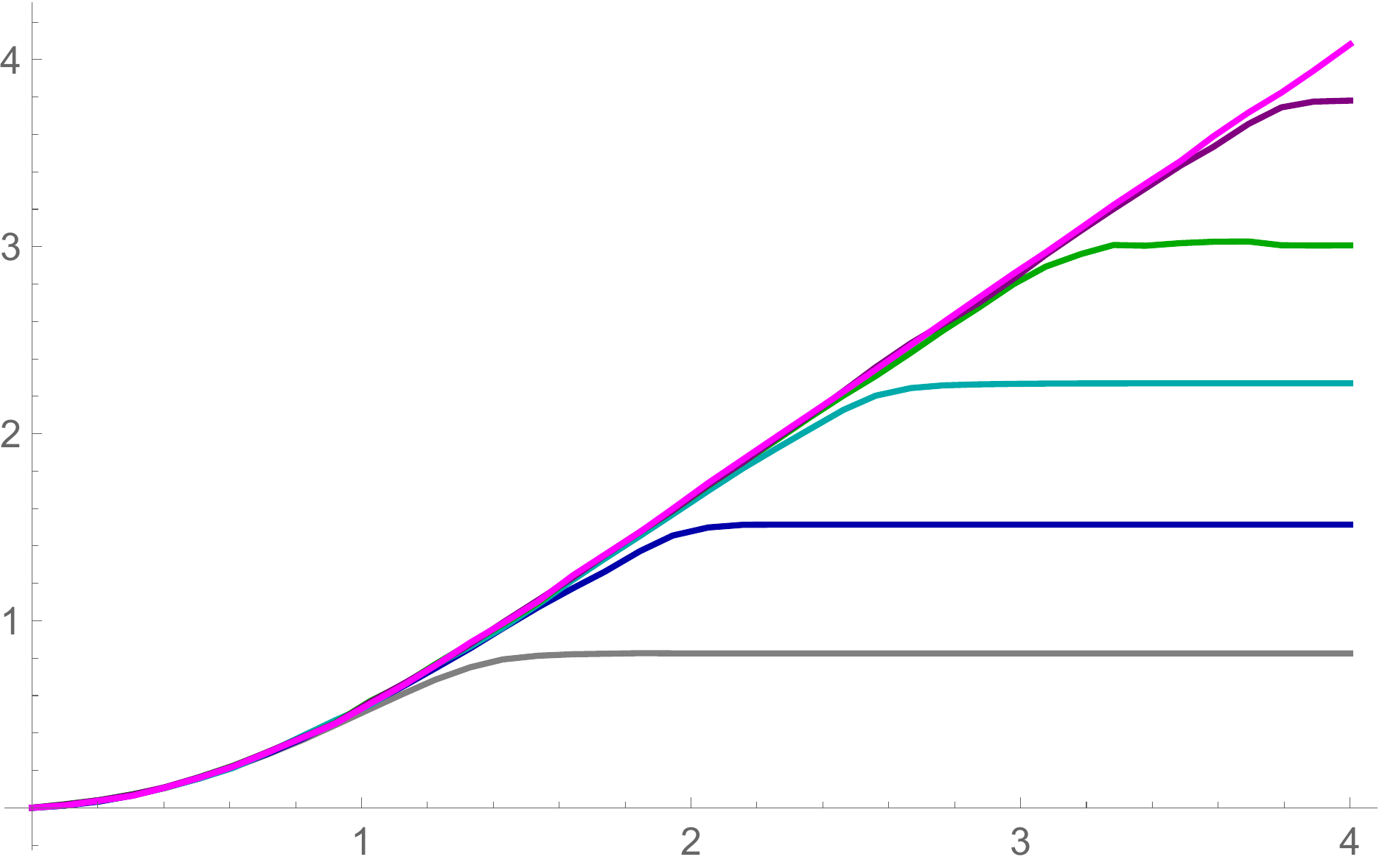}}
\put(90,0){$t$}
\put(-75,110){${\cal S}$}
\put(150,-3){\includegraphics[scale=0.27]{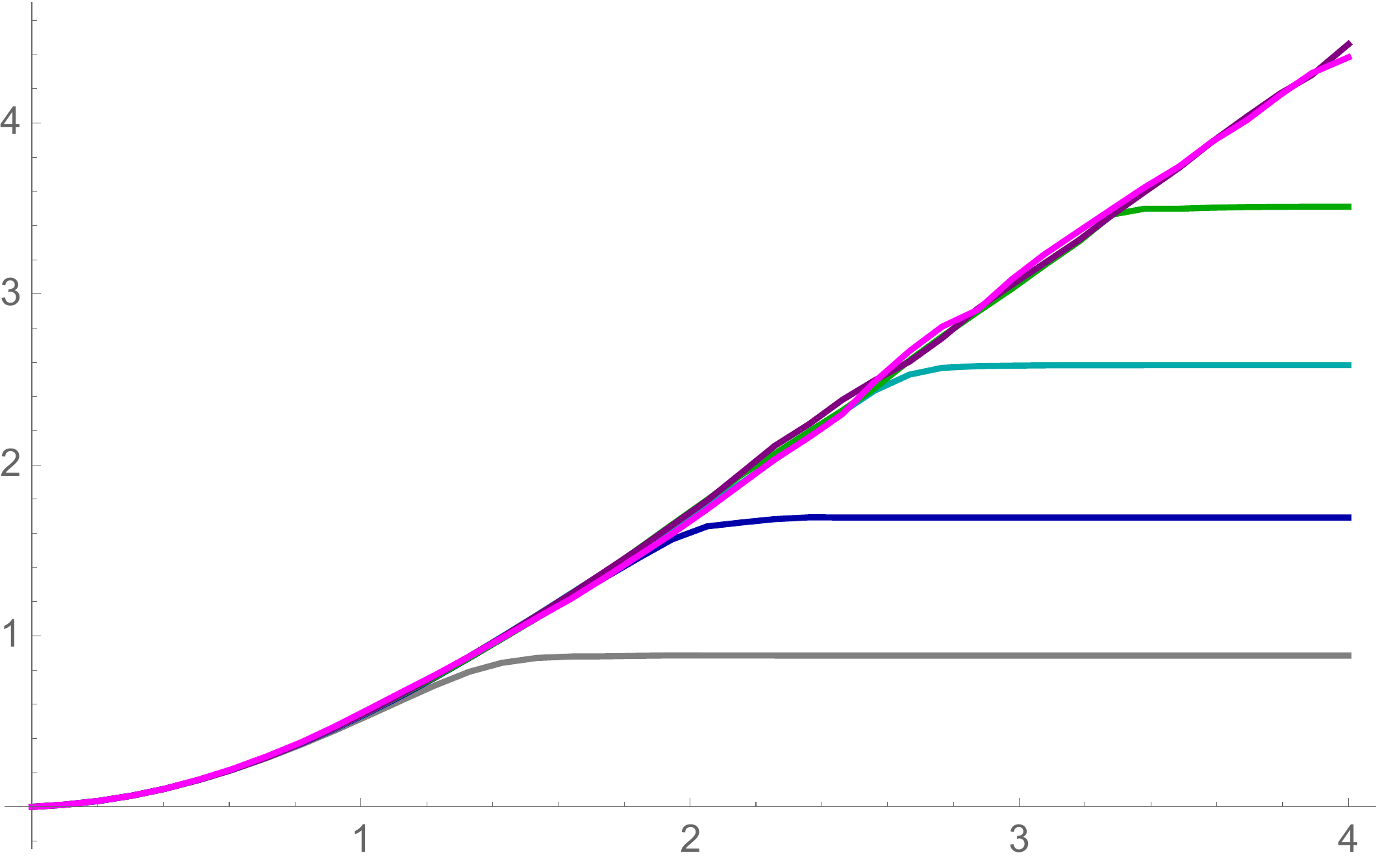}}
\put(330,0){$t$}
\put(150,110){${\cal S}$}\end{picture}
\\$\,$\\
 \caption{
 %{\it From file:Vaidya-AD-seek-23-march}. The top panels.
 The time dependence of the holographic entanglement entropy
  $\mathcal{A}_{ren,3}(\ell,t)$, after the corresponding initial state  subtraction,  for the Vaidya metric in the four dimensional black brane background with $f=f(z,v)$ ($m_0=0.25$, $m=1$ in the left panel and
  $m_0=0$, $m=1$  in the right panel)
 at fixed $\ell=1,1.5,2,2.5,3,3.5$ (gray, blue, darker cyan,
green, purple and magenta, respectively). For all cases $\alpha=0.2$
 }
 \label{Fig:Ent3-t}
\end{figure}
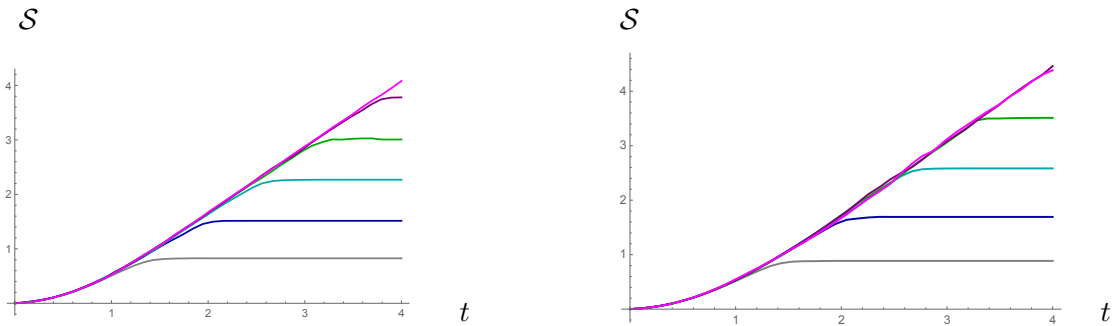

\begin{figure}[h!]
$$\,$$
\centering
\begin{picture}(250,100)
\put(-80,0){\includegraphics[scale=0.28]{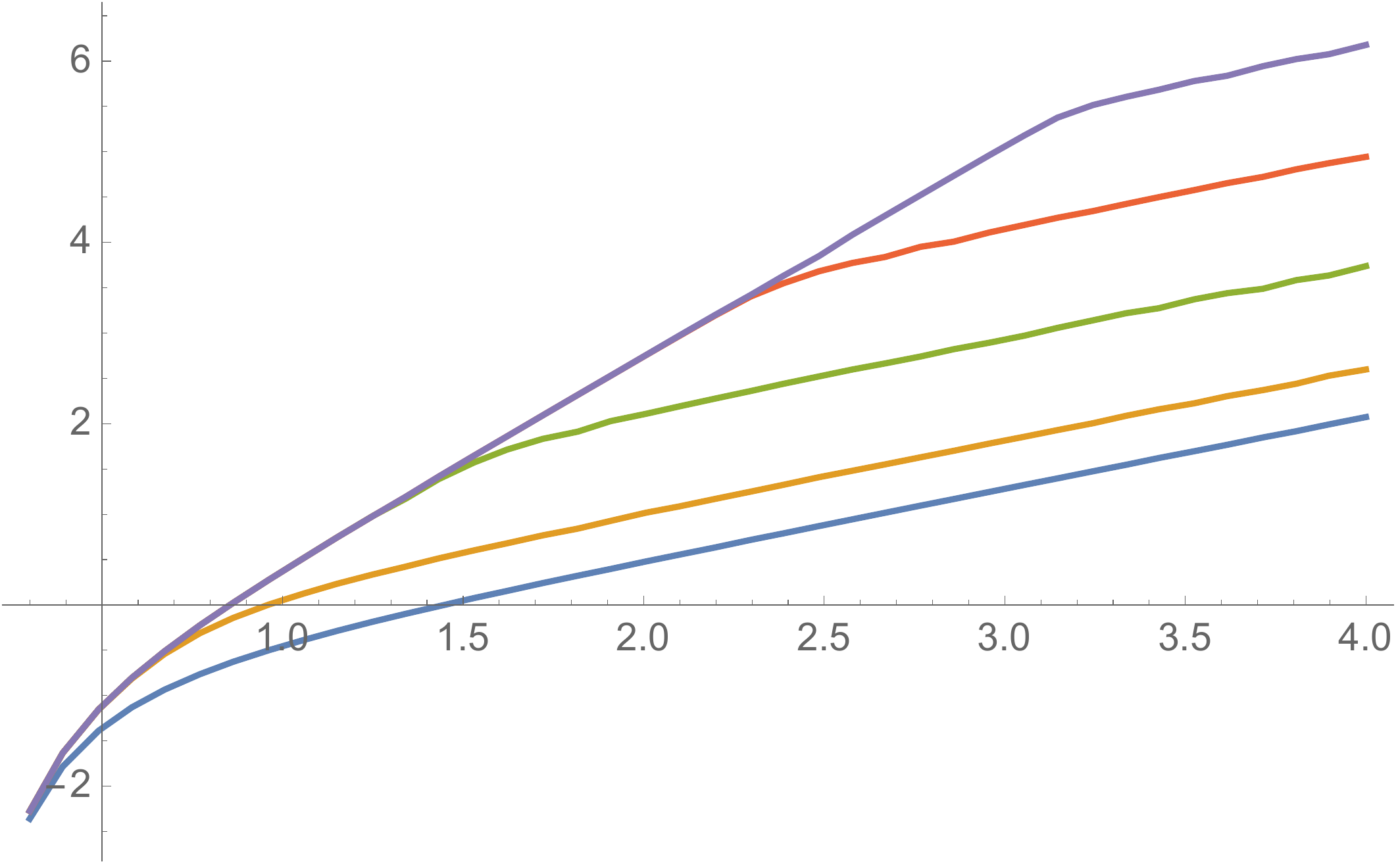}}
\put(96,25){$\ell$}
\put(-75,110){${\cal S}$}
\put(150,0){\includegraphics[scale=0.3]{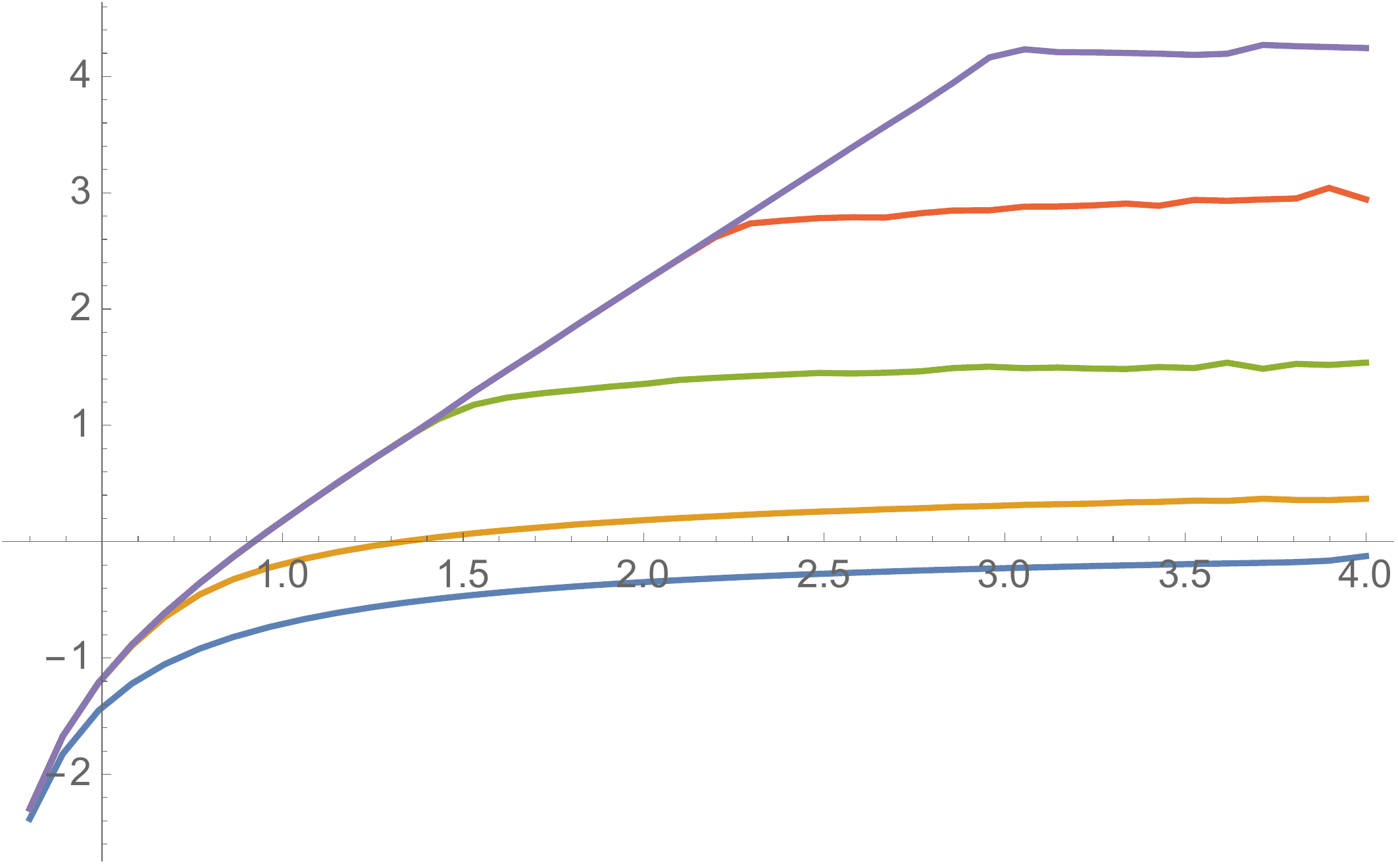}}
\put(340,25){$\ell$}
\put(150,110){${\cal S}$}\end{picture}
\\$\,$\\
 \caption{
 The renormalized holographic entanglement entropy
  $\mathcal{A}_{ren,3}(\ell,t)$, up to the normalizing factor, for the Vaidya metric in the four dimensional black brane background with $f=f(z,v)$ ($m_0=0.25$, $m=1$ in the left panel and
  $m_0=0$, $m=1$  in the right panel)
 at fixed $t=0,1,2,3,4$ (blue, orange, green, red  and darker blue, respectively) as function of $\ell$.
 For all cases $\alpha=0.2$ }
 \label{Fig:Ent3V}
\end{figure}
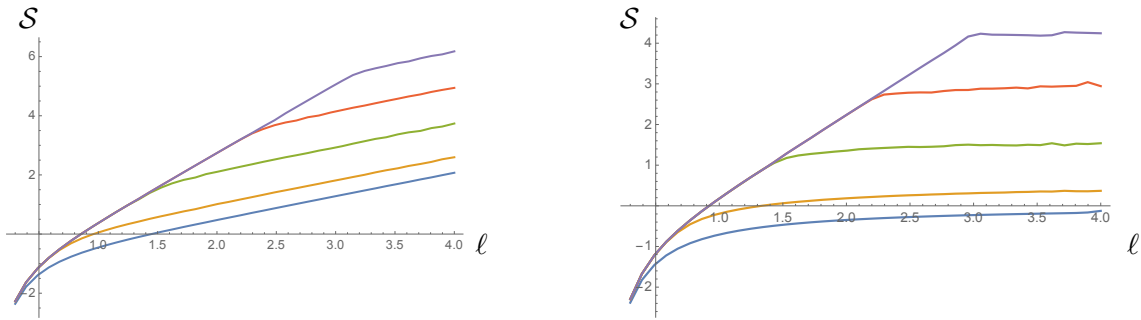

\begin{figure}[h!]
$$\,$$
\centering
\begin{picture}(250,100)
\put(-80,0){\includegraphics[scale=0.28]{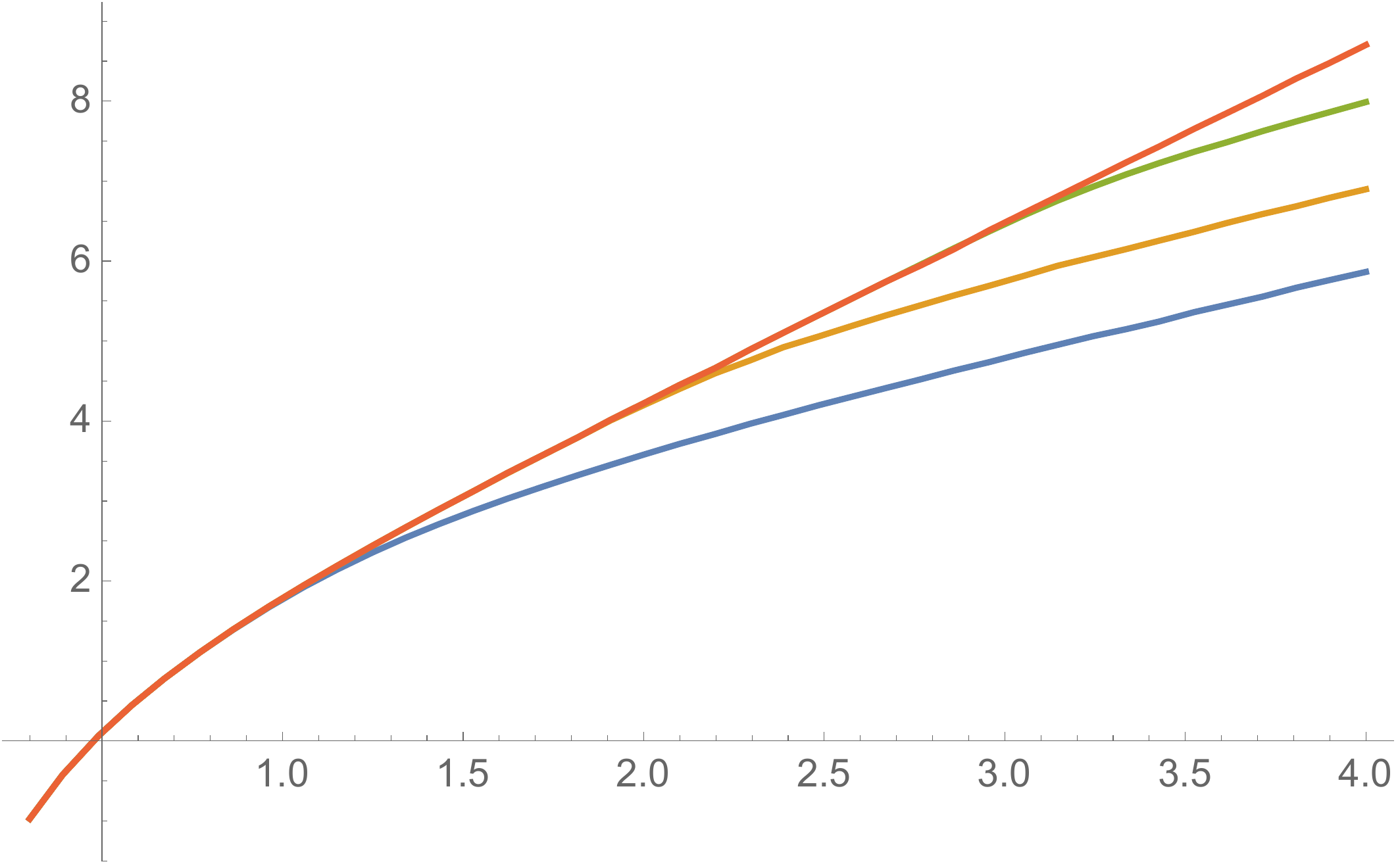}}
\put(96,5){$\ell$}
\put(-75,110){${\cal S}$}
\put(150,0){\includegraphics[scale=0.26]{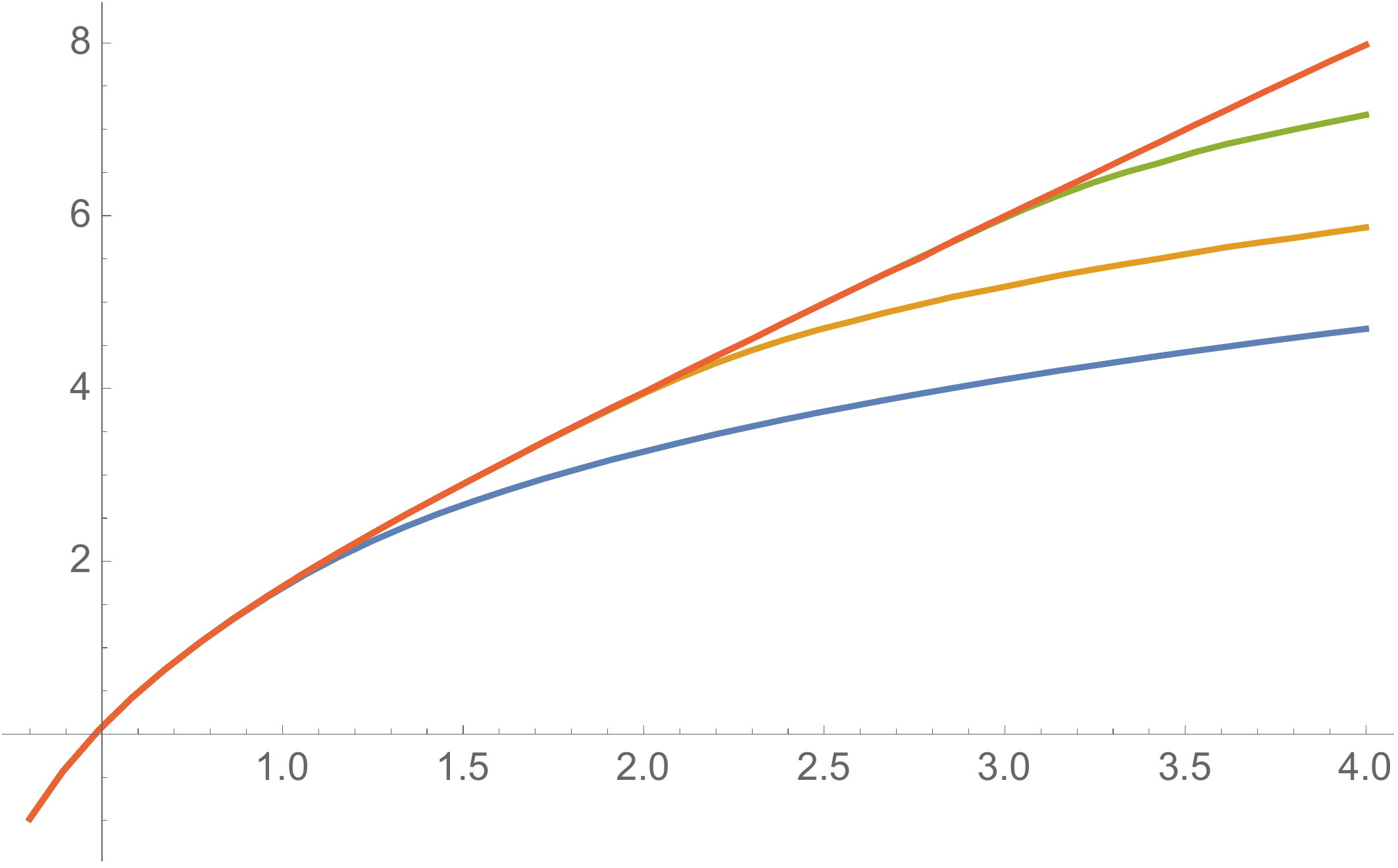}}
\put(340,5){$\ell$}
\put(150,110){${\cal S}$}\end{picture}
\\$\,$\\
 \caption{
 %{\it From file: Vaidya-AD-seek-11-march-d2-evening}.
 The renormalized holographic entanglement entropy
  $\mathcal{A}_{ren,2}(\ell,t)$, up to the normalizing factor, for the Vaidya metric in the three dimensional black brane background with $f=f(z,v)$
 given by \eqref{fvz} and \eqref{m-v-BB} ($m_0=0.25$, $m=1$ in the left panel and
  $m_0=0$, $m=1$  in the right panel) at fixed $t=0,1,2,3$ (blue, orange, green and red, respectively) as function of $\ell$. For all cases $\alpha=0.2$}
  \label{Fig:Ent2V}
\end{figure}
%%%%%%%%%%%%%%%%%%
%%%%%%%%%%%%%

Note that the similar technique has been intensively used  in study the thermalization processes, especially  for non-conformal invariant backgrounds, see \cite{Alishahiha:2014cwa,1401.6088,1503.02185,1512.05666,1601.06046} and  refs therein.

\newpage
\subsection{Iterative procedure to calculate  ${\cal S}(A_1\cup A_2\cup ...\cup A_n)$}\label{Decom}
In this section we present the rules that help to calculate the holographic entropy for n disjoint objects.
Here we consider for illustration a particular case of $n$-segments. This problem has been studied in several papers \cite{Balasubramanian:2011at,Hayden:2011ag,Allais:2011ys,Alishahiha:2014jxa,Ben-Ami:2014gsa,Mirabi:2016elb}. It is substantially simplified in the case of equal length strips and equal separation between them. However in order  to deal with holographic description of the FMO complex,  we have to find the entropy for non-equal  length strips.

To find the entropy one has to find the global minimum among all possible configurations.  To specify  all possible configurations corresponding to local minimum surfaces  it is convenient to use the diagrammatic language.  In special cases, it happens  that only primitive  diagrams contribute to the entropy.   The primitive diagrams are  diagrams that do not contain    cross-sections of the connected lines and   also do not contain    diagrams with "engulfed" sub-diagrams in the terminology used in  \cite{Ben-Ami:2014gsa}. Note that the class of primitive diagrams contains less diagram  then so-called rainbow diagrams in the  Boltzmann quantum field theory \cite{IAIV}.
 It is  possible that for more complicated backgrounds one has to take into account more  diagrams then only the primitive diagrams.

 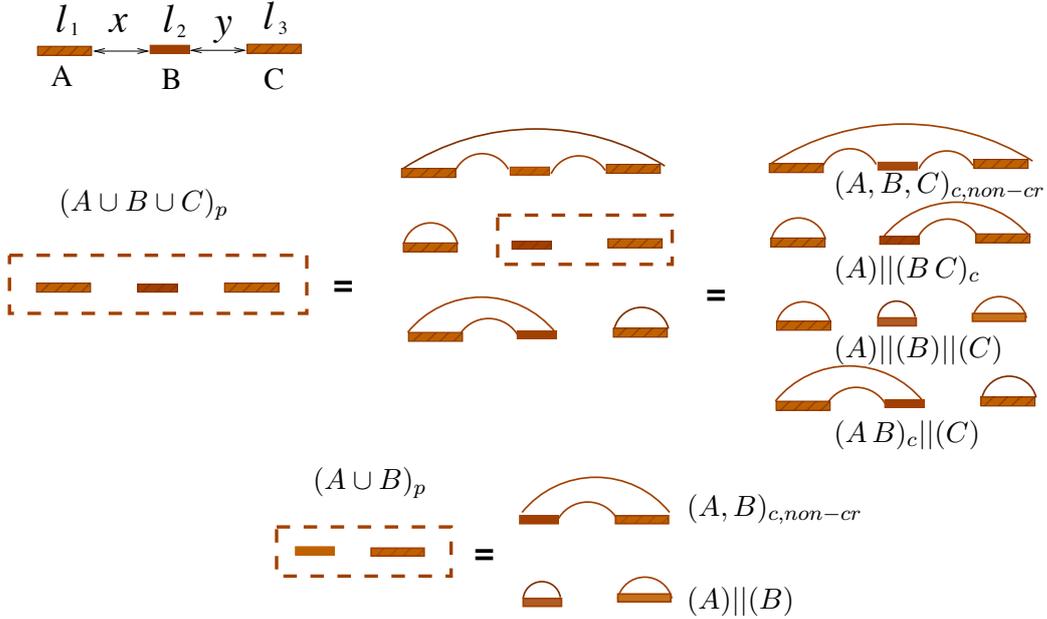
\begin{figure}[h!]
 $\,\,\,\,\,\,\,\,\,$ $\,\,\,\,\,\,\,\,\,$ $\,\,\,\,\,\,\,\,\,$
\begin{picture}(250,230)
\put(-50,110){\includegraphics[scale=0.7]{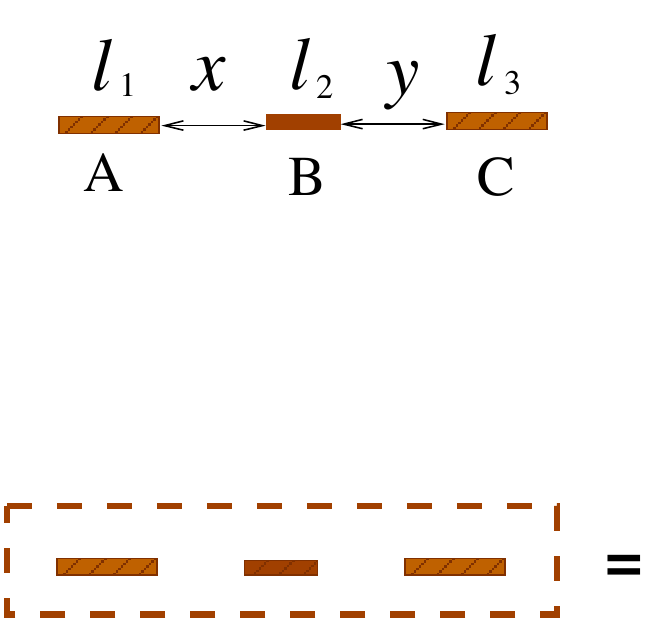}}
\put(65,45){$(A\cup B)_p$}
\put(260,157){$(A,B,C)_{c,non-cr}$}
\put(260,63){$(A\,B)_c||(C)$}
\put(260,125){$(A)||(B\,C)_c$}
\put(260,95){$(A)||(B)||(C)$}
\put(50,0){\includegraphics[scale=0.7]{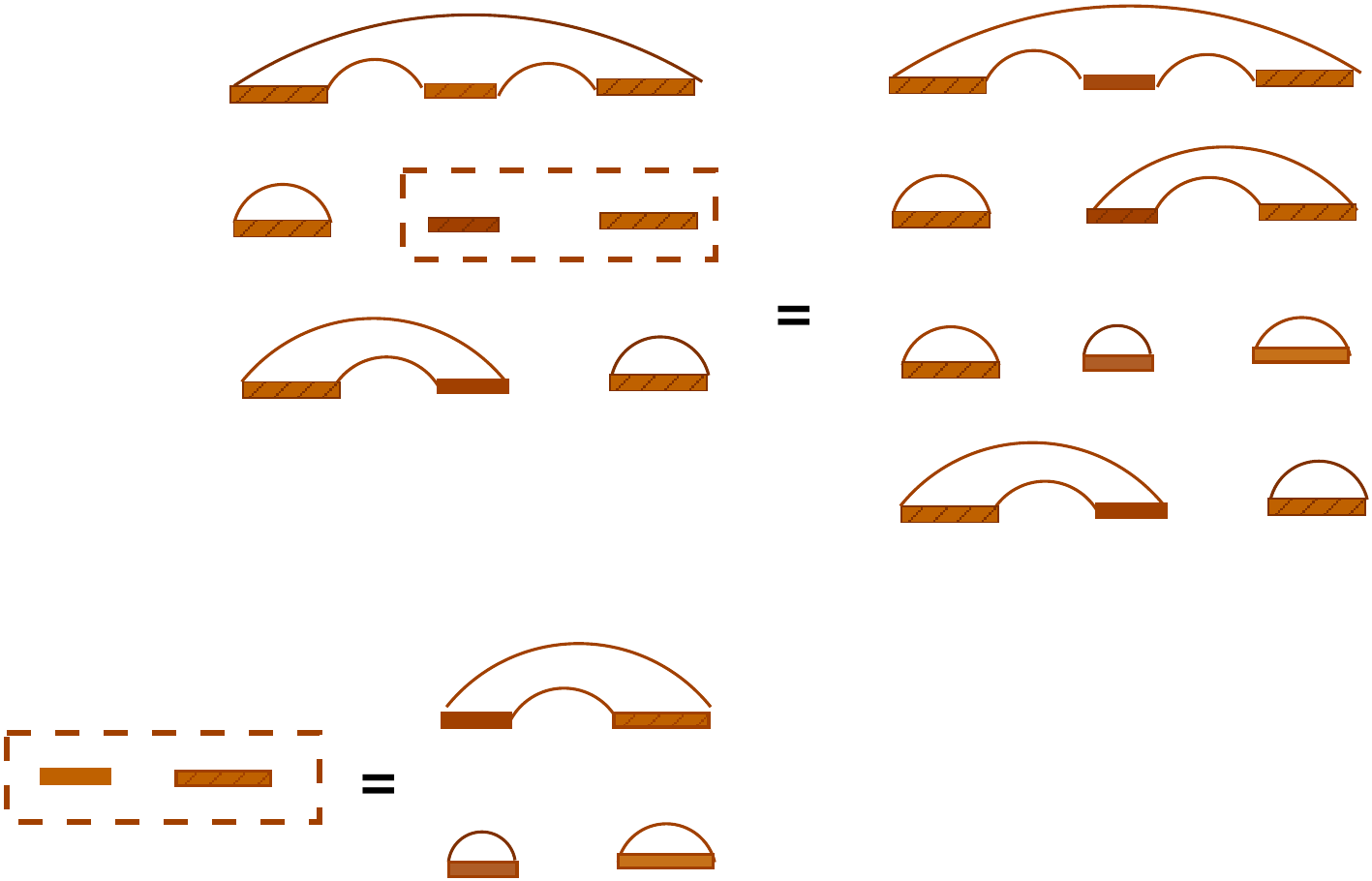}}
\put(-30,150){$(A\cup B\cup C)_p$}
\put(205,35){$(A,B)_{c,non-cr}$}
\put(205,0){$(A)||(B)$}
\end{picture}
\caption{The decompositions (\ref{decom}) for 3 and 2 segments that include only primitive diagrams. Selections of primitive diagrams are indicated by dashed lines.}
\label{Fig:Dec3p}
\end{figure}

In Fig.\ref{Fig:Dec3p}  we show primitive diagrams that contribute to calculation of the entanglement entropy for 2 and 3 segments.
There are three competing minimal surfaces  for 2 segments $A$ and $B$,  see  Fig.\ref{Fig:Dec2} below, one surface corresponds to a
 "disjoint" configuration $(A||B)$,
  the second to a "connected non-crossing"  one $(A,B)_{c,non-cr}$, the third to a "crossing" one
  $(A,B)_{c,cr}$, and we can write the symbolic representation
\be\label{decom2}
(A\cup B)=(A)||(B)+ (A,B)_{c,non-cr} +  (A,B)_{c, cr},\ee
here the brackets in $ (A\cup B)$  mean that we deal with corresponding diagrams and in the RHS of \eqref{decom2}
we list these diagrams.
Due to a possible  change of the leader in this competition the system may undergo  a phase transition under change the geometrical parameters, see  \cite{Alishahiha:2014jxa,Ben-Ami:2014gsa}.

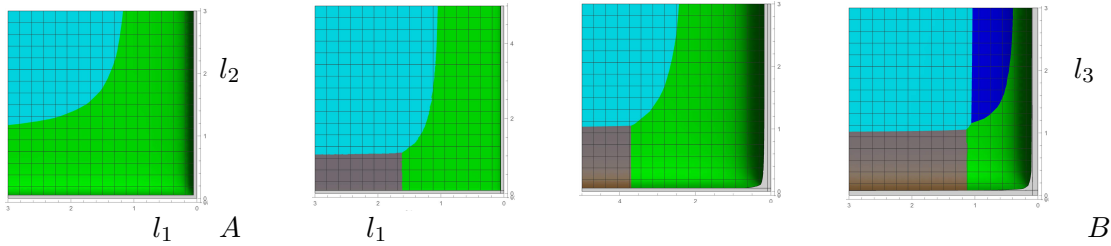
\begin{figure}[h!]
$$\,$$
\centering
\begin{picture}(250,100)
\put(-80,0){\includegraphics[scale=0.15]{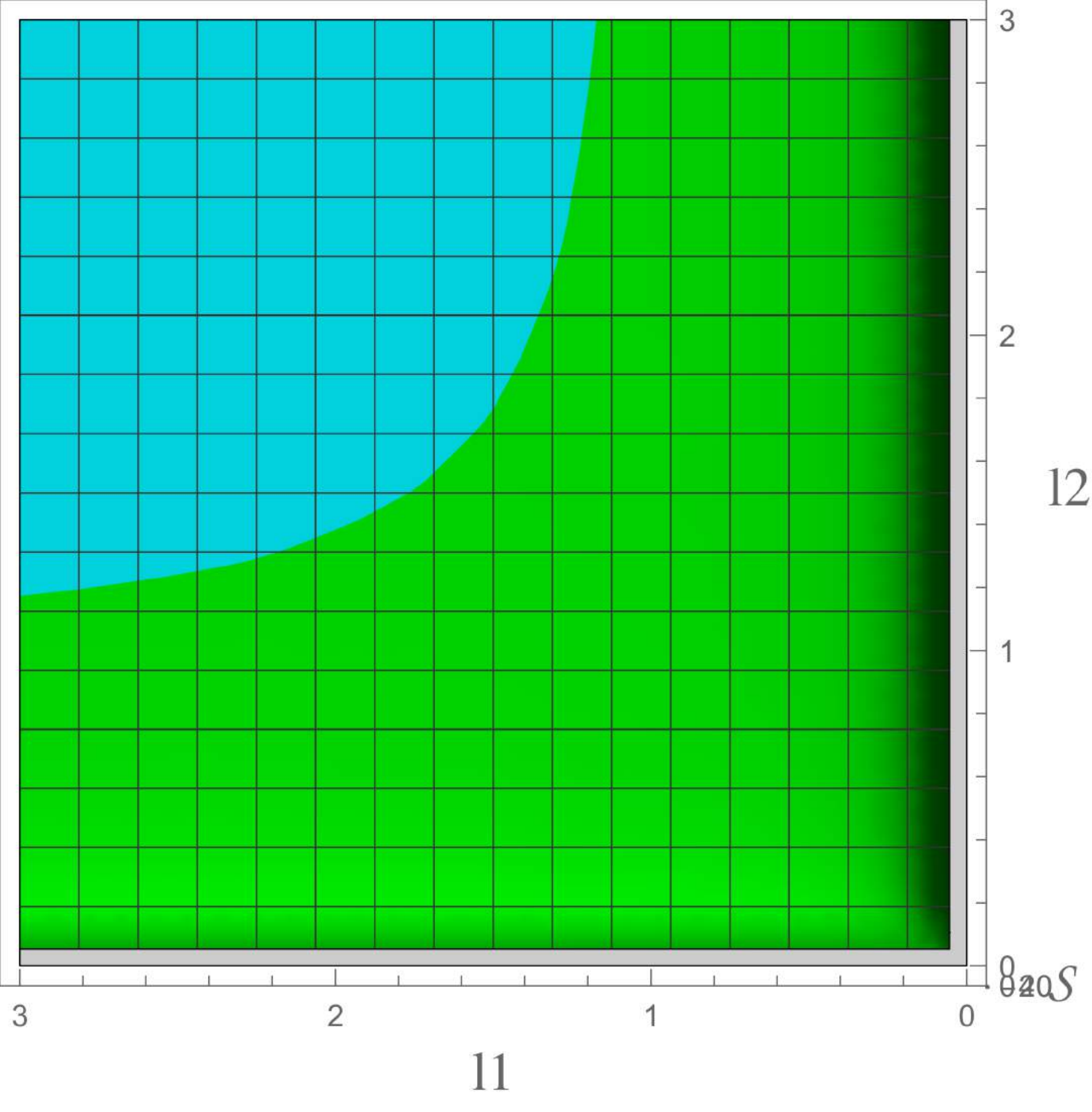}}
\put(-25,-10){$l_1$}
\put(0,50){$l_2$}
\put(0,-10){$A$}
\put(35,0){\includegraphics[scale=0.15]{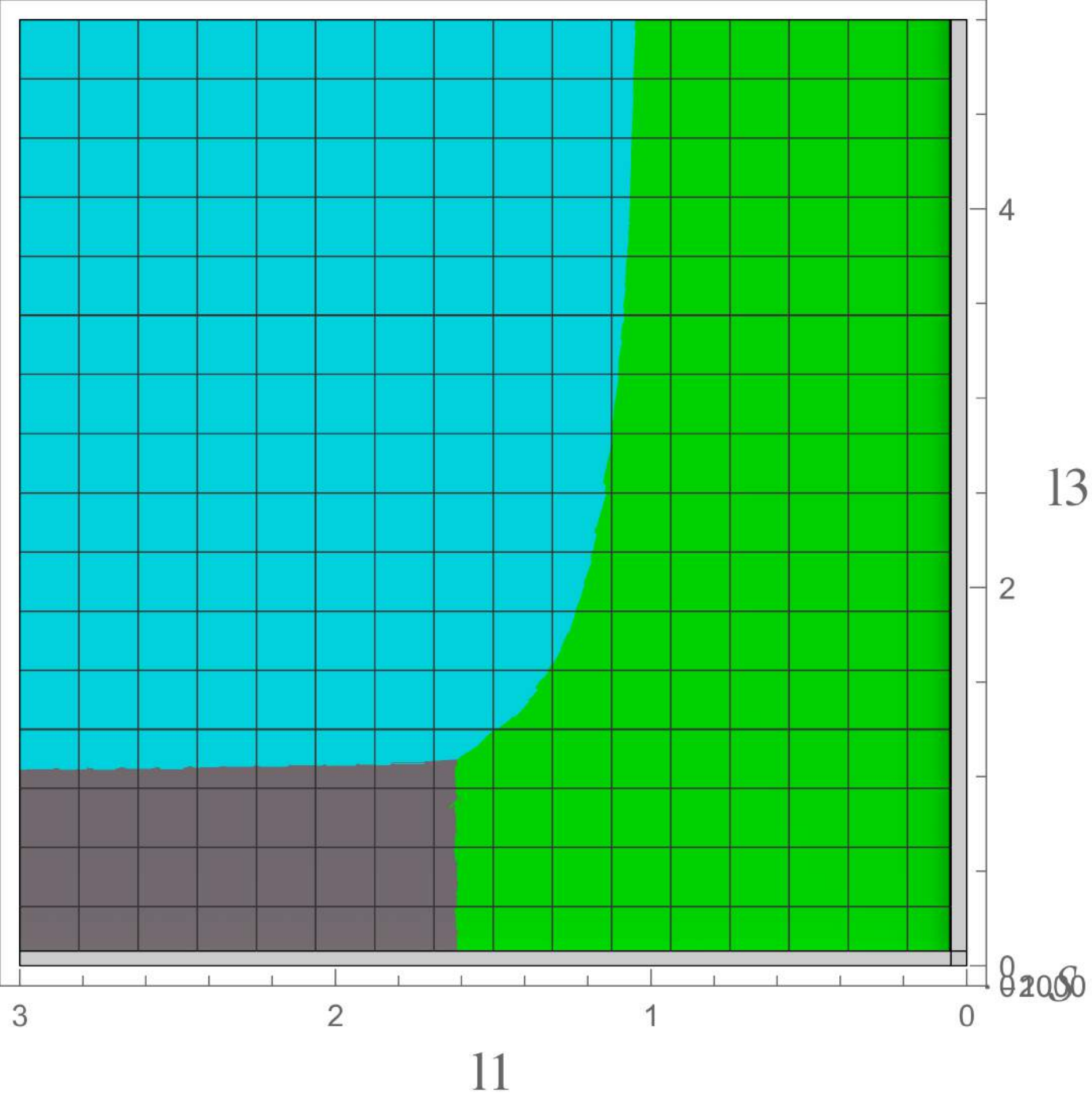}}
\put(135,3){\includegraphics[scale=0.15]{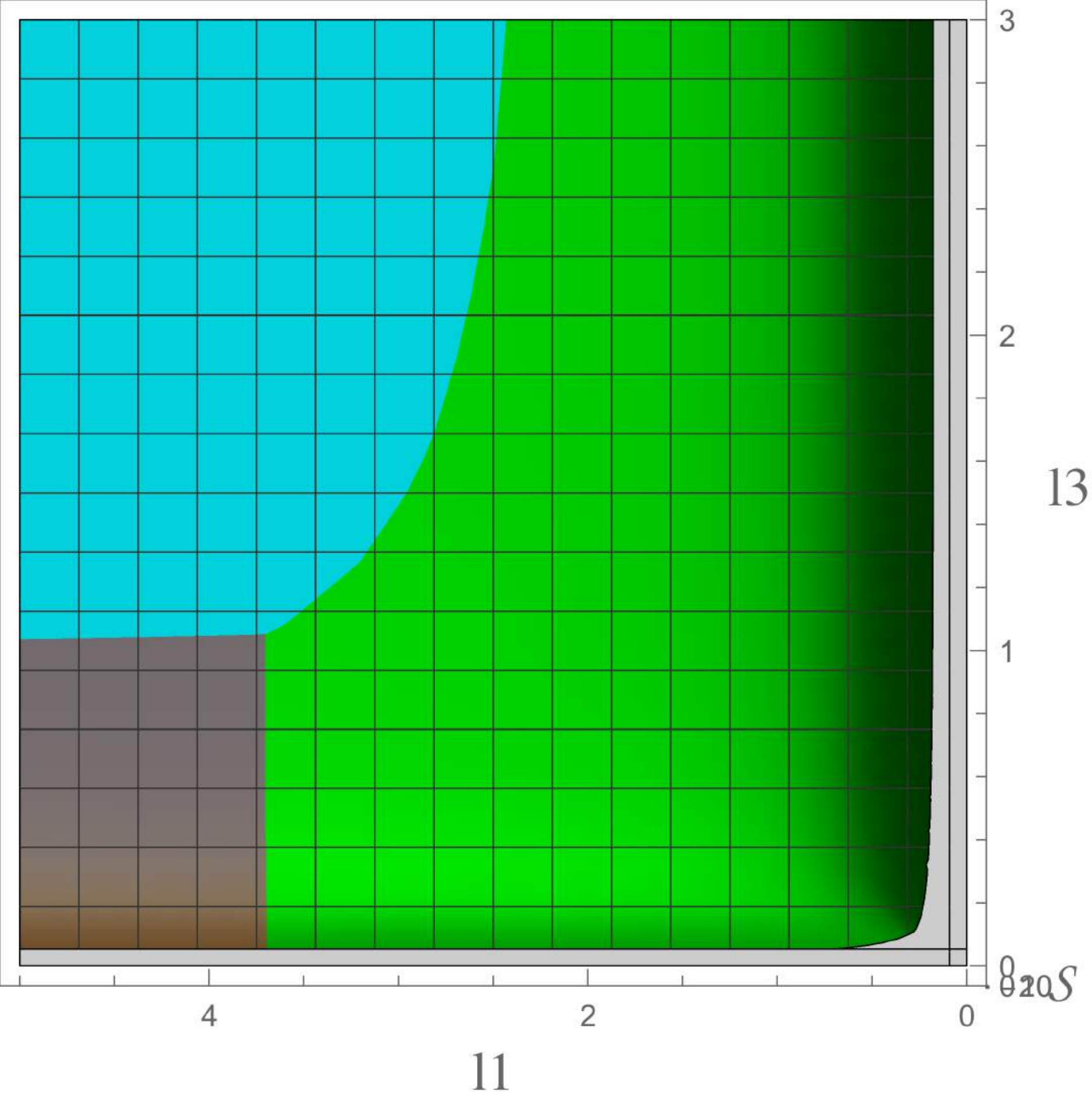}}
\put(235,0){\includegraphics[scale=0.15]{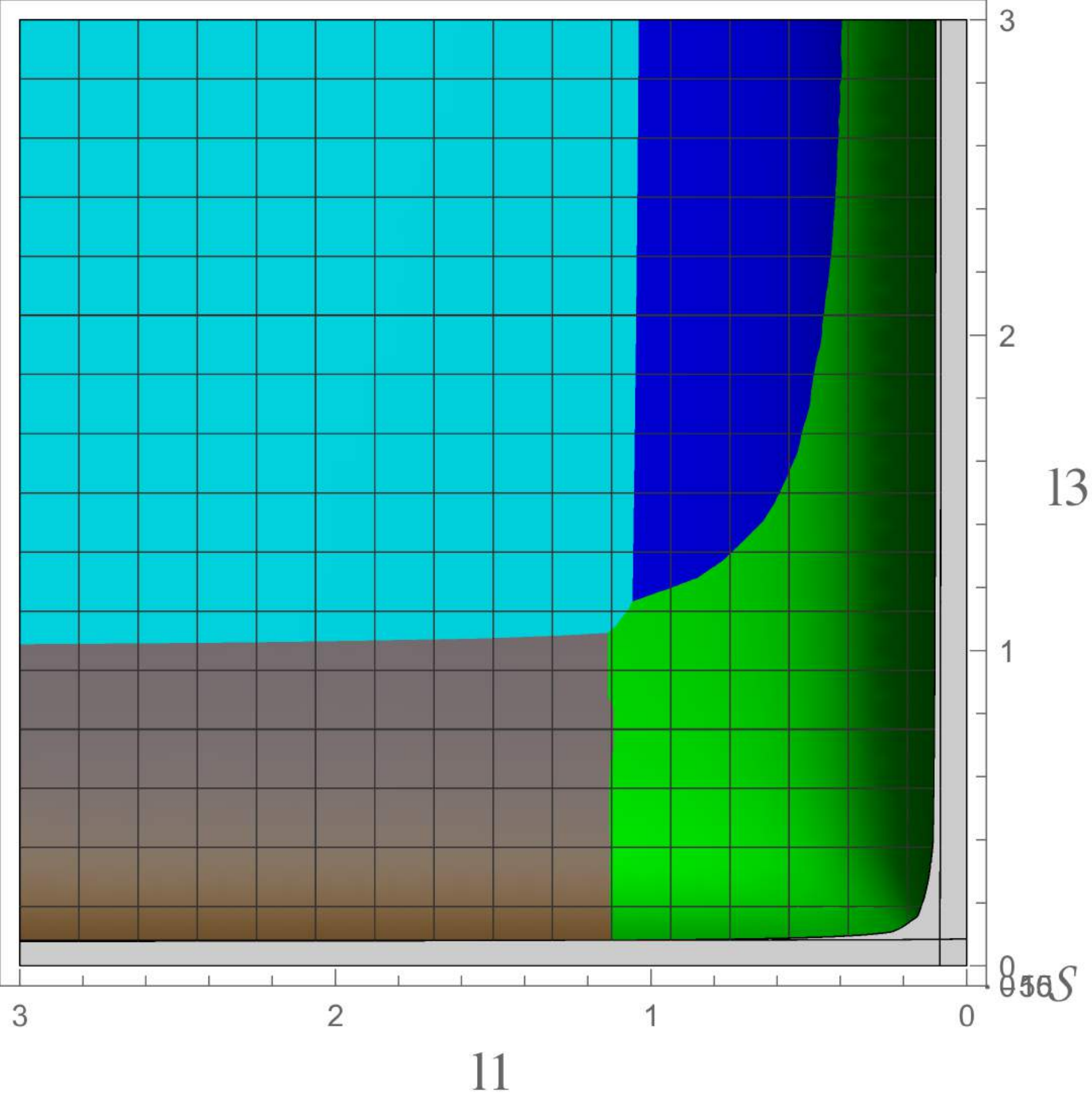}}
\put(55,-10){$l_1$}
\put(325,-10){$B$}
\put(320,50){$l_3$}\end{picture}
 \caption{{\bf A}. The phase diagrams for 2 strips, $A,B$ with unequal lengths $l_1,l_2$. The green color regions correspond to the bulk surface
  $(A)||(B)$, the cyan color regions correspond to the bulk surface $(A,B)_c$,  {\bf B}. The phase diagrams for 3 strips, $A,B,C,$ with unequal lengths $l_1,l_2,l_3$. The green color regions correspond to the bulk surface
  $(A)||(B)||(C)$, the blue color regions correspond to the bulk surface $(A)||(B\,C)_c$, the gray color regions correspond to the bulk surface $(A\,B)_c||(C)$
 and the cyan color corresponds to the bulk surface  $(A,B,C)_{c,non-cr}$. Different colors regions are separated by the curves that  are the transition lines. In the left plot $l_1=l_2$, in the middle plot $l_2=0.3\, l_1$ and  in the right plot $l_2=3\,l_1$
 and we vary $l_1$  and $l_2$ keeping the distances between segments fixed. All plots correspond to AdS$_4$.
 }
  \label{Fig:Ent3}
\end{figure}

For 3 segments one can write the full decomposition as
\bea
\label{decom3}
(A\cup B\cup C)=(A)||(B)||(C)+(A,B,C)_{c,cr}\\
+(A,B,C)_{c,non-cr}+(A,B)_c||(C)+
(A)||(B,C)_c+(A,\underbrace{{\bf B}}C)_c\nn\eea
The first term represents the totally disconnected diagram (the disconnected parts are separated by symbol $||$). The second and the third terms represent the crossing (`intersecting")  and non-crossing connected diagrams.
The  last three terms correspond to  diagrams with disjoint parts, wherein the last term corresponds to the rainbow diagram. The rainbow diagram contains  an "engulfed" subdiagram (in terminology used in \cite{Ben-Ami:2014gsa}), that is  indicated by writing the corresponding symbol  by the bold letter $ {\bf B}$ and the curl underbrace bracket.  We call the diagram corresponding to the  first, third, fourth and fifth terms as primitive diagrams. We use the same terminology also for $n$-segment cases.
Due to the competition between different terms in decomposition(\ref{decom3}) there are phase transitions. The number of
possible phases depends on number of diagrams that contribute to the entropy.
For 3  segments of equal  length and  with
unequal separations, the competition between  primitive diagrams gives rise to, generally speaking,  4 phases \cite{Allais:2011ys,Balasubramanian:2011at,Hayden:2011ag,Alishahiha:2014jxa,Ben-Ami:2014gsa}. The similar situation takes place for non-equal lengths
and in Fig.\ref{Fig:Ent3} we present the phase diagrams for $l_1\neq l_2\neq l_3$
and equal separations.

 \begin{figure}[h!]
$$\,$$
\centering
\begin{picture}(250,100)
\put(-50,0){\includegraphics[scale=0.2]{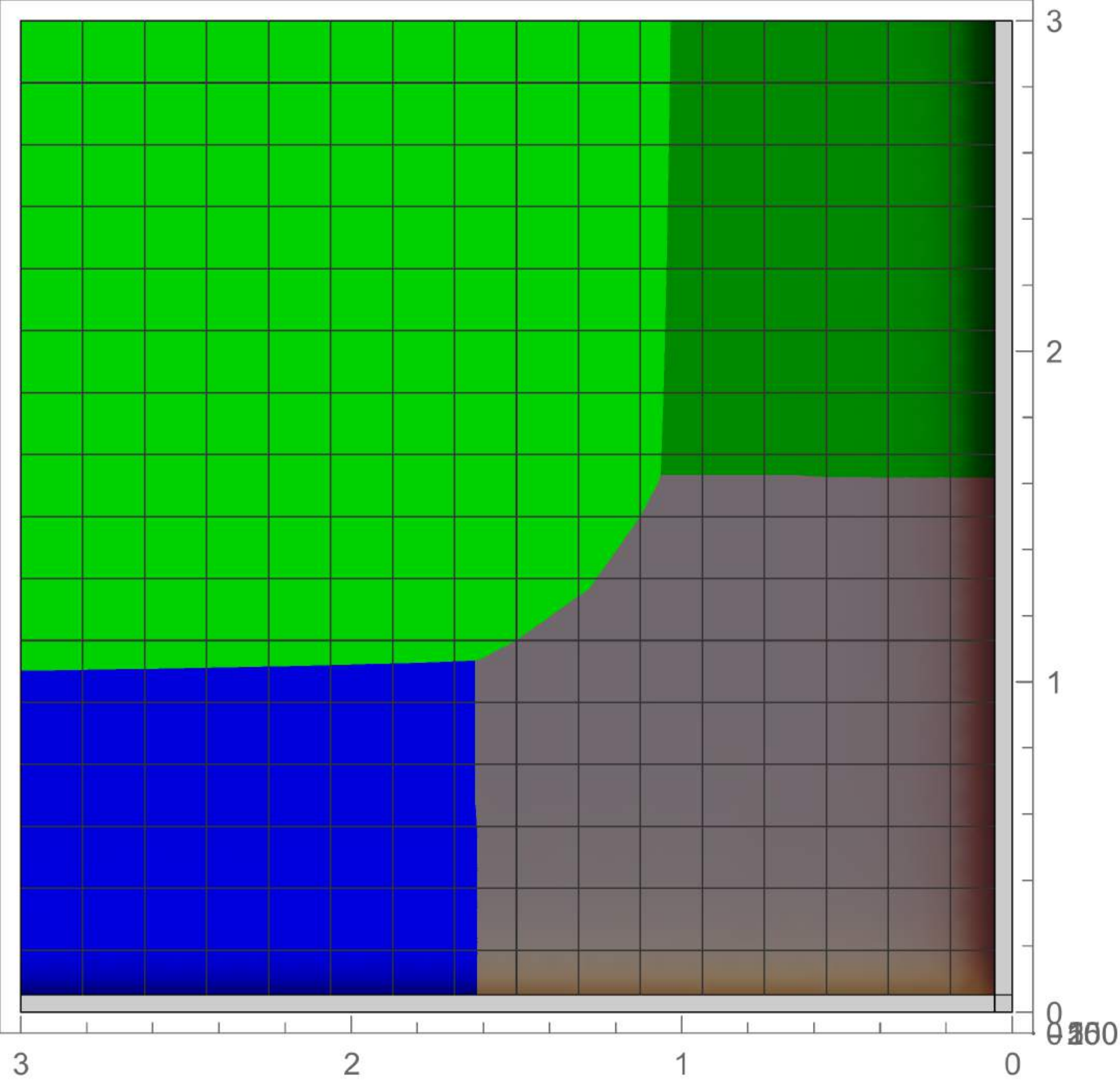}}
\put(75,0){\includegraphics[scale=0.2]{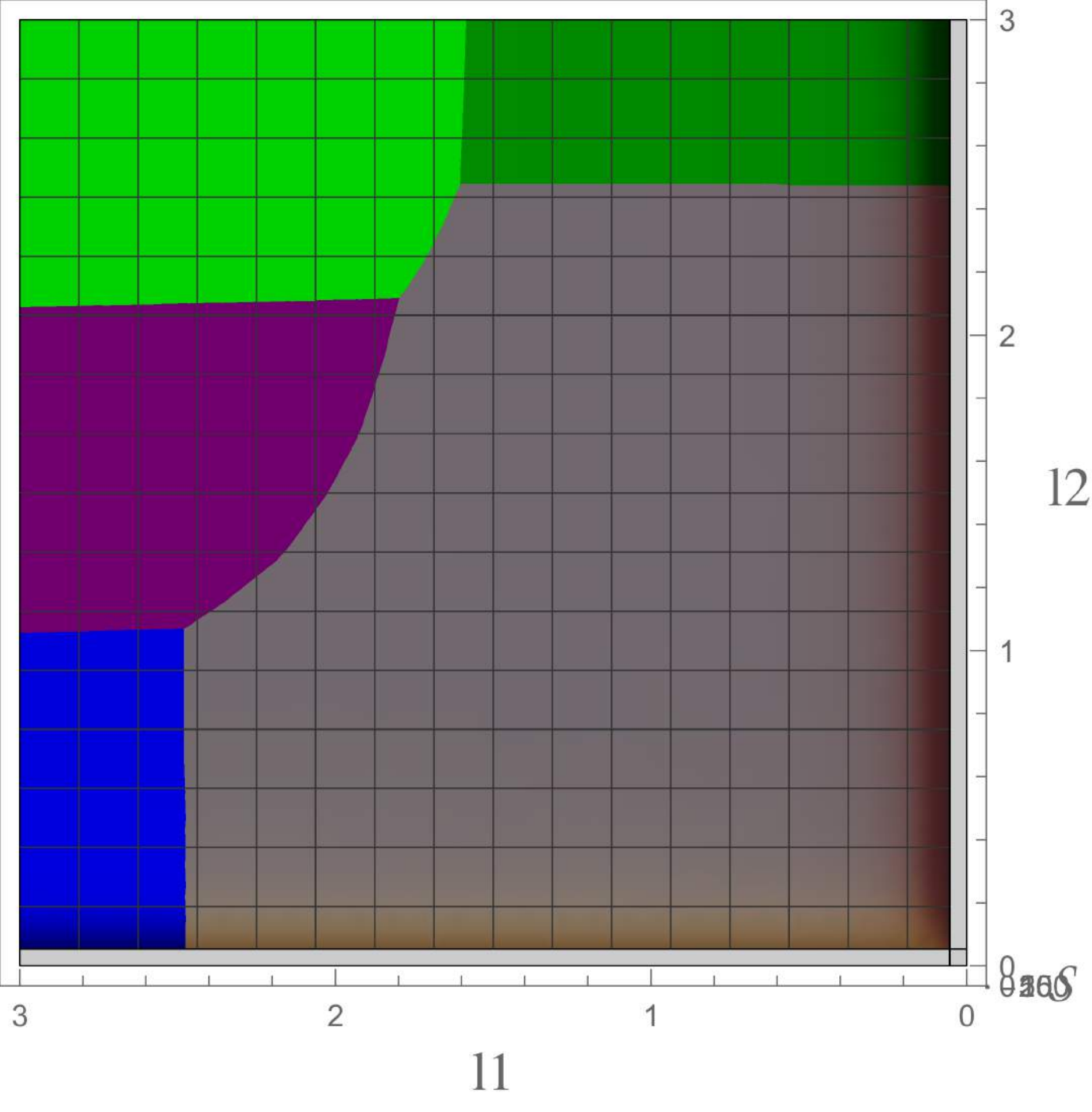}}
\put(190,0){\includegraphics[scale=0.2]{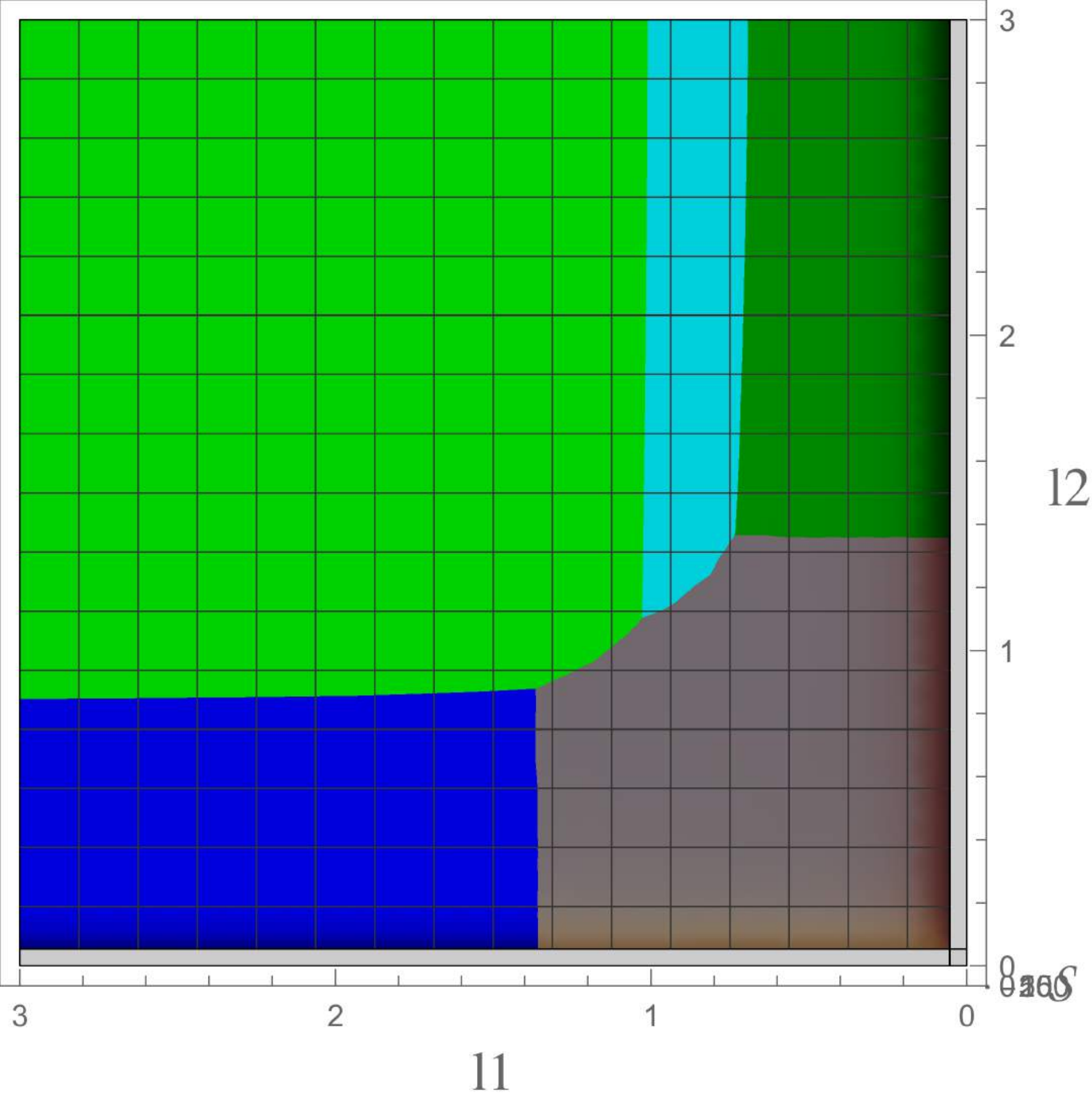}}
\put(0,-10){$l_1$}
\put(125,-10){$l_1$}
\put(310,50){$l_3$}\end{picture}\\
 \caption{ The phase diagrams for 4 strips, $A_1,A_2,A_3,A_4$ with unequal lengths $l_1,l_2,l_3,l_4$. The gray color regions correspond to the bulk surface
  $(A_1)||(A_2)||(A_3)||(A_4)$, the green color regions correspond to the bulk surface $(A_1,A_2,A_3,A_4)_{c,non-cr}$, the cyan color regions correspond to the bulk surface $(A_1)||(A_2\,A_3\,A_4)_c$
 and the purple color corresponds to the bulk surface  $(A_1A_2\,A_3)_c||(A_4)$.
 Dark green  and dark blue regions correspond to the bulk surface $(A_1)(A_2)||(A_3\,A_4)_c$
 and $(A_1,A_2)_c||(A_3)(A_4)$, respectively. Different colors regions are separated by the curves that  are the transition lines. In the left plot $l_1=l_2$, $l_3=l_4$, in the middle plot $l_2=0.5\, l_1$, $l_4=0.5\, l_3$ and  in the right plot $l_2=1.5\, l_1$, $l_4=1.5\, l_3$
 and we vary $l_1$  and $l_2$ keeping the distances between segments fixed. All plots correspond to AdS$_4$.
 }
  \label{Fig:Ent4}
\end{figure}
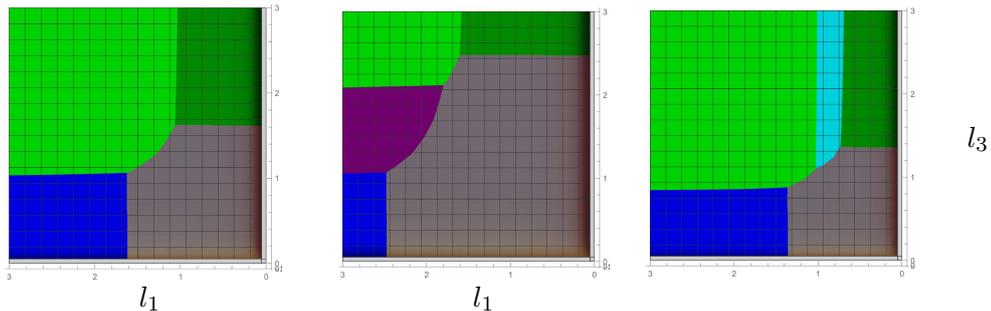

\begin{figure}[h!]
$$\,$$\\
\centering\includegraphics[scale=0.8]{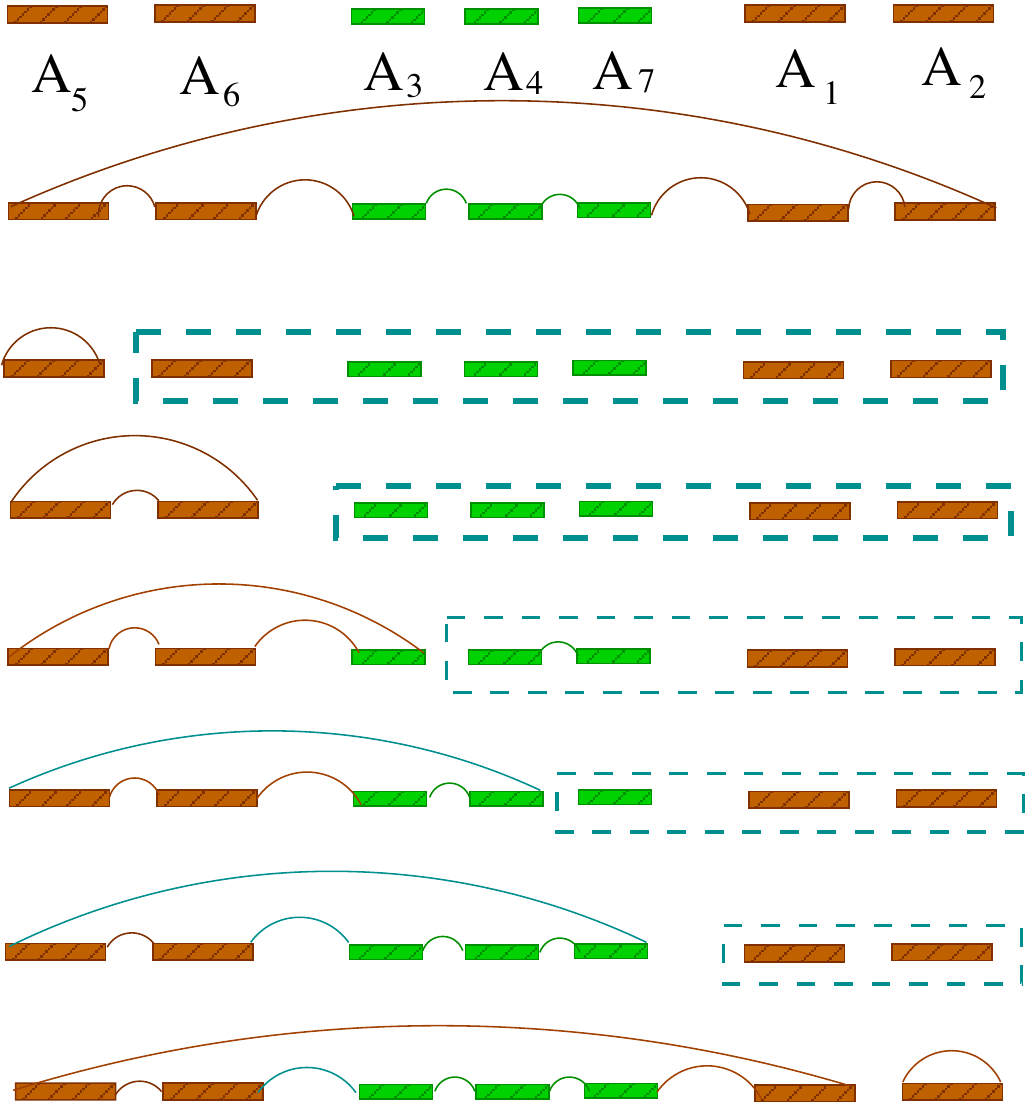}
\caption{Recursive procedure \eqref{decom}  to calculate $(A_5\cup A_6\cup A_3\cup A_4\cup A_7 \cup A_1\cup A_2 )_{p}$
}\label{Fig:EE-7-rec}
\end{figure}

 To calculate ${\cal S}(A_1\cup A_2\cup ...\cup A_n)$ taking into account only the primitive diagrams we use the following representation
\be\label{decom}
(A_1\cup A_2\cup ...\cup A_n)_{p}=(A_1, A_2, ...A_n)_c+\sum _{j=1}^{n-1}(A_1, A_2, ...A_j)_c\,\star\,
(A_{j+1}\cup A_2\cup ...\cup A_n)_{p}\ee
where $(A_1, A_2, ...A_j)_c$ is the bulk connected  surface ending on segments $A_1, A_2, ...A_n$
constructed without any crossing lines. There is only one such surface for a given number of segments. In \ref{decom}
there is the symbol $\star$ that means that  sub-diagrams on the
left and on the right of the symbol  compose the same n-segment diagram. We will omit this symbol in what follows.
 The decomposition (\ref{decom}) for 3 and 7 segments is shown in Fig. \ref{Fig:Dec3p} and Fig. \ref{Fig:EE-7-rec}, respectively (or the moment, the reader can ignore  the different  colors  for segments and connecting lines  in
 Fig.\ref{Fig:EE-7-rec}). In these pictures  contributions to $(A_1\cup A_2\cup ...\cup A_n)_{p}$  are shown  by the enveloping rectangles.

 To calculate ${\cal S}(A_1\cup A_2\cup ...\cup A_n)$ taking into account also the Boltzmann rainbow diagrams one can use the following representation
\be\label{decomB}
(A_1\cup A_2\cup ...\cup A_n)_{B}=(A_1, A_2, ...A_n)_{cwB}+\sum _{j=1}^{n-1}(A_1, A_2, ...A_j)_{cwB}\,\star\,
(A_{j+1}\cup A_2\cup ...\cup A_n)_{B}\ee
where $(A_1, A_2, ...A_n)_{cwB}$  are connected diagrams with Boltzmann insertions,
\bea\nn
(A_1, A_2, ...A_j)_{cwB}&=&\sum _{\substack{\{i_q\},\{k_q\} \\\nn
\sum _{q=1}^{l}(i_{q}+k_q)\leq j-2}}
\,\Big(A_1,... A_{i_1},\,\underbrace{
({\bf A}_{i_{1}+1}\cup {\bf A}_{i_{1}+2}\cup...{\bf A}_{i_{1}+k_{1}})_B}\,A_{i_{1}+k_{1}+1},...\\ \nn\\
&\,&A_{i_{1}+i_{2}+k_{1}},\,\underbrace{({\bf A}_{i_{1}+i_{2}+k_{1}+1}\cup {\bf A}_{i_{1}+i_{2}+k_{1}+2}\cup...{\bf A}_{i_{1}+i_{2}+k_{1}+k_{2}})_B}\,A_{i_{1}+i_{2}+k_{1}+k_{2}+1},
 ... , \nn\\\nn\\
&\,&\,\underbrace{{\bf A}_{1+\sum _{q=1}^{l}i_{q}+\sum _{q=1}^{l-1}k_q}\cup ...{\bf A}_{\sum _{q=1}^{l}(i_{q}+k_q)})_B}\,
A_{1+\sum _{q=1}^{l}(i_{q}+k_q)},...A_{j}\Big)_c\nn\\\label{cwB}\eea
Here the Boltzmann insertions are indicated by  the bold letters and curl underbrace brackets. Note, that $cwB$ diagrams are not connected, but if we remove the Boltzmann insertions
we are left with the connected diagrams. In particular, if we remove the Boltzmann insertions in the RHS of (\ref{cwB}) we left with connected diagrams
\bea\nn
\,(A_1,... A_{i_1},\,A_{i_{1}+k_{1}+1},...
&\,&A_{i_{1}+i_{2}+k_{1}},\,A_{i_{1}+i_{2}+k_{1}+k_{2}+1},\,
A_{1+\sum _{q=1}^{l}(i_{q}+k_q)},...A_{j})_c\eea
Comparing (\ref{decom}) with (\ref{decomB})  and (\ref{cwB})  we see that the Boltzmann  diagrams are composed by insertions of low order Boltzmann  diagrams
to primitive diagrams.

 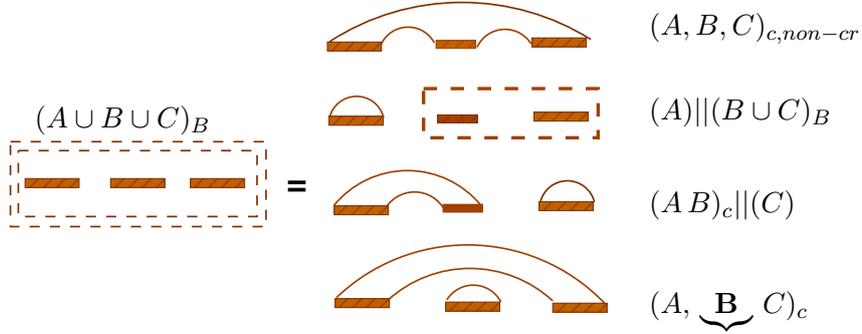
\begin{figure}[h!]
\begin{picture}(250,230)
\put(290,130){$(A,B,C)_{c,non-cr}$}
\put(290,63){$(A\,B)_c||(C)$}
\put(290,98){$(A)||(B\cup C)_B$}
\put(50,25){\includegraphics[scale=0.7]{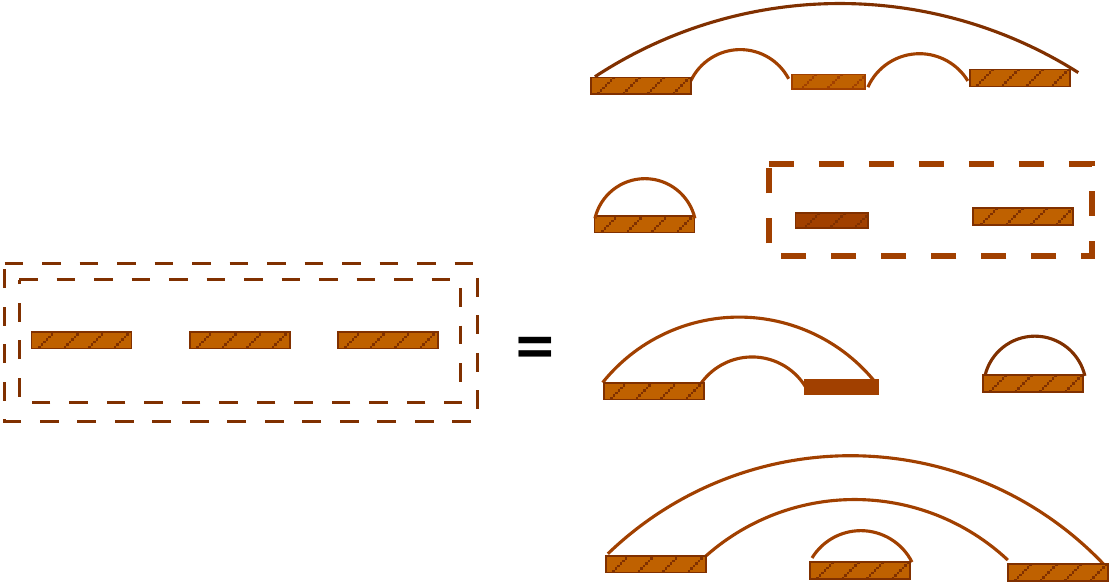}}
\put(60,95){$(A\cup B\cup C)_B$}
\put(290,25){$(A,\,\underbrace{{\bf B}}\,C)_c$}
\end{picture}
\caption{The decomposition (\ref{cwB}) for 3  segments. Selections of Boltzmann diagrams are indicated by double dashed lines.
There are no difference between the primitive and the Boltzmann selections  for two segments, but there is the difference for 3 and more segments. }
\label{Fig:Dec3B}
\end{figure}

\newpage
\subsection{Holographic  mutual information }\label{HMI-phases}

The mutual information $I(A;B)$ of two entangling regions $A$ and $B$ is defined by \eqref{MI-general}:
\be\label{HMI}
I(A;B)=S(A)+S(B)-S(A\cup B),\ee
where $S(A\cup B)$ is the entropy of the union of two regions.
Holographic definition of the mutual information $I(A;B)$ just means that we calculate
all terms in the RHS of (\ref{MI}) holographically.

To find $S(A\cup B)$ one has to find the minimal surface between two competing ones, "joint" and "disjoint" surfaces, as shown in Fig.\ref{Fig:Dec2}. The winner, in the case of vacuum background, depends only on the ratios of segments to the distance between them. It is obvious that
$I(A;B)$  is equal to zero for the case when the minimal surface is realized on the "disjoint" configuration.
 Generally speaking, $I(A,B)$ undergoes a first order phase transition as one increases the
 distance between two strips \cite{Hubeny:2007xt,Headrick:2010zt,Fischler:2012uv}

\bea\label{MI}
I(A;B)&=&\left\{
\begin{array}{ccc}{\cal I}(A;B),
  &  \,\,{\mbox {if}}  \,\,\,& {\cal I}(A;B)\geq 0  \\
 0 & {\mbox {if}}  &  {\cal I}(A;B) \leq 0
\end{array}
\right.\\\nn\\
{\cal I}(A;B)&\equiv&S(A)+S(B)-S(A+x+B)-S(x),
\eea
where $S(A)$ are areas of surfaces entangling the belt (stript) with segment $A$. In the case of the AdS$_d$ background ${\cal I}(A;B)$  is given by
\be
\frac{{\cal I}(A;B)_{vac}}{2L^{d-2}c_0}=-\frac{1}{\ell_1^{d-2}}-\frac{1}{l_1^{d-2}}+\frac{1}{(\ell_1+\ell_2+x)^{d-2}}+\frac{1}{x^{d-2}},
\ee
here $d>2$.
The region where the mutual information  is nonzero depends on the ratios of parameters  $\ell_1/x,\ell_2/x$ in  the case of $AdS_d$,
and on  the ratios of parameters  $\ell_1/x,\ell_2/x$ and  $z_h/x$, $z_h=m^{-1/d}$, in the case of AdS$_d$ black brane, see \ref{Fig:Reg-d3-static} and Fig.\ref{Fig:MI-d3-static}. From these plots
we clearly see that the mutual information of two static belts is maximal when the system is in the vacuum state.

\begin{figure}[h!]
\centering
\includegraphics[scale=0.3]{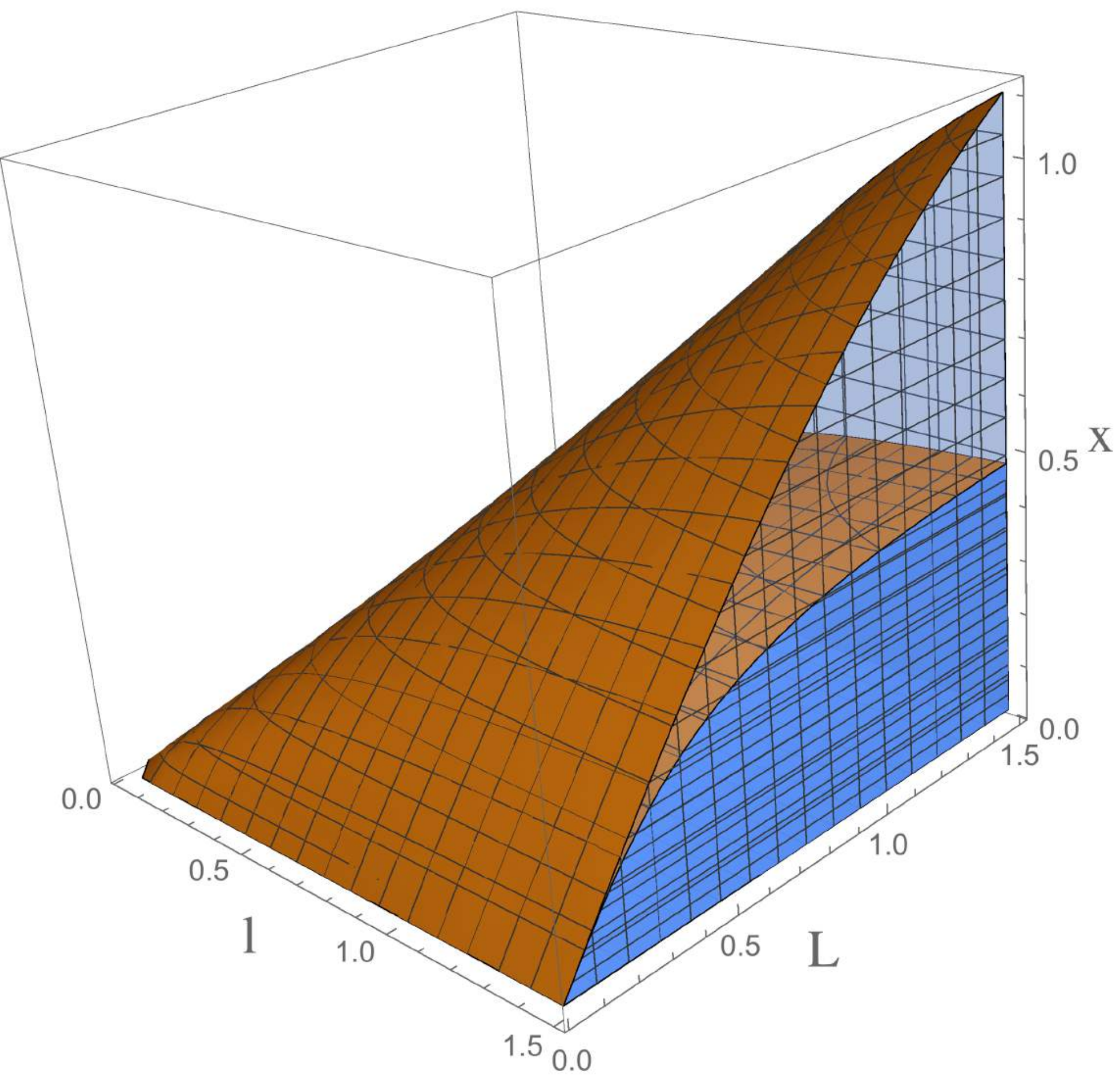}
 \caption{The regions of non-zero  holographic mutual information for two strips  with the widths $\ell$
 and $L$ and distance $x$ for  two thermal states whose gravity duals are given by the AdS$_4$ black brane metric (\ref{BB}) with $m=0.25$ and $m=1$. The region corresponding to $m=1$ is inside  the region corresponding to  $m=0.25$}
 \label{Fig:Reg-d3-static}
 \end{figure}
 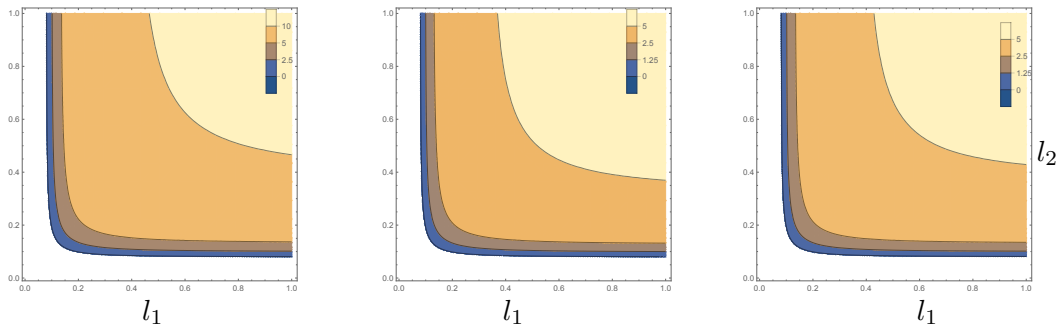
\begin{figure}[h!]
$$\,$$
\centering
\begin{picture}(250,100)
\put(-50,0){\includegraphics[scale=0.2]{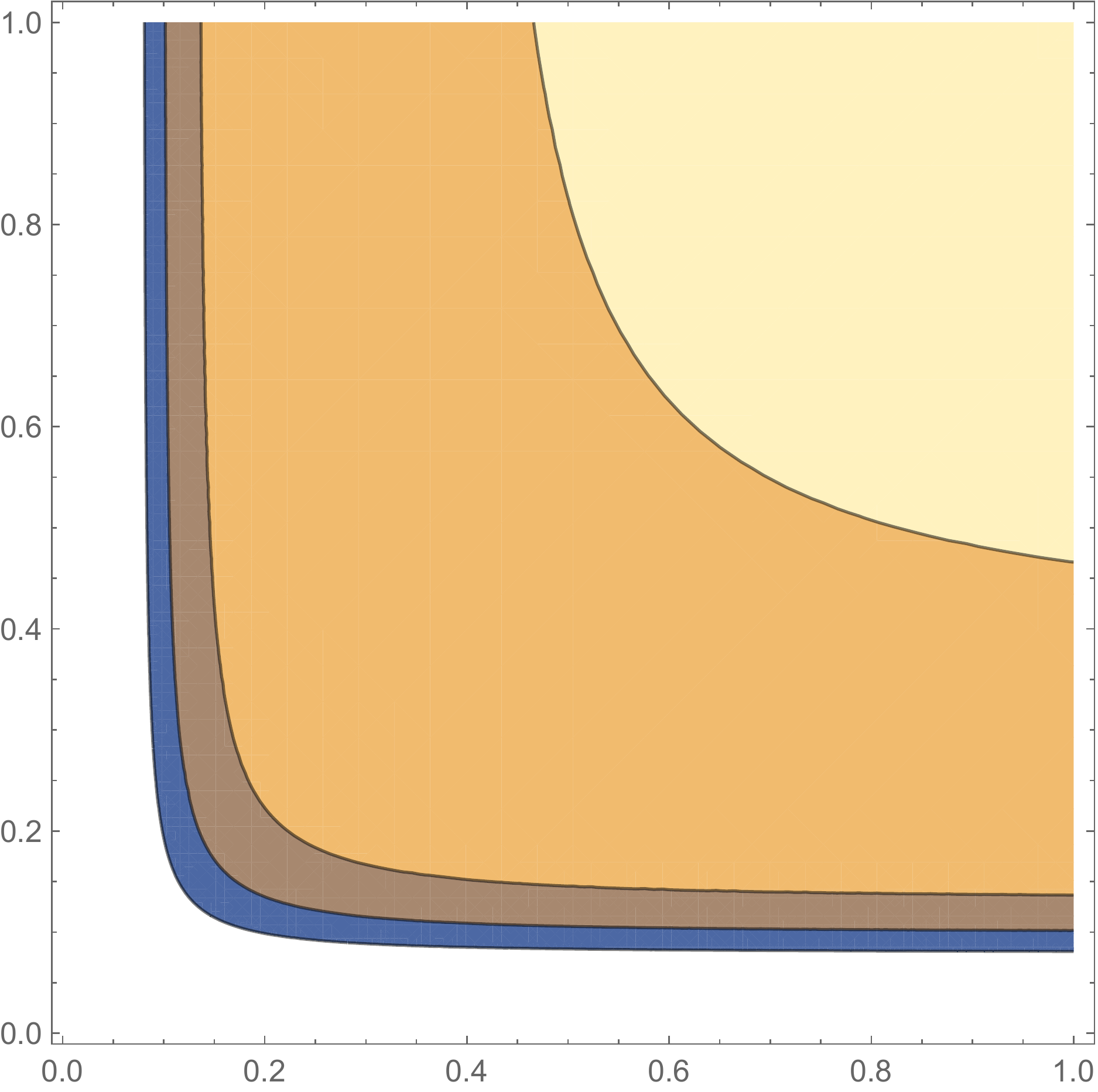}}
\put(45,70){\includegraphics[scale=0.2]{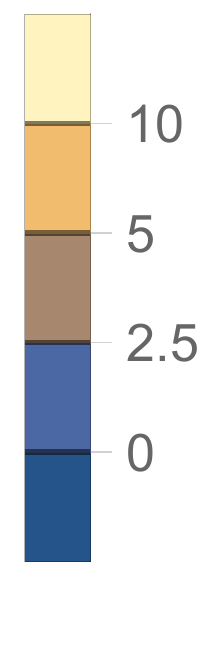}}
\put(90,0){\includegraphics[scale=0.2]{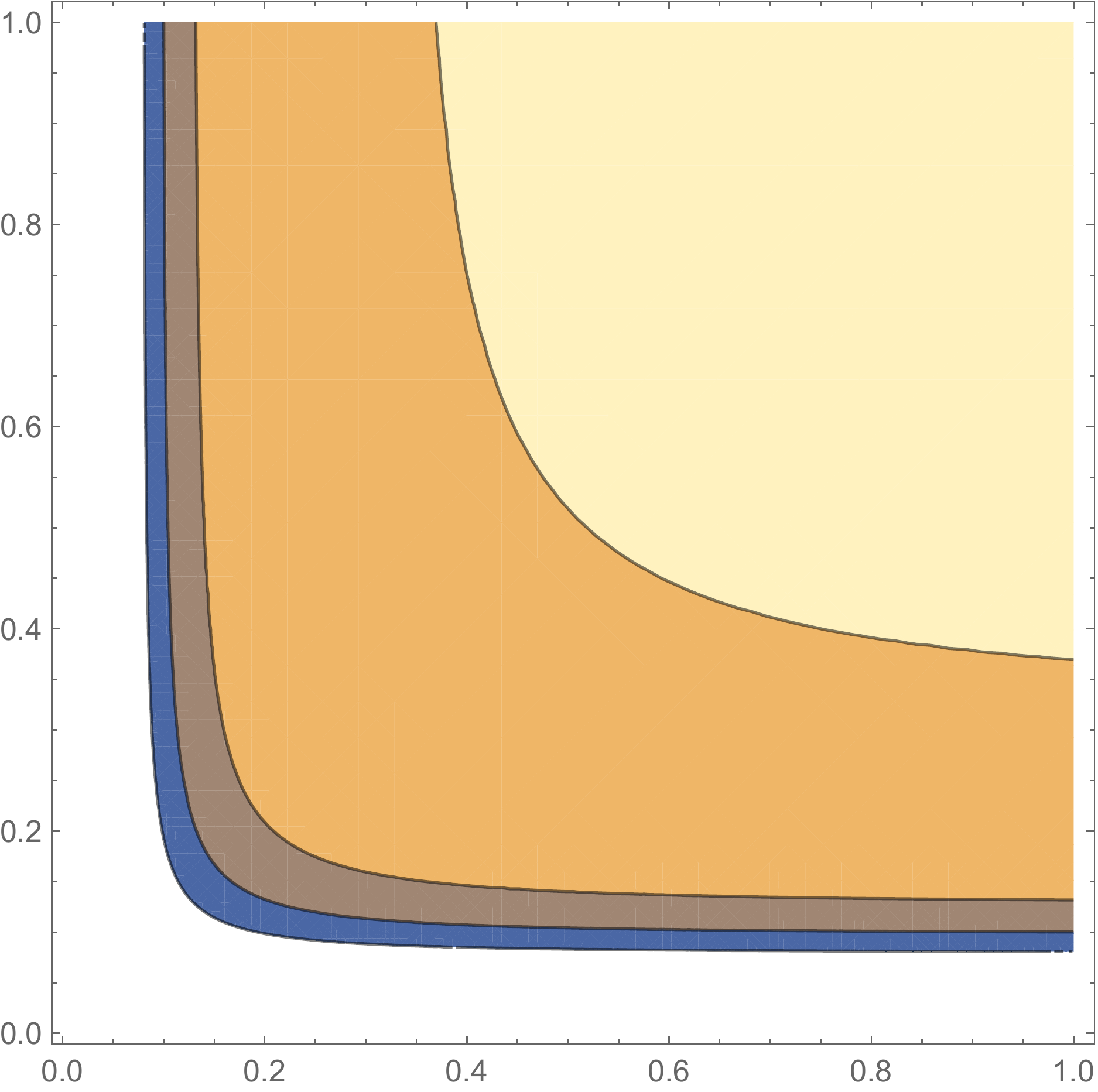}}
\put(180,70){\includegraphics[scale=0.2]{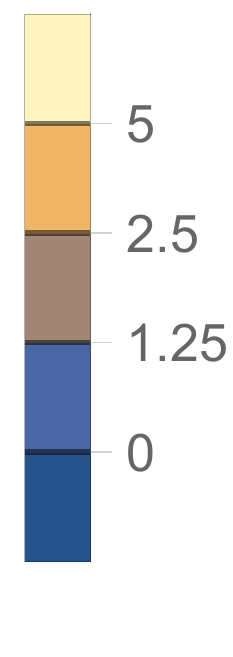}}
\put(225,0){\includegraphics[scale=0.2]{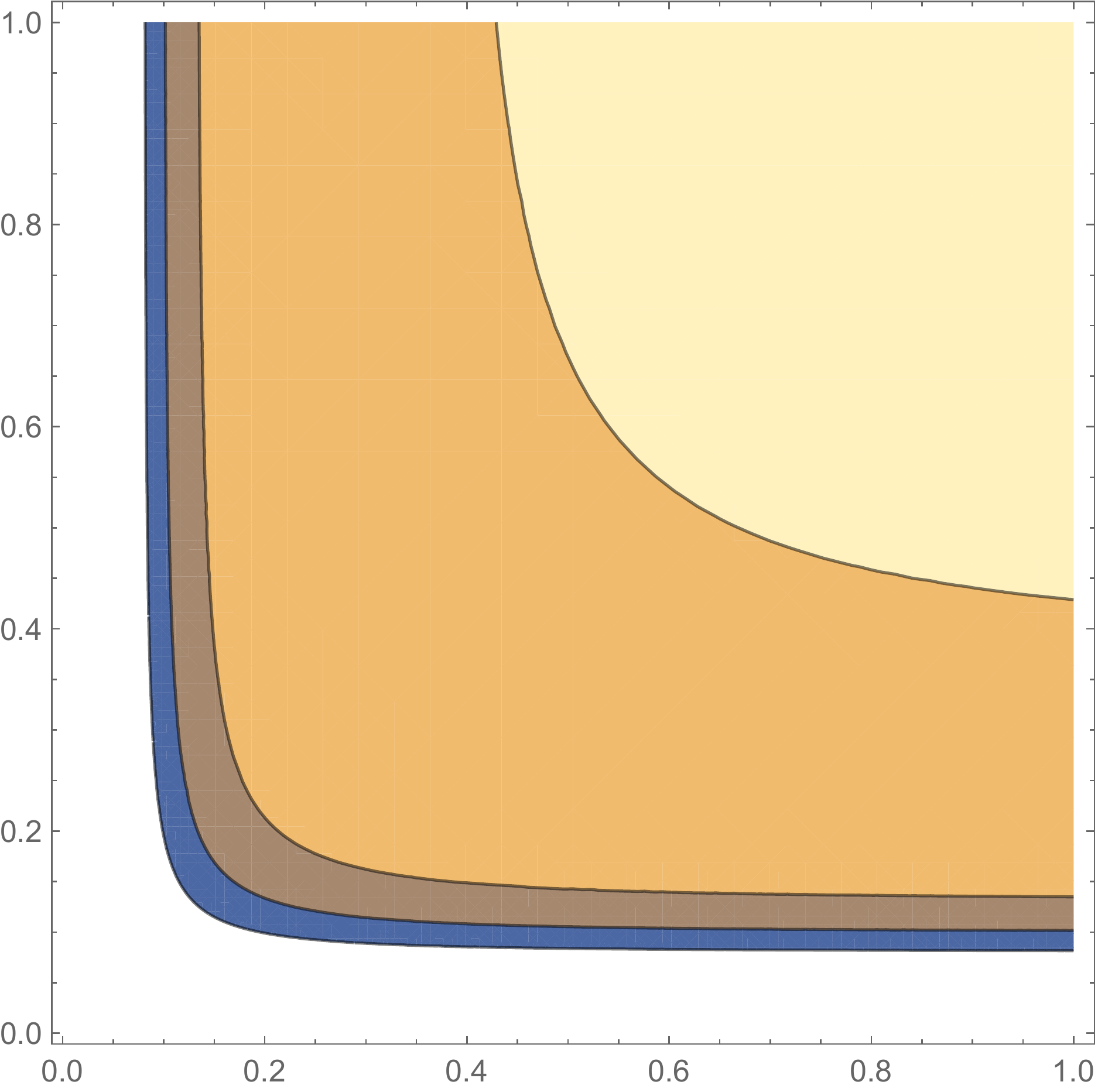}}
\put(320,65){\includegraphics[scale=0.2]{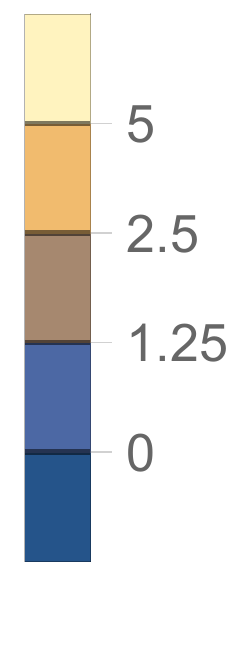}}
\put(0,-10){$l_1$}
\put(135,-10){$l_1$}
\put(290,-10){$l_1$}
\put(335,50){$l_2$}\end{picture}\\
 \caption{The density plots of the holographic mutual information for two strips  with the widths $\ell_1$
 and $\ell_2$ and fixed distance $x=0.08$ for the static vacuum state whose gravity dual is the AdS$_4$ metric (the left plot) and two thermal states whose gravity duals are provided by the AdS$_4$ black brane metric (\ref{BB}) with $m=0.25$ and $m=1$, respectively (the middle and the right plots)}
 \label{Fig:MI-d3-static}
 \end{figure}

%%%%%%%%%%%%%%%%%%%%%%
%%%%%%%\input{MI-static.tex}
%%%%%%%%%%%%%%%%%%%

\section{Holographic studies  of  simplest FMO complex}\label{HMI-simple-FMO}
\subsection{Holographic entanglement entropy for two site system during a quench at nonzero temperature.}
\label{HMI-simple-FMO-m0}
\begin{figure}[h!]
\centering
photon\\
 \includegraphics[scale=0.6]{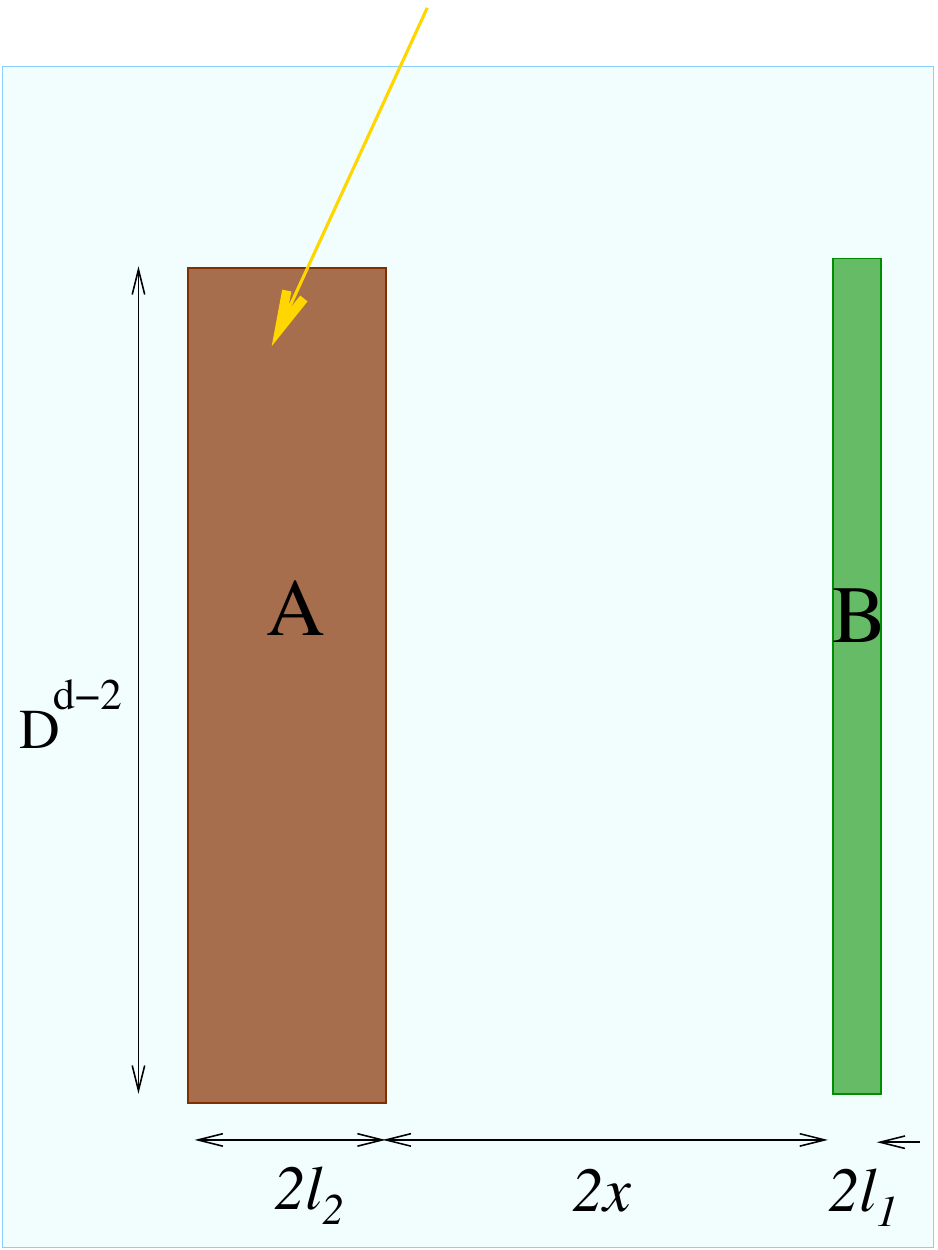}
 \caption{Schematic picture of locations of the excited complex (A) and the out site (B), that sends the information to  the reaction center,  and the infalling photon.}
 \label{Fig:2segm}
\end{figure}

As has been mention in Sect.\ref{Sect:reductions},  the simplest model of LHC is the two systems model. We can  consider two parallel infinite strips with non-equal widths $2\ell_1$ and $2\ell_2$
separated by a distance $2x$ in a $d$-dimensional
field theory as depicted in Fig.\ref{Fig:2segm} as a simplest model of LHC and find the mutual information induced by
the coming photon. This coming photon we model by the Vaidya shell.

The  holographic  mutual information for two
parallel strips in static backgrounds and in the Vadya AdS has been studied in several papers \cite{Hubeny:2007re,Hayden:2011ag,Balasubramanian:2011at,Allais:2011ys,Alishahiha:2014jxa,1602.07307}. In particular for equal widths of segment in certain limits the analytical results have been obtained \cite{Alishahiha:2014jxa}.
The specificity of our consideration here is that we want to consider the time dependence of the mutual information  during a global quench  of a  medium that already has  high temperature. The corresponding holographical model
is given by the Vaidya shell  propagating in the  AdS$_{d+1}$ black brane background. We also compare the obtained time dependence with the  time dependence of the thermalization process started from the zero temperature.

  \begin{figure}
$$\,$$
\centering
  \includegraphics[scale=0.9]{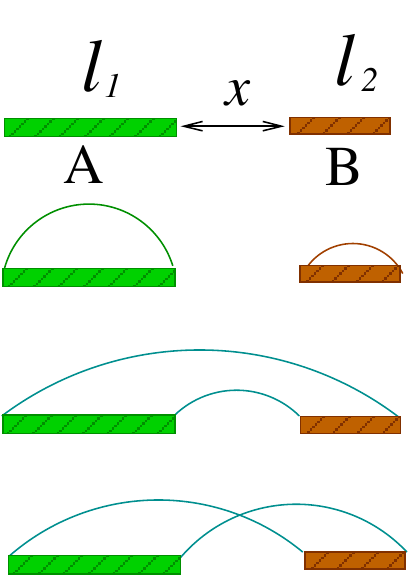}
 \caption{Illustration of bulk surfaces anchored to the boundaries of disjoint regions A and B. The surfaces started and ended on the boundaries of the same region are shown by the same color as the region, meanwhile the surfaces connecting the ends of different regions are shown by blue lines. In the case of
 considered here holographic models we do not obligate to take into account the diagram with crossing lines  depicted in the last line.}\label{Fig:Dec2}
 \end{figure}

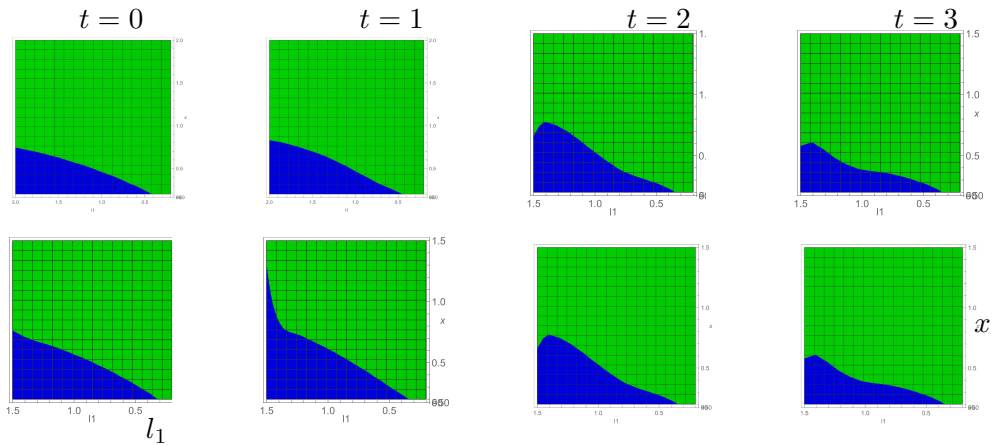
\begin{figure}[h!]
\centering
$\,\,\,\,\,\,$$\,\,\,\,\,\,$$\,\,\,\,\,\,$$\,\,\,\,\,\,$$\,\,\,\,\,\,$
\begin{picture}(250,200)
\put(-80,80){\includegraphics[scale=0.12]{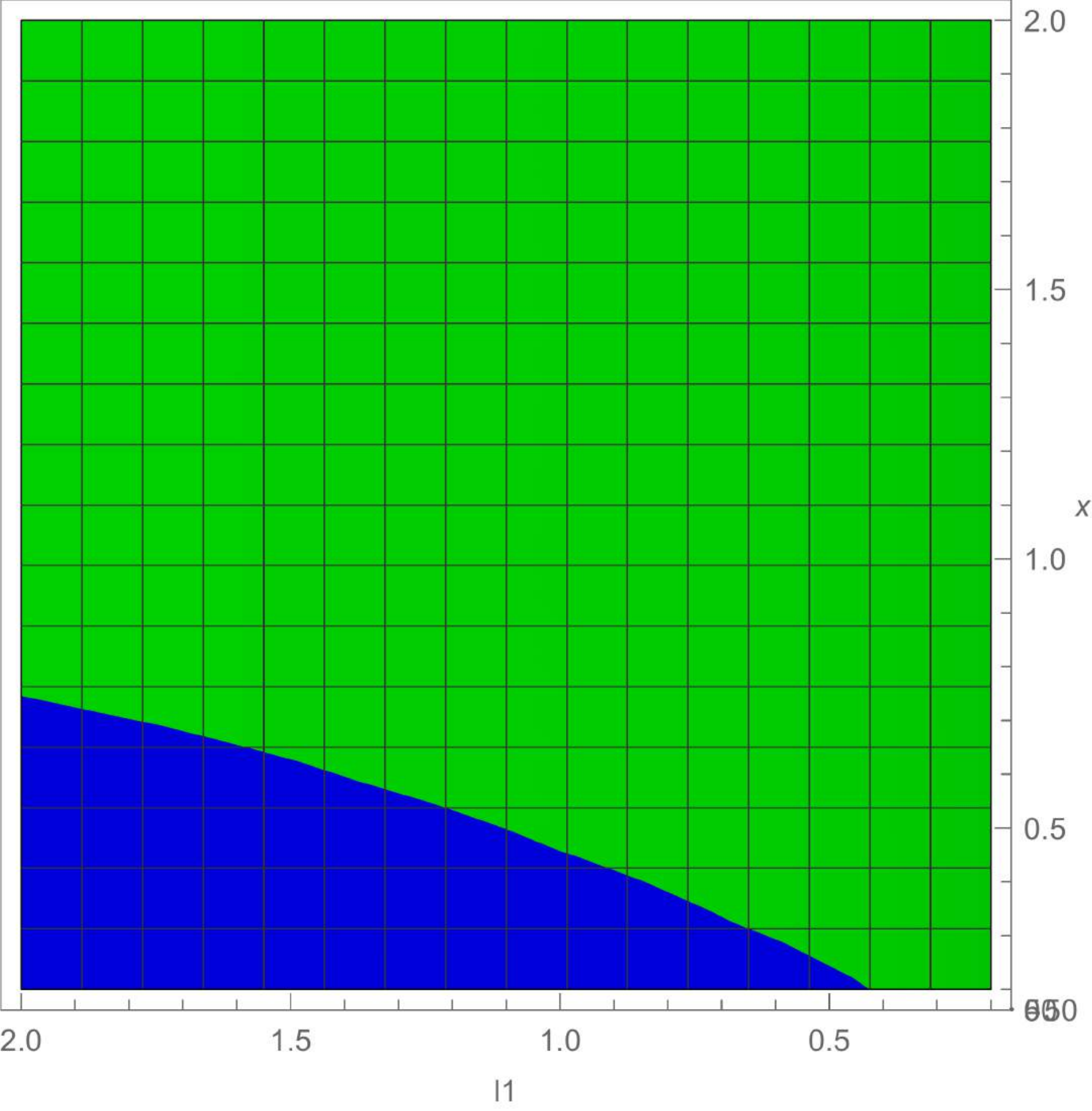}}
\put(15,80){\includegraphics[scale=0.12]{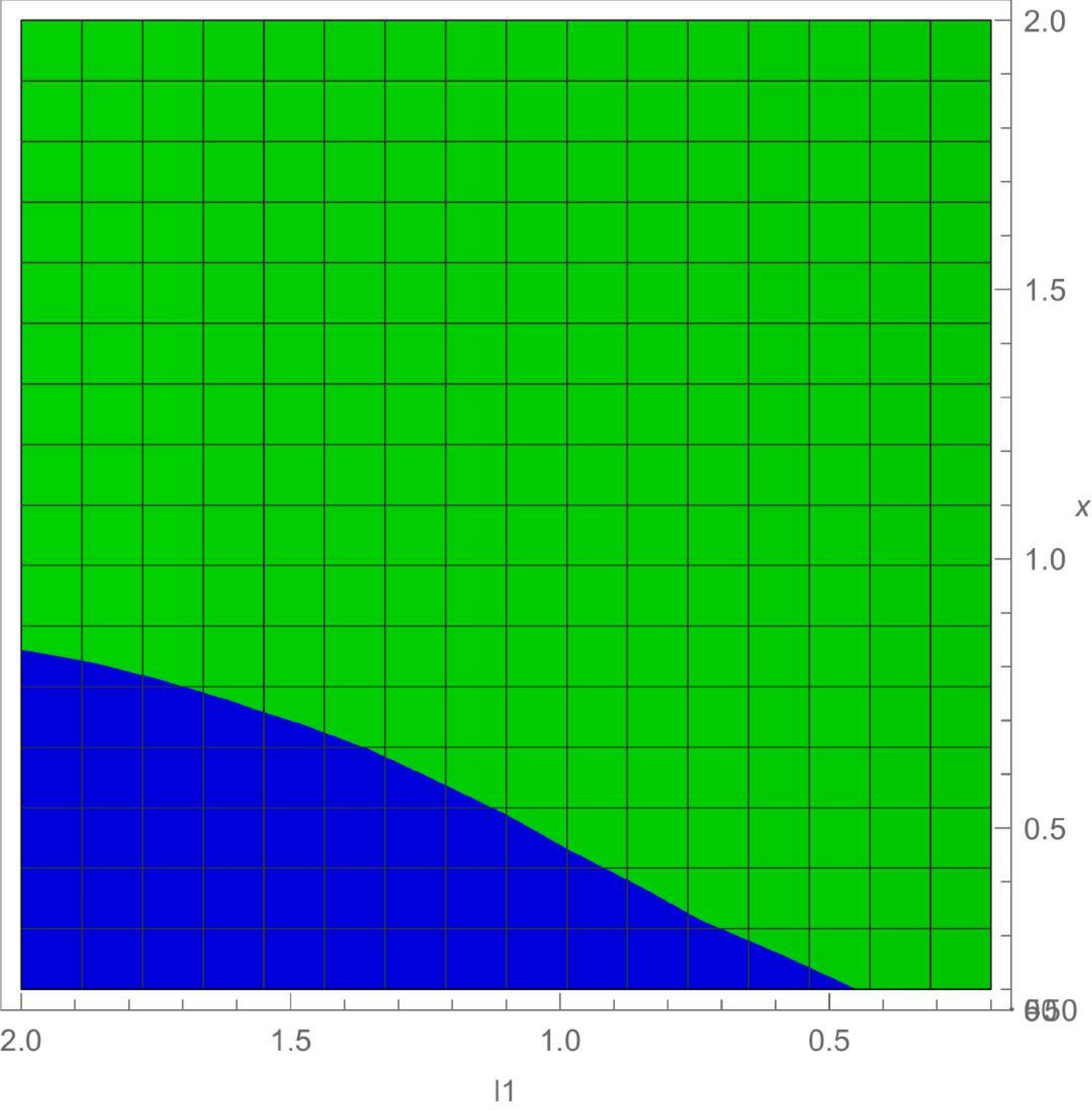}}
\put(80,78){\includegraphics[scale=0.25]{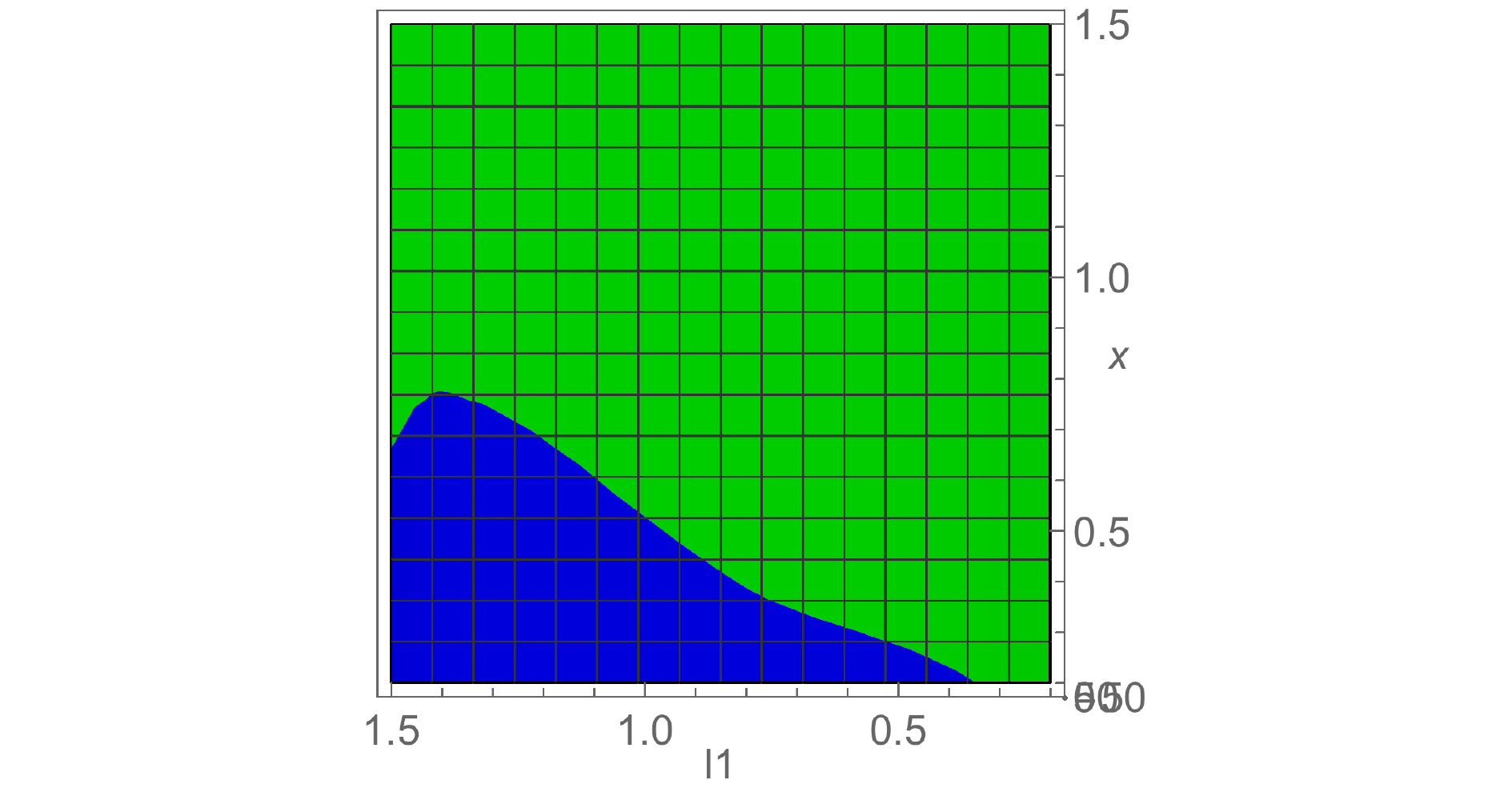}}
\put(180,78){\includegraphics[scale=0.25]{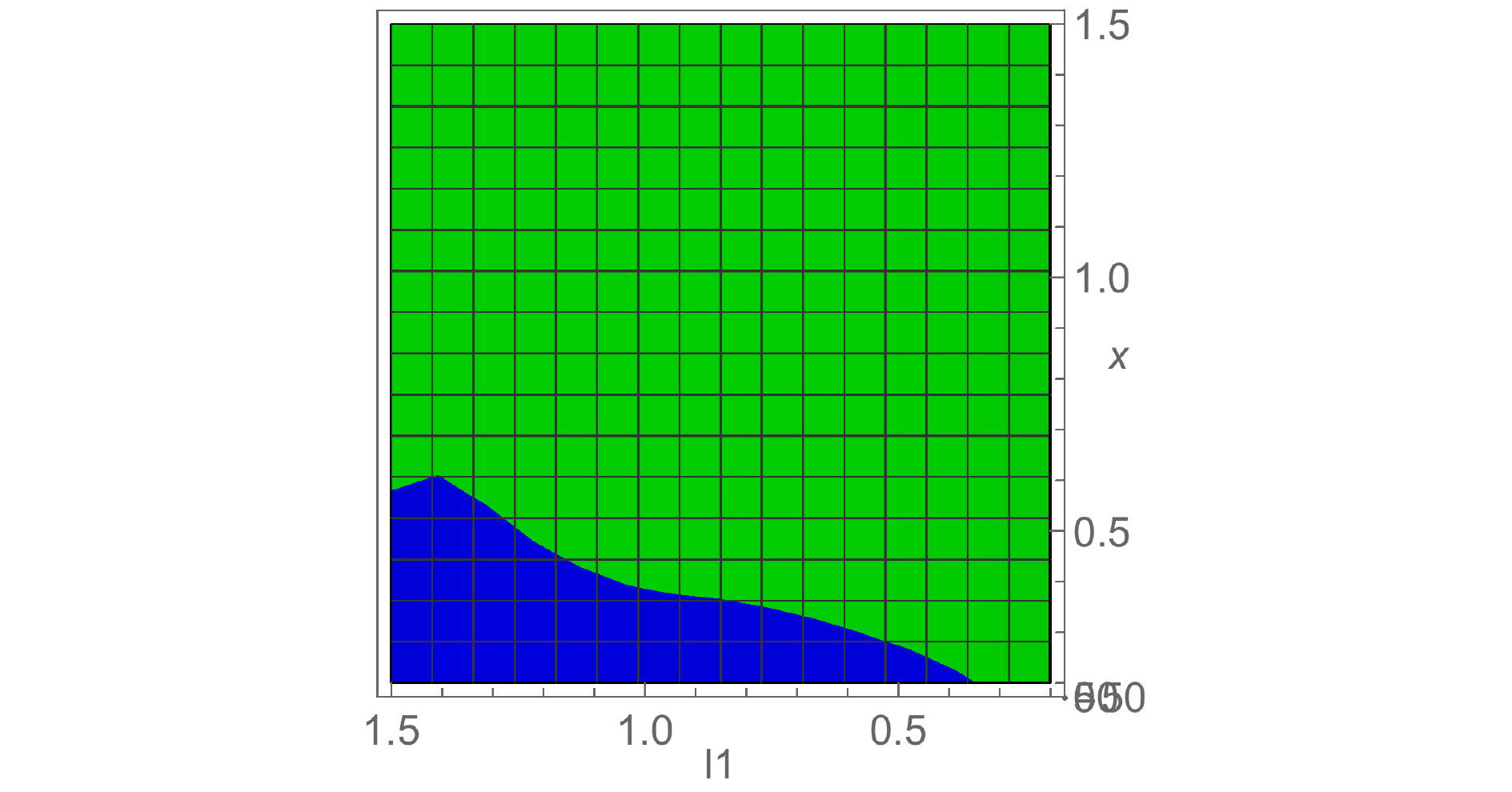}}
\put(-115,0){\includegraphics[scale=0.25]{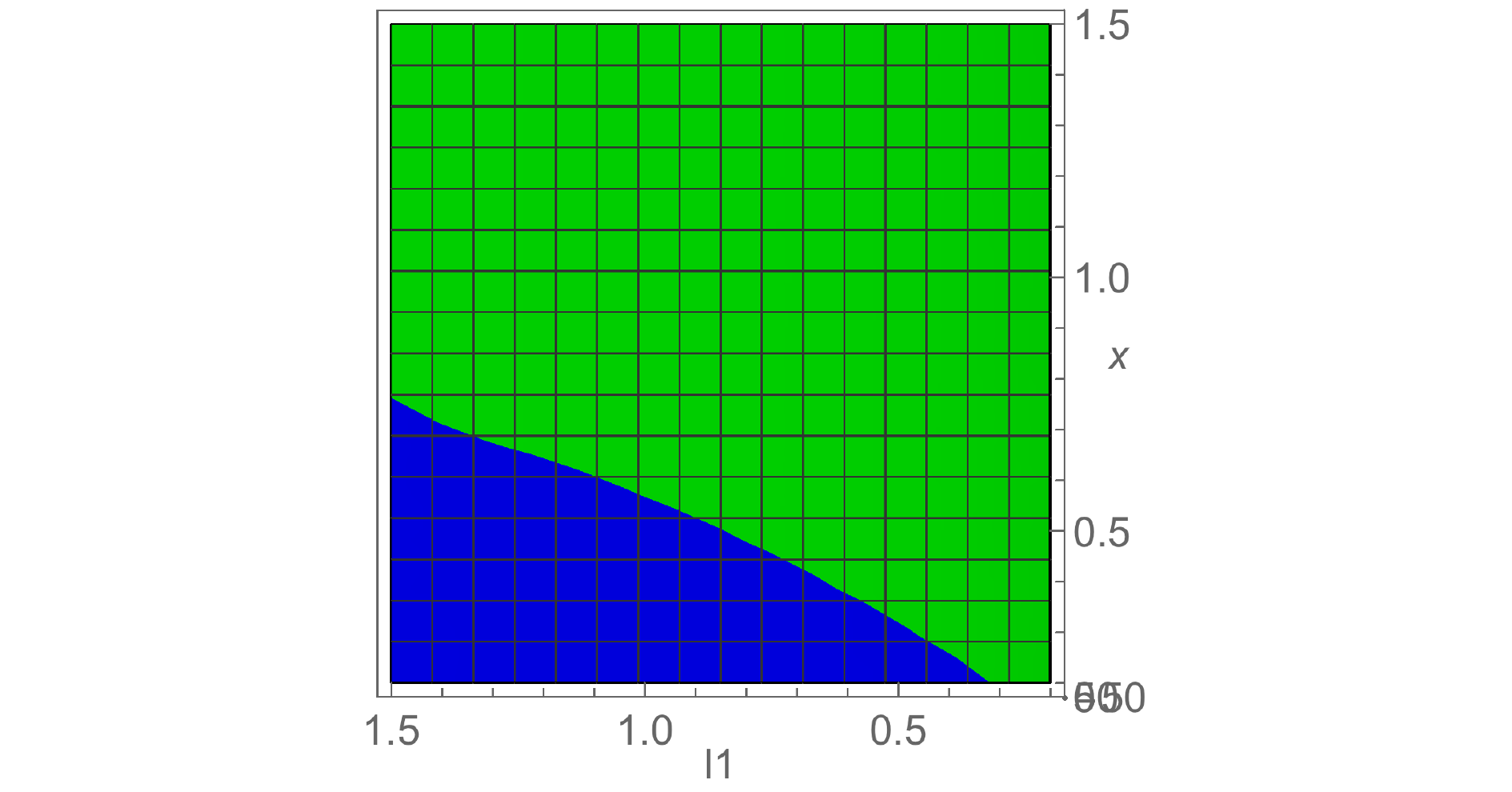}}
\put(-20,0){\includegraphics[scale=0.25]{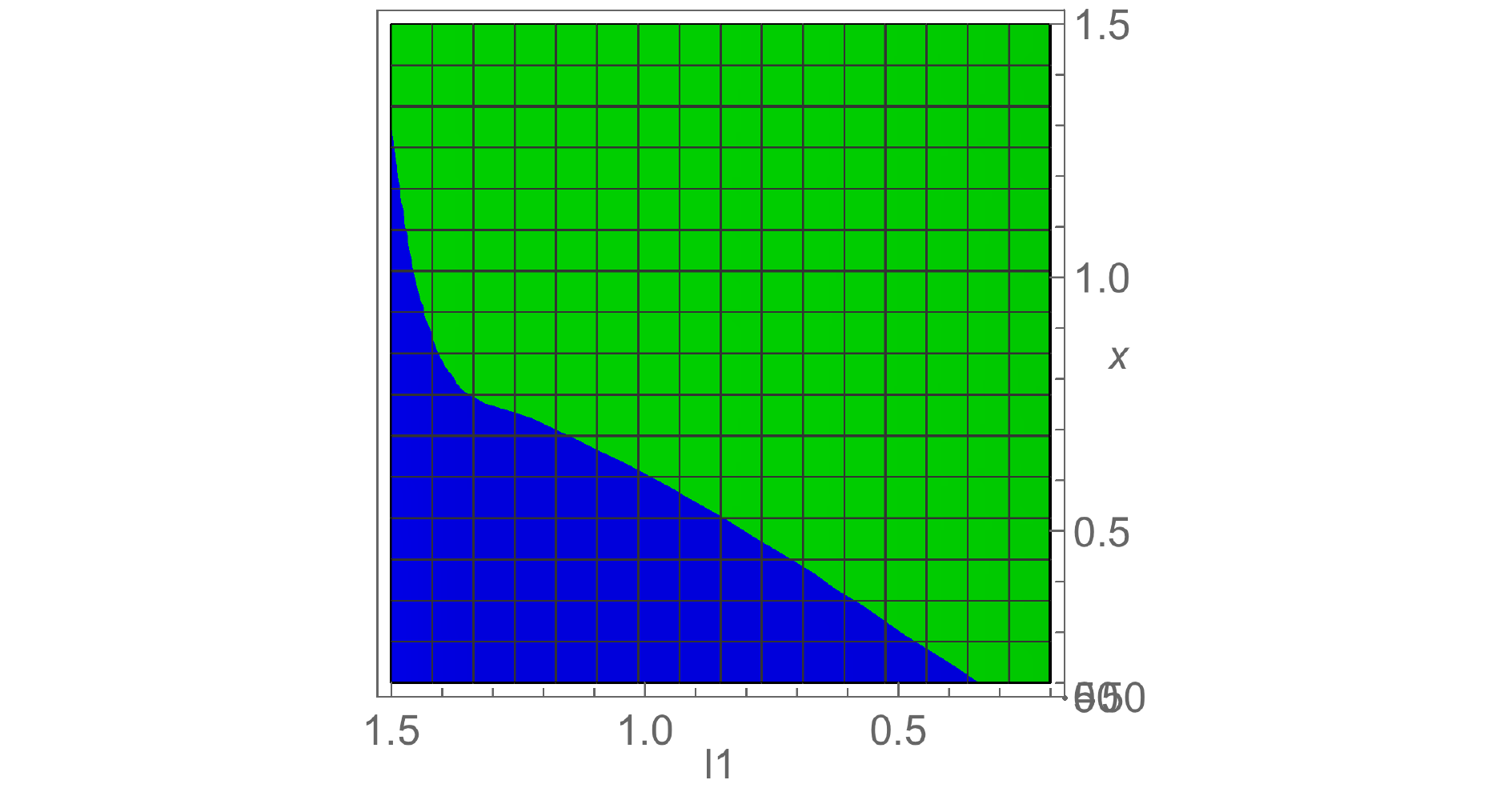}}
\put(105,0){\includegraphics[scale=0.15]{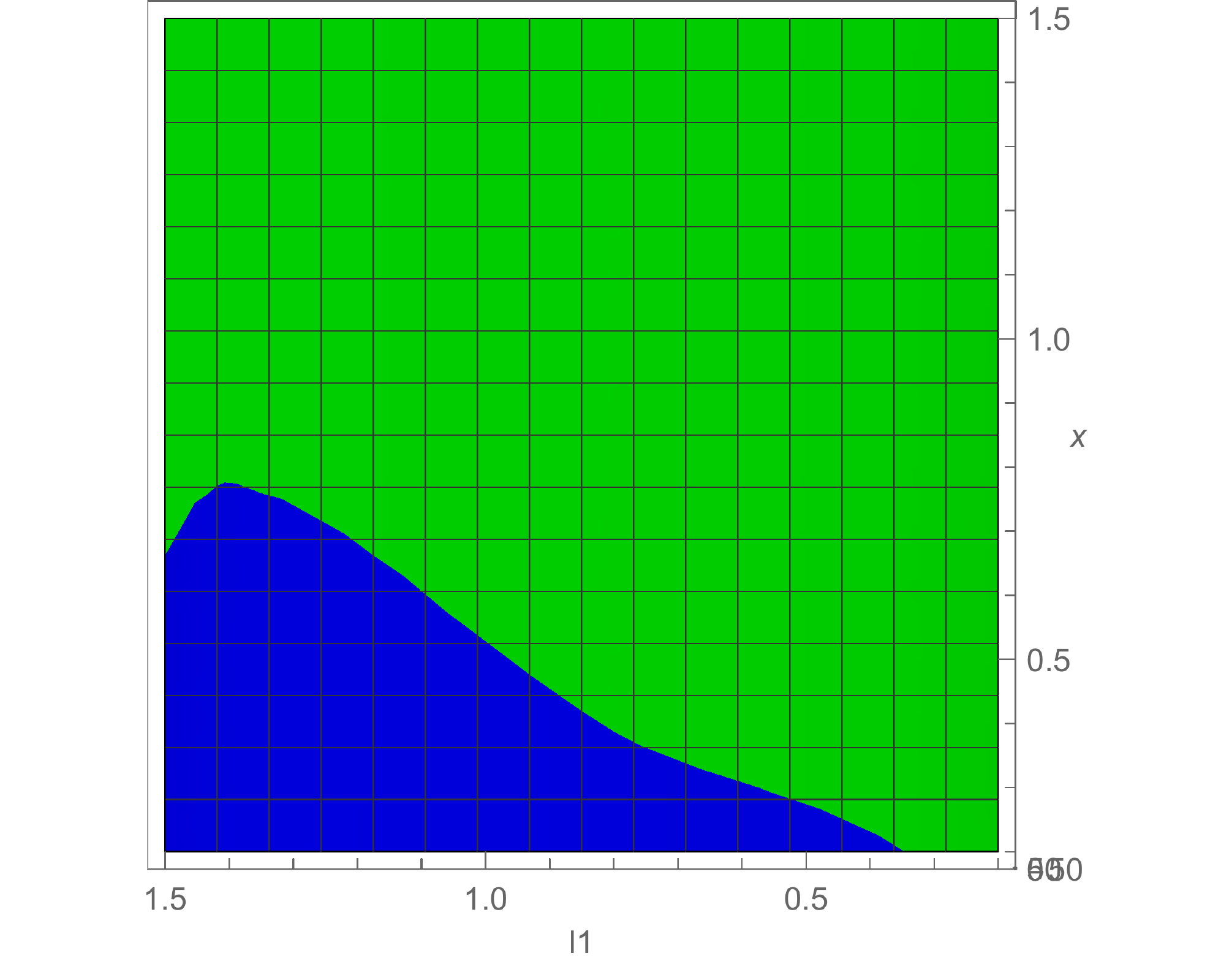}}
\put(205,0){\includegraphics[scale=0.15]{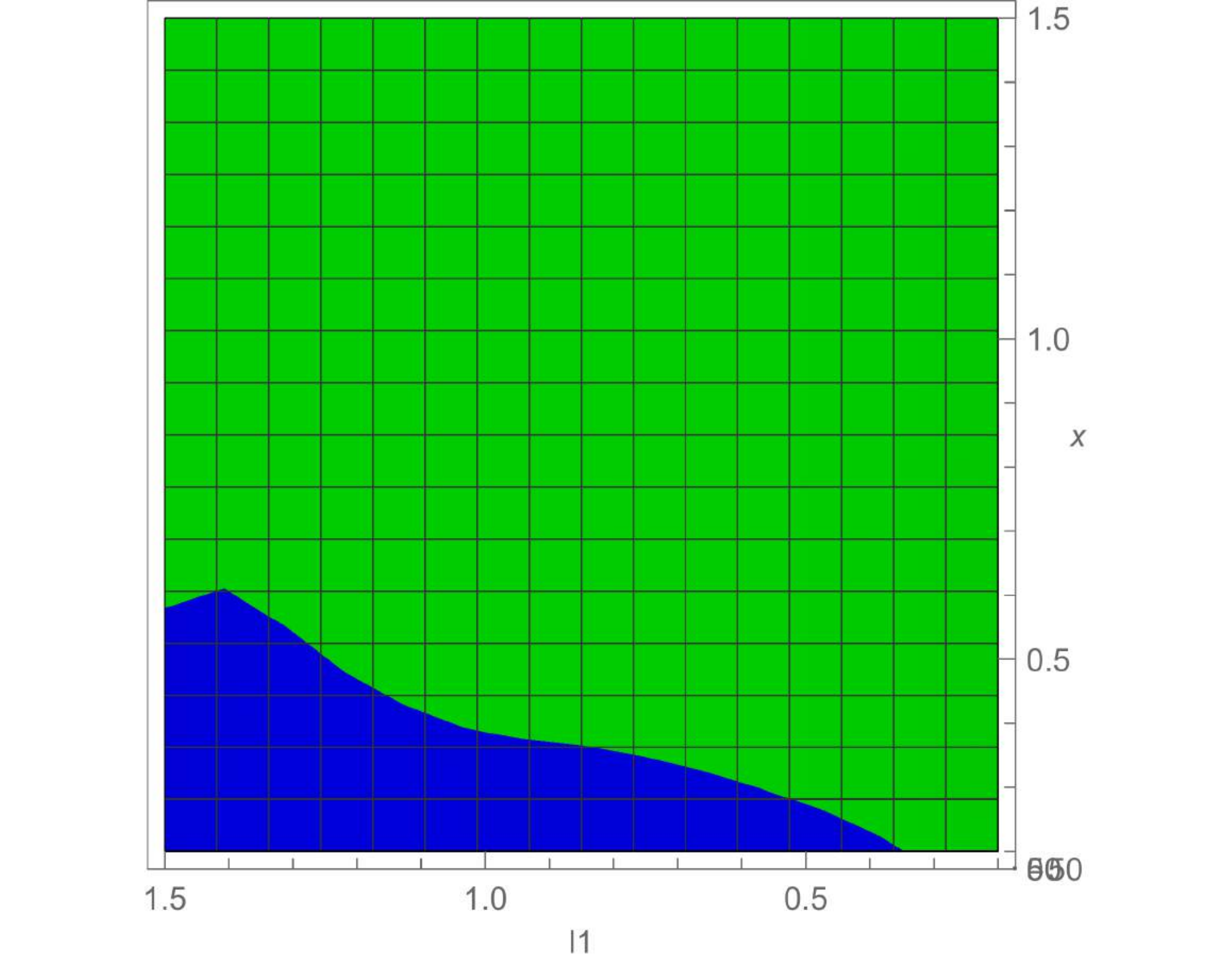}}
\put(-30,-5){$l_1$}
\put(280,35){$x$}
\put(-55,150){$t=0$}
\put(50,150){$t=1$}
\put(150,150){$t=2$}
\put(250,150){$t=3$}\end{picture}
\caption{Cartoon diagrams that show  contributions of various diagrams to the entanglement entropy for 2 strips, $A$ and $B$, with unequal lengths $l_1,l_2,$ at different times during the transition  from the  black brane configuration with the small mass $m_0$ to the large mass $m$ in AdS$_3$. The green color regions correspond to the bulk surface
  $(A)||(B)$, the blue color regions correspond to the bulk surface $(B\,C)_c$, the red color regions correspond to the bulk surface $(A\,B)_{c,cr}$ (we do not see  red regions). Different colors regions are separated by the curves that  are the transition lines. In the top plots $l_2=l_1$, and $t=0,1,2,3,$ in the bottom plots $l_2=2l_1$,
 and we vary $l_1$  and $x$. All plots correspond to $\alpha=0.2$.
 }
  \label{Fig:Cartoon-d2-2seg}
\end{figure}

As already has been mentioned in Sect.\ref{Decom},  equation  \ref{decom2}, given the two disconnected regions $A$ and $B$ in the boundary, there are three configurations of hypersurfaces  extending in the bulk whose boundaries coincide with $\partial (A\cup B) =\partial A \cup \partial B$: the "disjoint" configuration, the ``connected''  configuration, given by a bridge connecting $A$ and $B$ through the bulk and the "crossing" configuration, that looks as a configuration with cross-sections in the projection on  the $(x,z)$-plane, Fig.\ref{Fig:Dec2}.
The "crossing" configurations  do not contribute to the holographic entropy calculations \cite{Alishahiha:2014jxa,Ben-Ami:2014gsa},
in the static AdS and AdS black brane backgrounds. They also do not contribute to the same calculations during the thermalization process described by
the Vaidya AdS metric \cite{Allais:2011ys}. We have checked numerically that they do not contribute in the
case of the Vaidya AdS black brane metric  during the transition from the black brane with mass $m_0$ to the black brane with mass $m_0+m$,
see cartoon plots in Fig.\ref{Fig:Cartoon-d2-2seg}. In these plots
we show the leading contributions to the entanglement entropy for 2 strips, $A$ and $B$, with unequal lengths $l_1,l_2,$ at different times during the transition  from the  black brane configuration with the small mass $m_0$ to the large mass $m$ in AdS$_3$.
 The green color regions correspond to the bulk surface
  $(A)||(B)$, the blue color regions correspond to the bulk surface $(B\,C)_c$. In our program we have used the red color to indicate
  contribution of the bulk surface $(A\,B)_{c,cr}$, but  we have not seen  red regions for all variety of parameters $0.2<l_1,l_2,x<1.5$, $0<t<4$.  Transition lines
separate the different color region and as usual are interpreted as the lines of topological phase transitions. Note that one can solve this problem analytically for the case of   infinitely thin shell   in AdS$_3$ black brane.

\subsection{Holographic mutual information for two site system at nonzero temperature.}\label{Sect:HMI}

Let us first consider the holographic mutual information  $S(A\cup B)$ for the Vaidya shell in the three dimensional AdS black brane background with $f=f(z,v)$
 given by \eqref{fvz} and \eqref{m-v-BB},
and two segments shown in Fig.\ref{Fig:Dec2}, as function of the boundary time t at fixed $l_1,l_2,x$.
This quantity depends on many parameters, $l_1,l_2,x$, $m,m_0$ and $\alpha$ and we perform our analysis  numerically.
On the right in Fig.\ref{Fig:Dec2} we show the case $m_0=0.25$ and on the left we show for comparison the case $m_0=0$ considered before in \cite{Balasubramanian:2011at,Allais:2011ys}.
 The same color  curves correspond to the same 
 fixed values of $l_1$  and $x$,
 herewith  different style lines (solid, dashed and dotted)  correspond to  different values of
$l_2$ for both cases, $m_0=0.25$ and $m_0=0$.

 \begin{figure}[h!]
\centering
\includegraphics[scale=0.25]{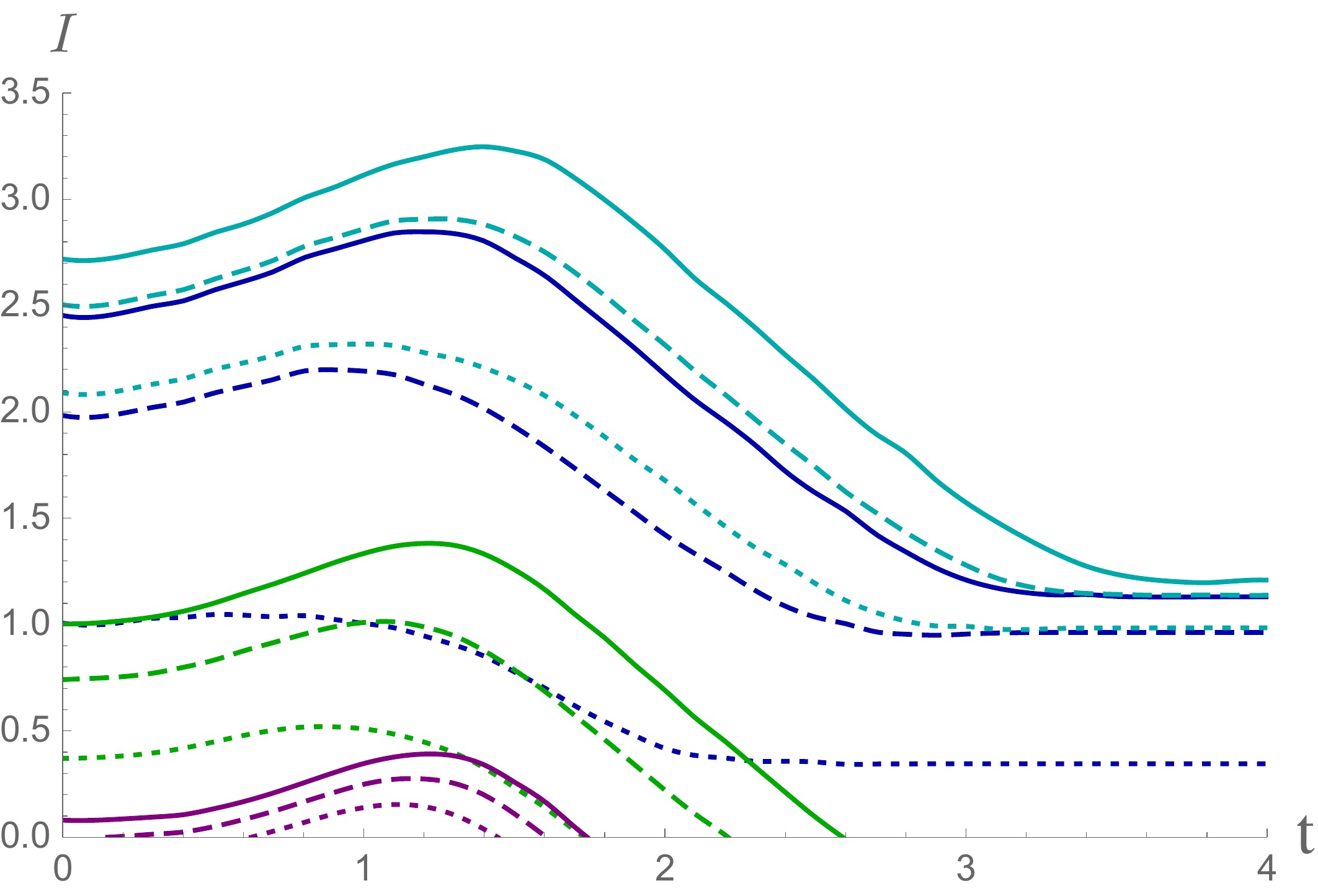}$\,\,\,A\,\,\,\,$$\,\,\,\,\,\,\,$
\includegraphics[scale=0.25]{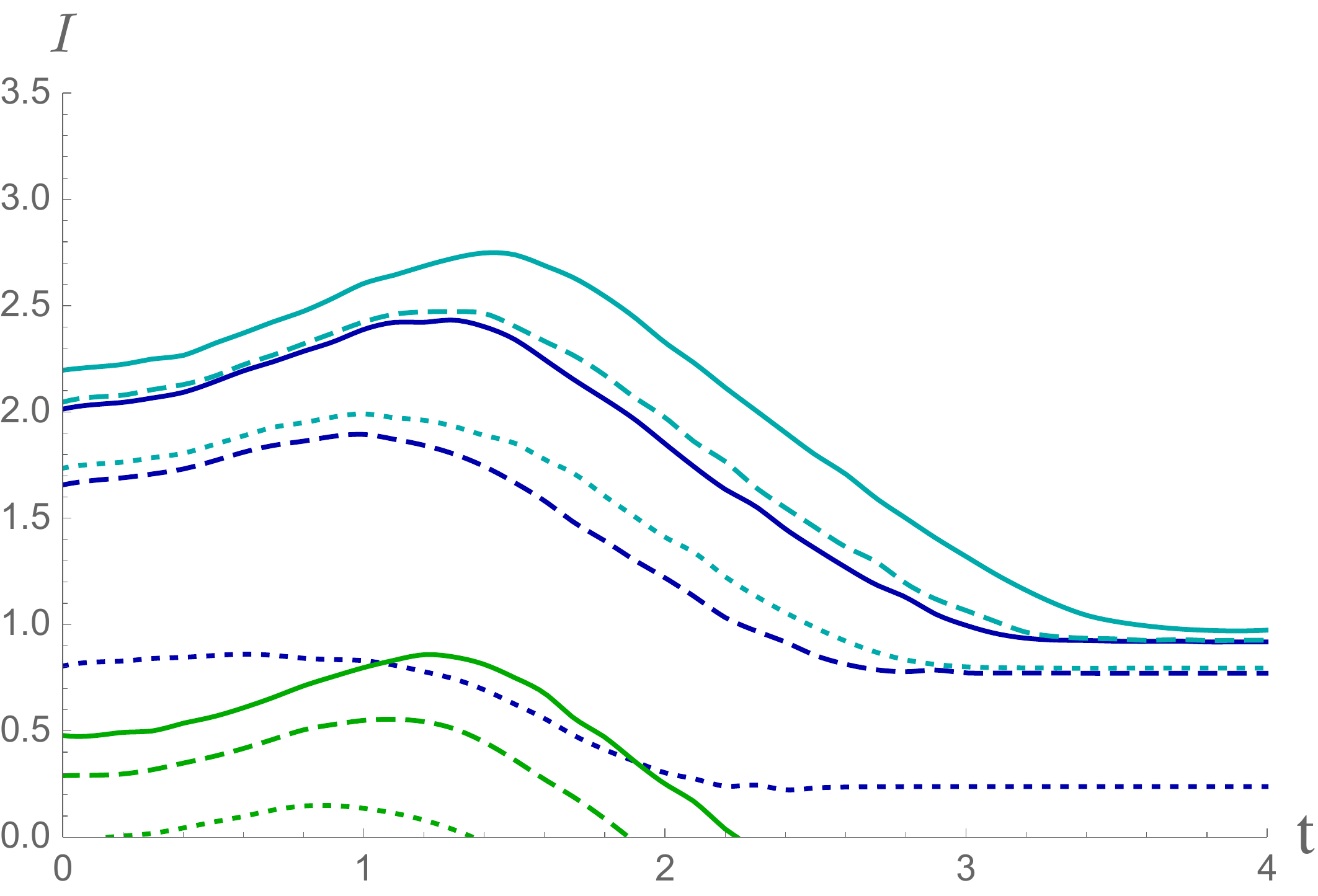}$\,\,B$
 \caption{
Holographic mutual information $I(A,B)$, up to the normalizing factor, for the Vaidya metric in the three dimensional black brane background with $f=f(z,v)$
 given by \eqref{fvz} and \eqref{m-v-BB} ($m_0=0,m=1$ for the  panel {\bf A} and $m_0=0.25,m=1$ for the  panel {\bf B})
and two segments shown in Fig.\ref{Fig:Dec2}, as function of the boundary time t at fixed $l_1,l_2,x$. The same color  curves are characterized
 by fixed values of $l_1$  and $x$
and varying  $l_2$.  Blue  lines correspond to $l_1=1.5$ and $x=0.2$ and
solid, dashed and dotted styles  correspond to
$l_2= 1.5, 1$ and $ 0.5 $, respectively;  green  lines correspond to $l_1=1.5$ and $x=0.4$ and
solid, dashed and dotted styles  correspond to
$l_2= 1.5, 1.2$ and $ 0.9 $, respectively; purple lines   correspond to  $l_1=1.5$  and $x=0.6$ and
solid,  dashed and dotted lines
correspond to
$l_2= 1.5, 1.4, 1.3 $; dark cyan lines   correspond to  $l_1=1.7$  and $x=0.2$ and
solid,  dashed and dotted lines
correspond to $l_2= 1.5, 1.2, 0.9 $.  For all cases $\alpha=0.2$. We see that there are no purple lines in plots B.}
\label{Fig:Ent2s}
\end{figure}

We see that some color and style lines present in both, left and right plots in Fig.\ref{Fig:Ent2s},  but some  color and style lines present only  in the left plot in Fig.\ref{Fig:Ent2s}. In particular,
there are no purple lines in Fig.\ref{Fig:Ent2s}.B.  Notice that  the character of time dependences shown in the left and right plots are the same. There are curves that start from non-zero value of the mutual information $I_{0}$, then  increase
during some time $t_{max}$  up to some maximal value, $I_{max}$ and then during the saturation time $t_s$ go to other constant value $I_{s}$. Increasing the starting temperature we decrease the corresponding $t_{r}$ and $t_s$.
There are also curves that start from $I_{0}\neq 0$, reach $I_{max}$ and then at the scrambling time $t_{sc}$ become zero.
At the scrambling time any sort of correlations, in particular the exchange of information, is erased.
 There is also the regime when the mutual information starts from zero at wake up time,
 $t_{wu}$, then it increases up to the maximum and when becomes zero at  $t=t_{sc}$. In this  regime the time dependence has the bell form and it  is more interesting for us, see Sect.\ref{Comparison}. There are also geometrical configurations, when $I$ is zero at all times. These configurations correspond to the cases with large distances $x$.

\begin{figure}[h!]
\centering
$$\,$$
\begin{picture}(250,100)
\put(-80,0){\includegraphics[scale=0.28]{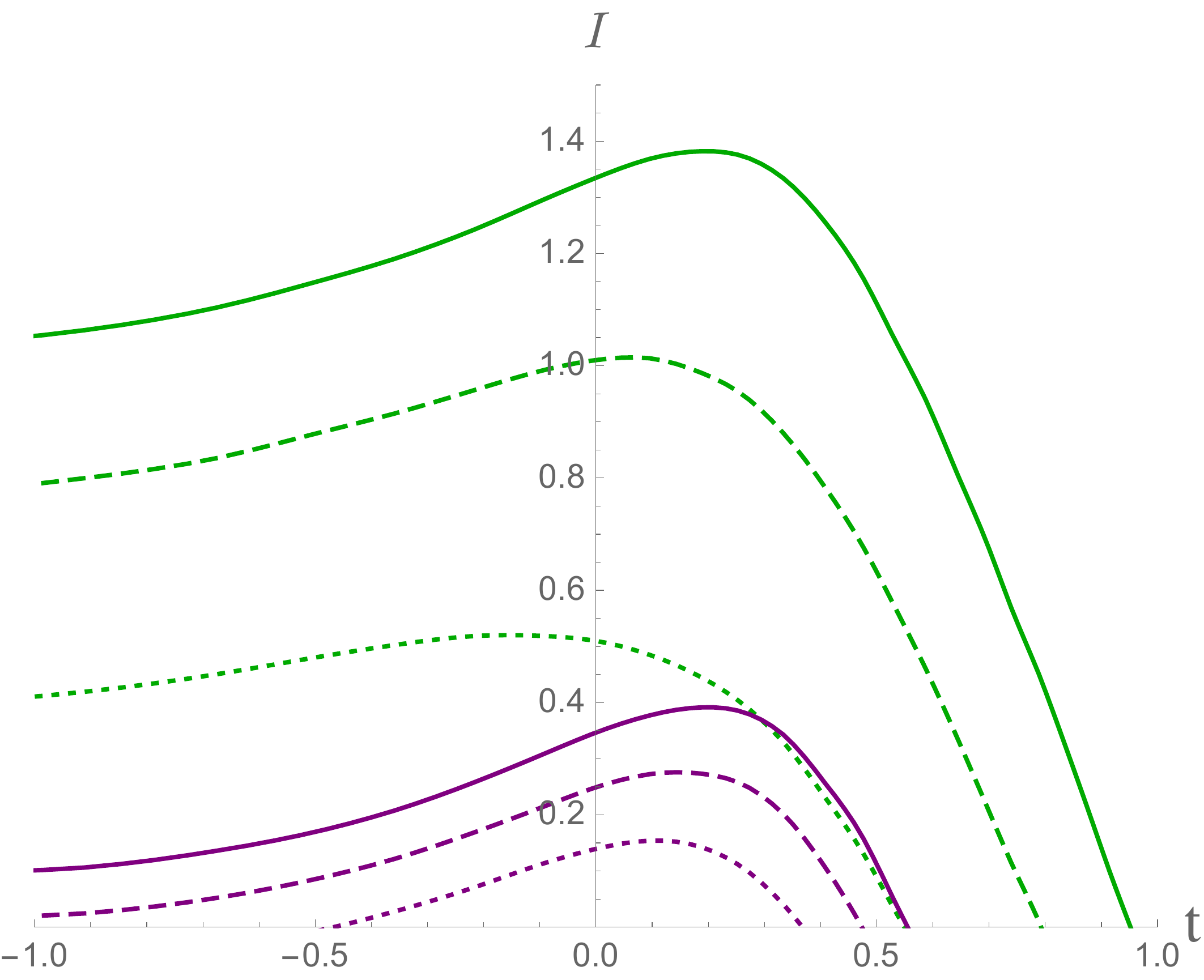}}
\put(89,8){$_{\log t}$}
\put(150,0){\includegraphics[scale=0.3]{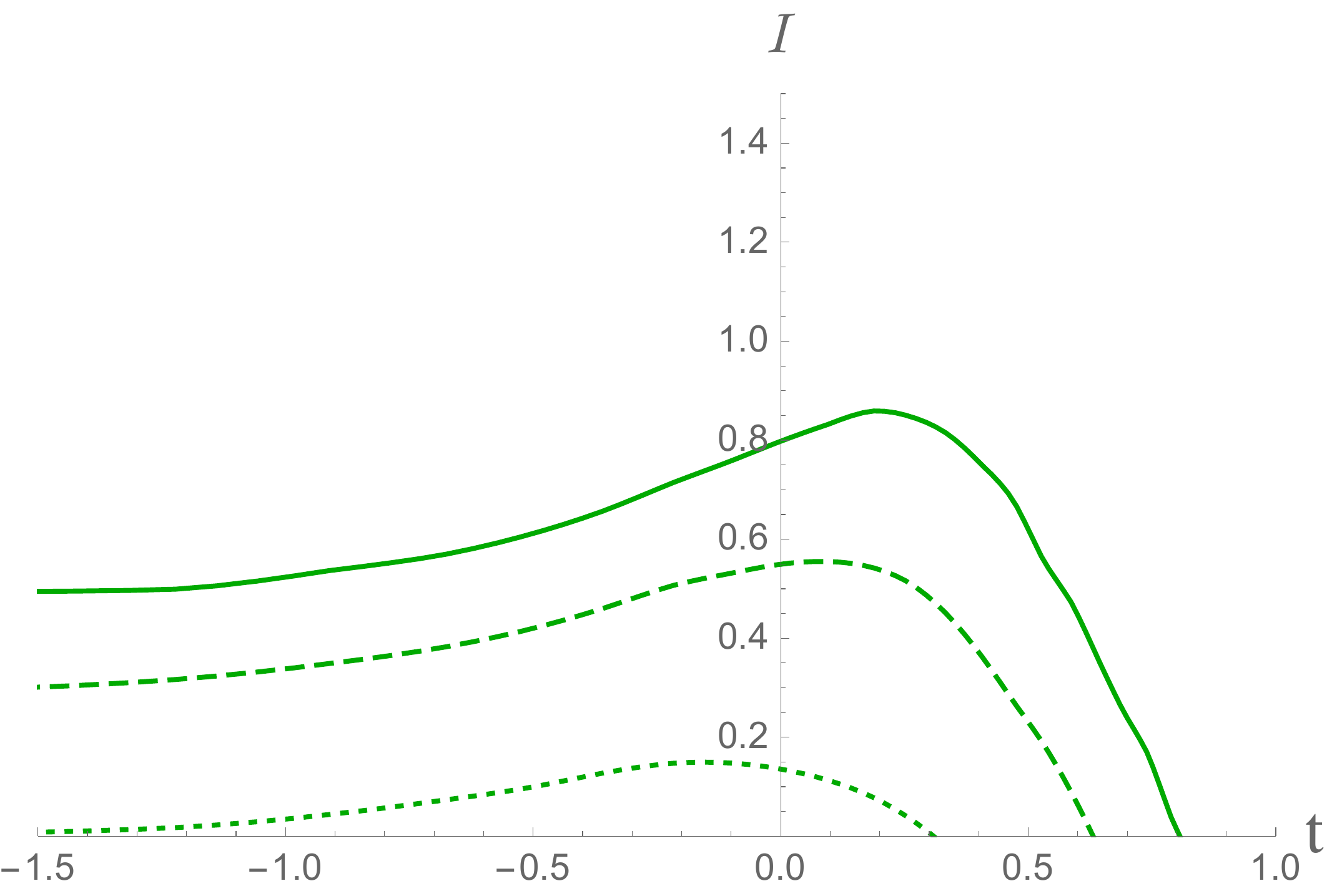}}
\put(331,8){$_{\log t}$}\end{picture}\\$\,$\\
\caption{
%A. {\it From file: Vaidya-AD-seek-13-march-d2-AdSBH}. B. {\it From file: Vaidya-AD-seek-13-march-d2-BHBH}
Holographic mutual information $I(A,B)$, up to the normalizing factor, for the Vaidya metric in the four dimensional black brane background with $f=f(z,v)$
 given by \eqref{fvz} and \eqref{m-v-BB} ($m_0=0,m=1$ for the left panel and $m_0=0.25,m=1$ for the right panel)
and two belts shown in Fig.\ref{Fig:2segm},
as function of the boundary time t (in the logarithmical scale) at fixed $l_1,l_2$ and $x$.
The same color  curves are characterized
 by fixed values of $l_1$  and $x$
and varying  $l_2$.  For all green lines $l_1=1.5$, $x=0.2$ and  the solid, dashed and dotted
green lines correspond to
$l_2= 1.5, 1.2,0.9 $, respectively; for all  purple  lines $l_1=1.5$, $x=0.6$ and the solid, dashed and dotted purple lines correspond to
$l_2= 1.5, 1.4, 1.3 $, respectively.  For all cases $\alpha=0.2$}
\label{Fig:Ent3a}
\end{figure}

 The time dependence, in the logarithmic time scale,  of the holographic mutual information  $S(A\cup B)$ for the Vaidya shell in the four  dimensional AdS black brane background is presented in Fig.\ref{Fig:Ent3a}. Comparing Fig.\ref{Fig:Ent3a} and
 Fig.\ref{Fig:Ent2s} we see that the qualitative features of the curves are the same.

\subsection{Matching holographic calculations to the simulation results.}\label{Comparison}
\begin{figure}[h!]
\centering
\includegraphics[scale=0.5]{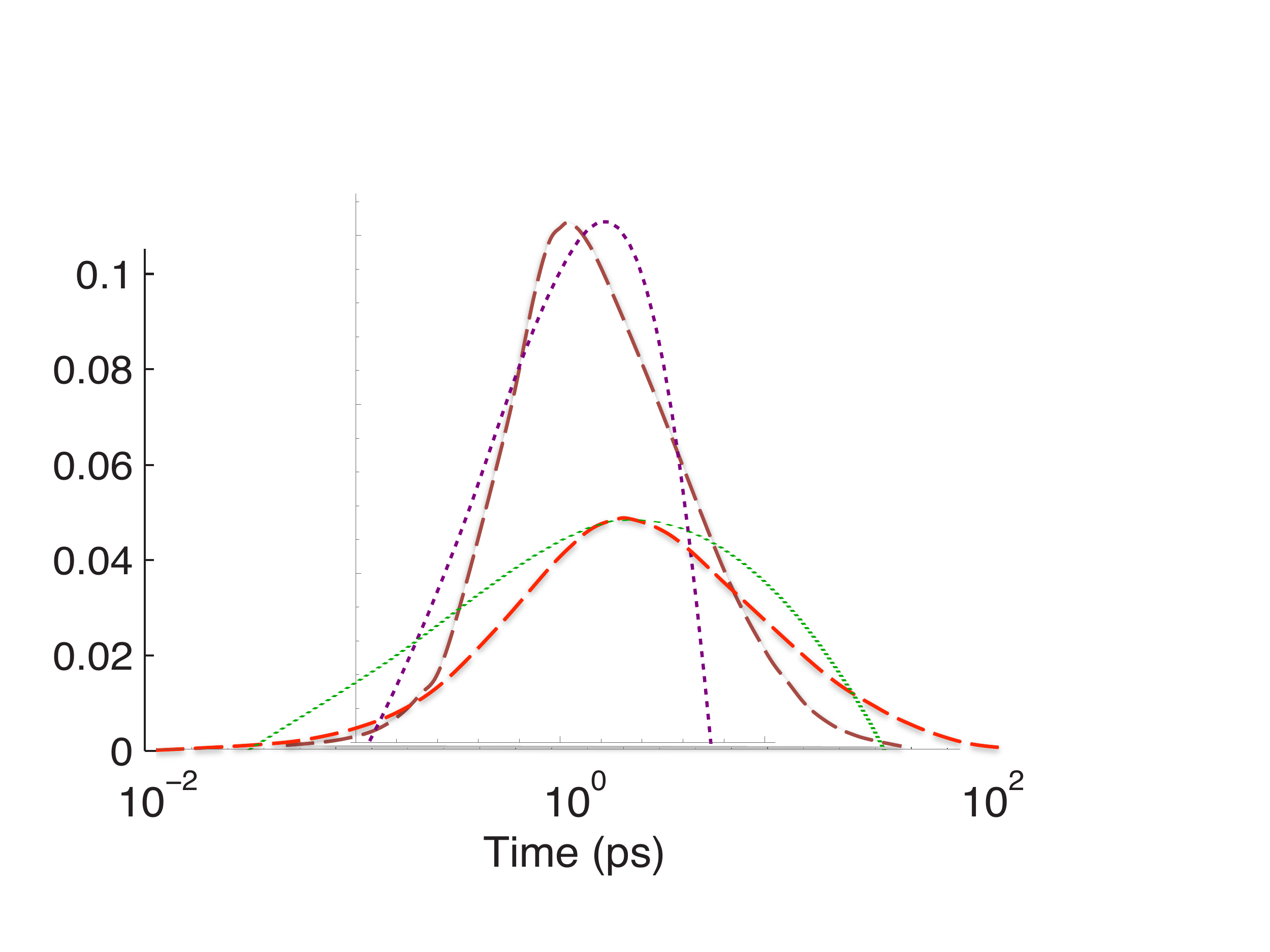}
\caption{
The purple and red dashed lines show the time dependence of the mutual information of the reduced FMO system calculated   in  \cite{0912.5112} at 70${{}^\circ}$K and  300${{}^\circ}$K temperature, respectively.
The green and purple  solid lines show the rescaled time dependence of the holographical mutual information presented in Fig.\ref{Fig:Ent3a} for the cases of the global quench in AdS$_4$ and AdS$_4$ black brane, respectively.
}
\label{Fig:Compa}
\end{figure}

It is obvious that we find the scaling behavior of the holographic mutual information only up to normalization factors.
To fit the mutual information for a given physical system there are 3 numerical parameters in our disposal: the (d+1)- dimensional gravity constant, the radius of the AdS and the scale, that in our case can be identified with the black brane mass.
We  compare the results of our calculations with the mutual information
 $I\left(  A;B\right)  $ calculated for the reduced FMO system in \cite{0912.5112},
 where system A consists of site three and  system B consists of site one, see Fig.3a in  \cite{0912.5112}.
 We reproduce this curve in Fig.\ref{Fig:Compa}
by the purple dashed line. In Fig.4a of \cite{0912.5112}  this quantity at physiological
temperature (300${{}^\circ}$K) is presented. We reproduce this curve in Fig.\ref{Fig:Compa}
by the red dashed line. We see that in this particular case the purple curve has a maximum approximately twice higher
when the red one. As it is shown in Fig.\ref{Fig:Compa}  these two curves fit two curves that we take from
Fig.\ref{Fig:Ent3a}
for  particular choices of the geometry of two strips.

It is interesting to note, that the mutual information calculated in  \cite{0912.5112} for the
 the reduced FMO system,
 where system A consists of site three and  system B consists of site six, or two, see Fig.1b and 3b in  \cite{0912.5112},
 has the two-humped form. We can reproduce this form of the time dependence of the mutual information taking two thin shells
 Vaidya AdS$_{d+1}$ metric with
 \be\label{m-v2}
m(v)=\frac{m_1}{2}\left(1+\tanh \frac{v}{\alpha}\right)+\frac{m_2}{2}\left(1+\tanh \frac{v-v_0}{\alpha}\right),
\ee
cf.\cite{AV}.

\subsection{Wake up  and scrambling times}\label{Sect:wakeup}
\begin{figure}[h!]
\centering
\includegraphics[scale=0.25]{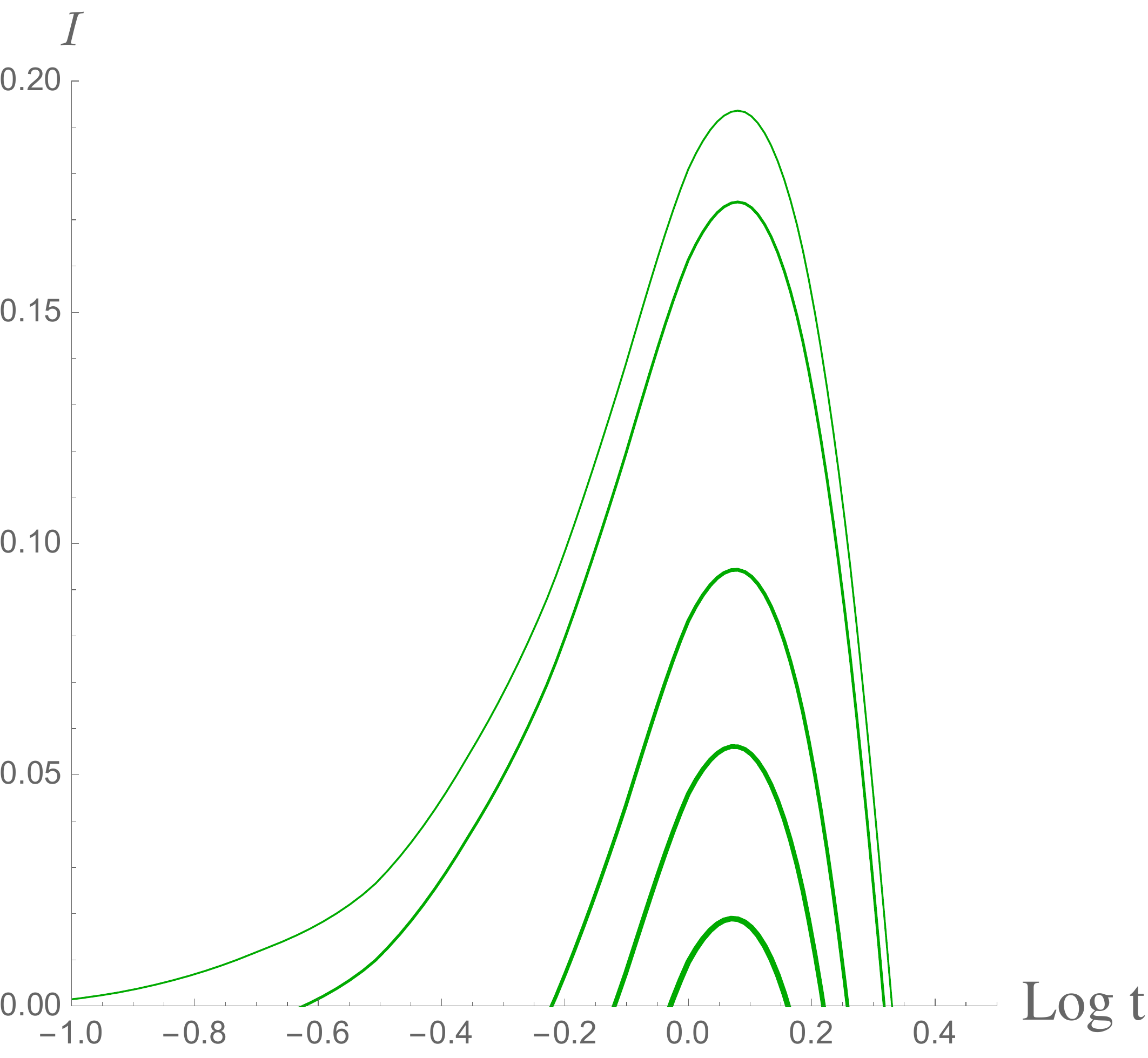}$\,\,\,\,\,\,$$\,\,\,\,\,\,$$\,\,\,\,\,\,$$\,\,\,\,\,\,$$\,\,\,\,\,\,$
\includegraphics[scale=0.25]{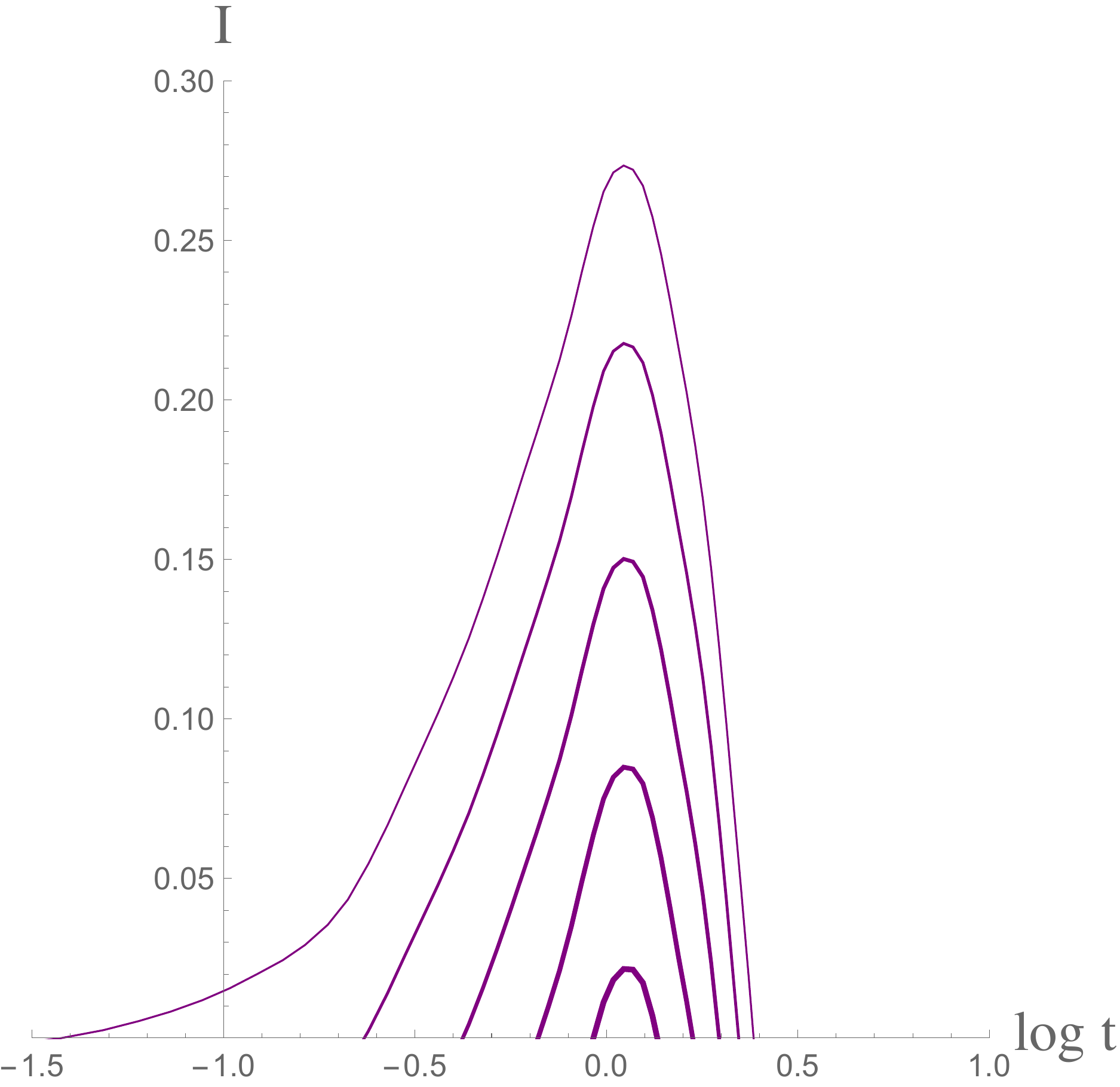} \caption{
We show typical bell like profiles of the time dependence of the mutual information
for   Vaidya quenches in the AdS$_4$ (the left plot) and in the AdS black brane (the right plot).
}
\label{Fig:Bell}
\end{figure}

As we have seen in Sect.\ref{Sect:HMI} there are configurations for which the  time dependence of the mutual information has the bell form, see also plots in Fig.\ref{Fig:Bell}. We see that for some time the mutual information is equal to zero (or very small) then at the wake up time
$t_{wu}$
it starts to increase, reaches the maximum at time $t_{max}$ and then decreases and vanishes at the scrambling time $t_{scr}$. This behaviour can be explained by a causality argument \cite{Balasubramanian:2011at}.
It is supposed that before the quench there is a finite correlation length in the system. Therefore after the quench only excitations created
at nearest points will be entangled. At later times, such entangled pairs
will only contribute to the mutual information between two
intervals, if each interval contains one of the two particles. 

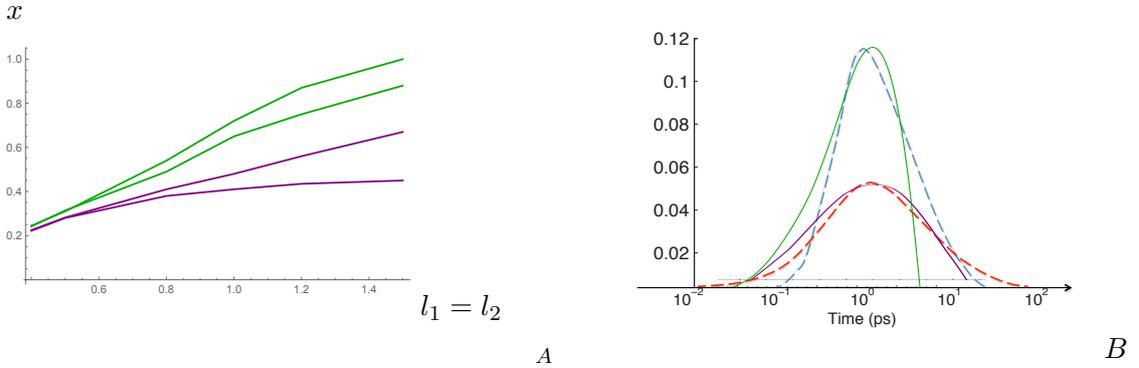
\begin{figure}[h!]
\centering
$$\,$$\\
\begin{picture}(250,100)
\put(-80,15){\includegraphics[scale=0.28]{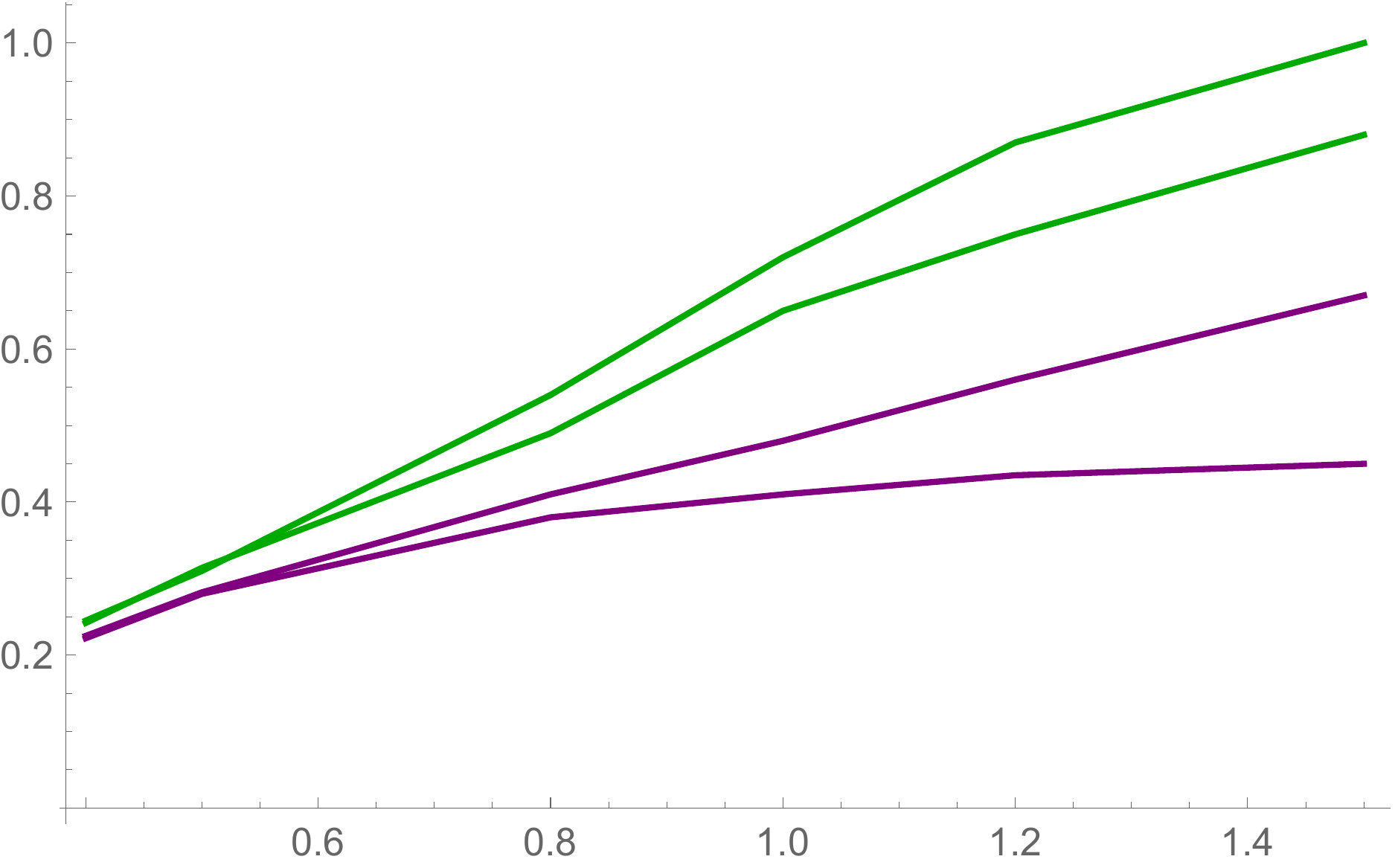}}
\put(75,8){${l_1=l_2}$}
\put(-80,120){$x$}\put(118,-8){$_{A}$}
\put(150,0){\includegraphics[scale=0.3]{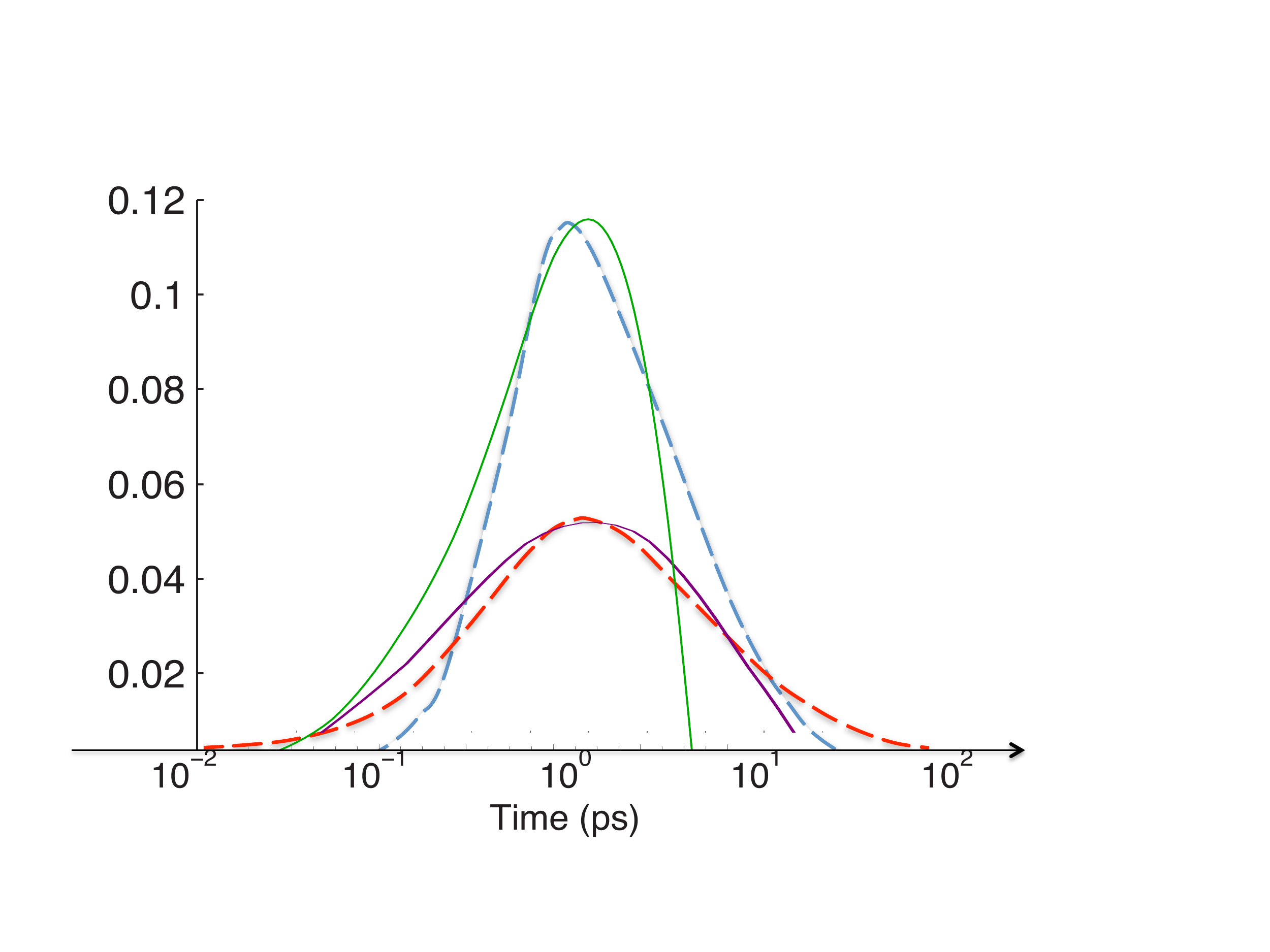}}
\put(331,-8){$B$}
\end{picture}\\$\,$\\
 \caption{
{\bf A}. The domains corresponding to the bell profiles  of  time dependence of the mutual information
after the Vaidya quench in the AdS$_4$ (the domain between green lines) and in the AdS black brane
(the domain between purple lines) for equal width  belts and different distances between them.
{\bf B}. The blue and red dashed lines show the time dependence of the mutual information of the reduced FMO system calculated   in  \cite{0912.5112} at 70${{}^\circ}$K and  300${{}^\circ}$K temperature, respectively.
The green and purple  solid lines show the rescaled time dependence of the holographical mutual information for
$l_1=0.8, l_2=0.8$ and  $x=0.5$ and $x=0.4$ for cases of the global quench in AdS$_4$ and AdS$_4$ black brane, respectively.}
\label{Fig:Bell-reg}
\end{figure}

The scrambling time is the time scale at which all correlations in a system have been destroyed after introducing
a perturbation \cite{1602.07307}.
One defines the scrambling time as the time $t_{scr}$ at which the mutual information of any two subsystems will be vanishing or small. It was conjectured that black holes are fastest scramblers in nature \cite{Sekino:2008he}.

Numerical results show that all these time scales depend on the geometry as well as on the form of the quench.
Plots in Fig.\ref{Fig:Bell}  show that the wake up time is very sensitive to the change of distances between segments,
while the scrambling time is not so sensitive.  In Fig.\ref{Fig:Bell-reg}.A we show the domains corresponding to the bell profiles  of  time dependence of the mutual information
after the Vaidya quench in the AdS$_4$ (the domain between green lines) and in the AdS black brane
(the domain between purple lines) for equal width  belts and different distances between them.
We see that the gaps where the bell profiles are available  are quite narrow. The gaps corresponding to initial vacuum state and the state with non-zero temperature are not crossing. By this reason our fitting with results  \cite{0912.5112} have been performed for different geometries. 
In the case of infinitely thin shell these time scales  admit  analytical studies \cite{Liu:2013qca}.

\section{Holographic studies  of the FMO complex with one composite part}\label{Sect:HMI-comp}
 In this section we consider holographic description of the quantum mutual information for the  reduction of the FMO complex with one composite part.
 \subsection{Phase structure of  holographic entanglement entropy  $S(A\cup B\cup C)$ for the Vaidya shell in the AdS$_d$ black brane background}
 \label{Sect:HEE-comp}

\begin{figure}[h!]
\centering
\begin{picture}(250,200)
\put(-100,80){\includegraphics[scale=0.15]{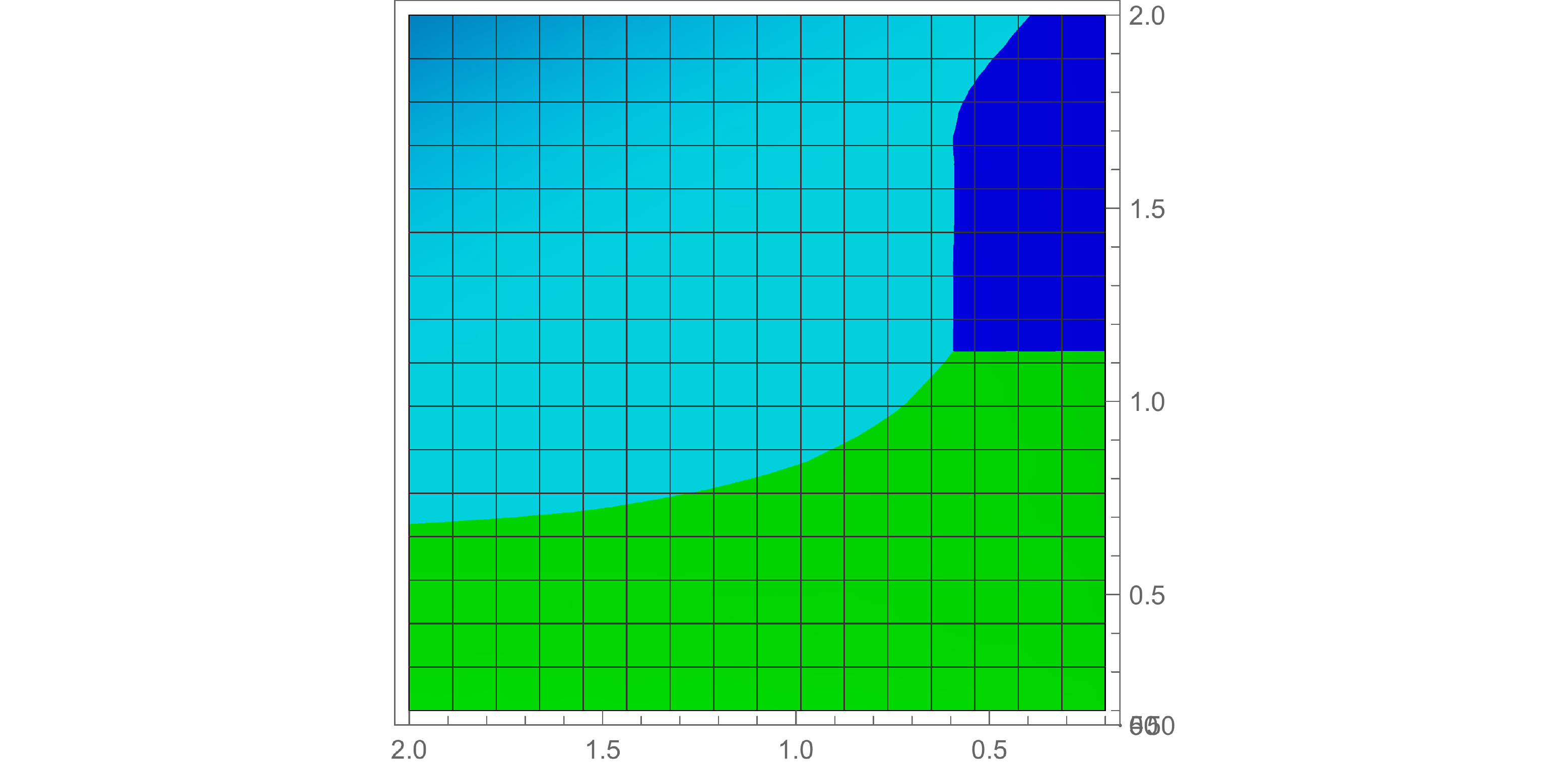}}
\put(0,80){\includegraphics[scale=0.15]{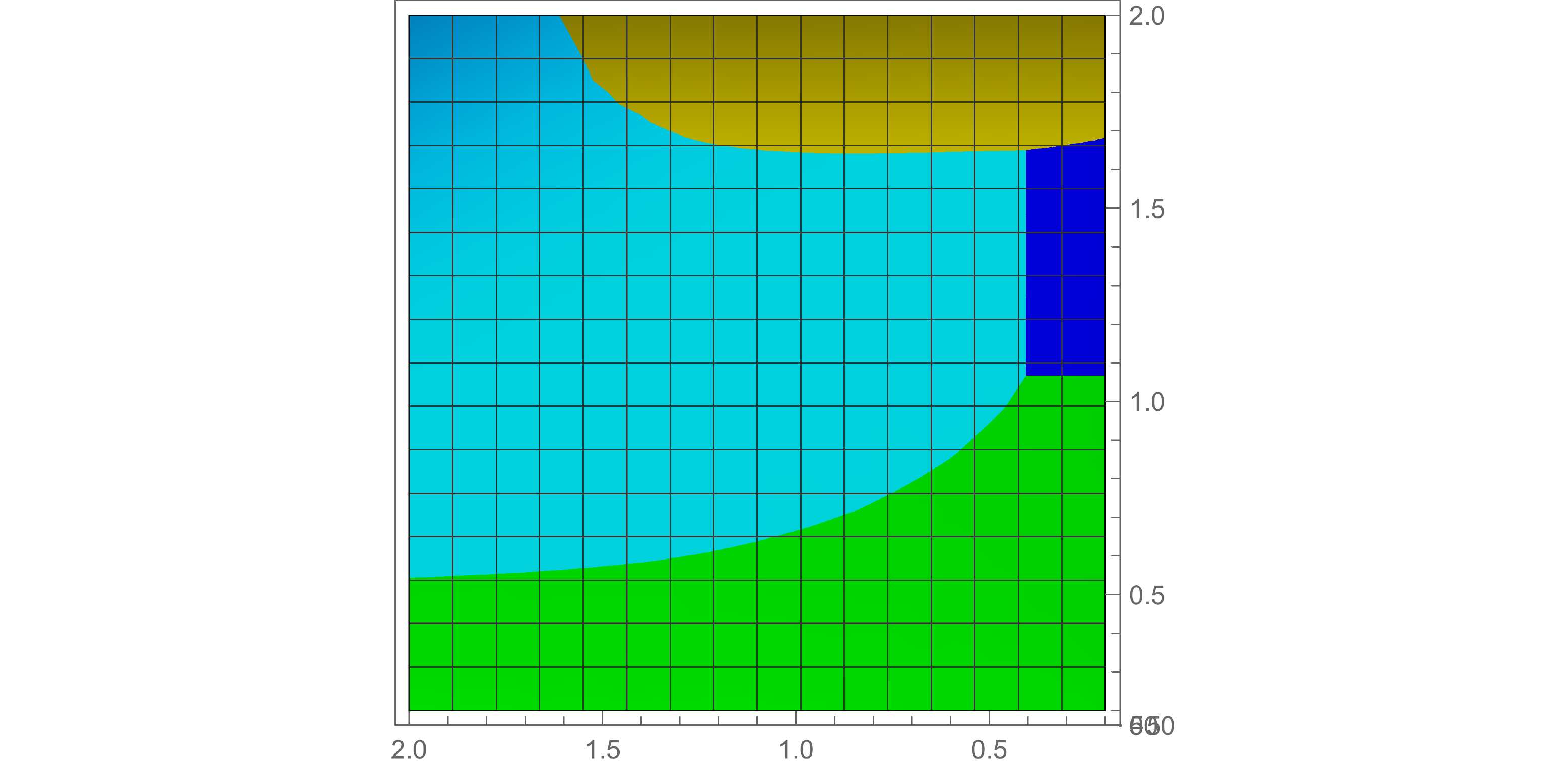}}
\put(100,80){\includegraphics[scale=0.15]{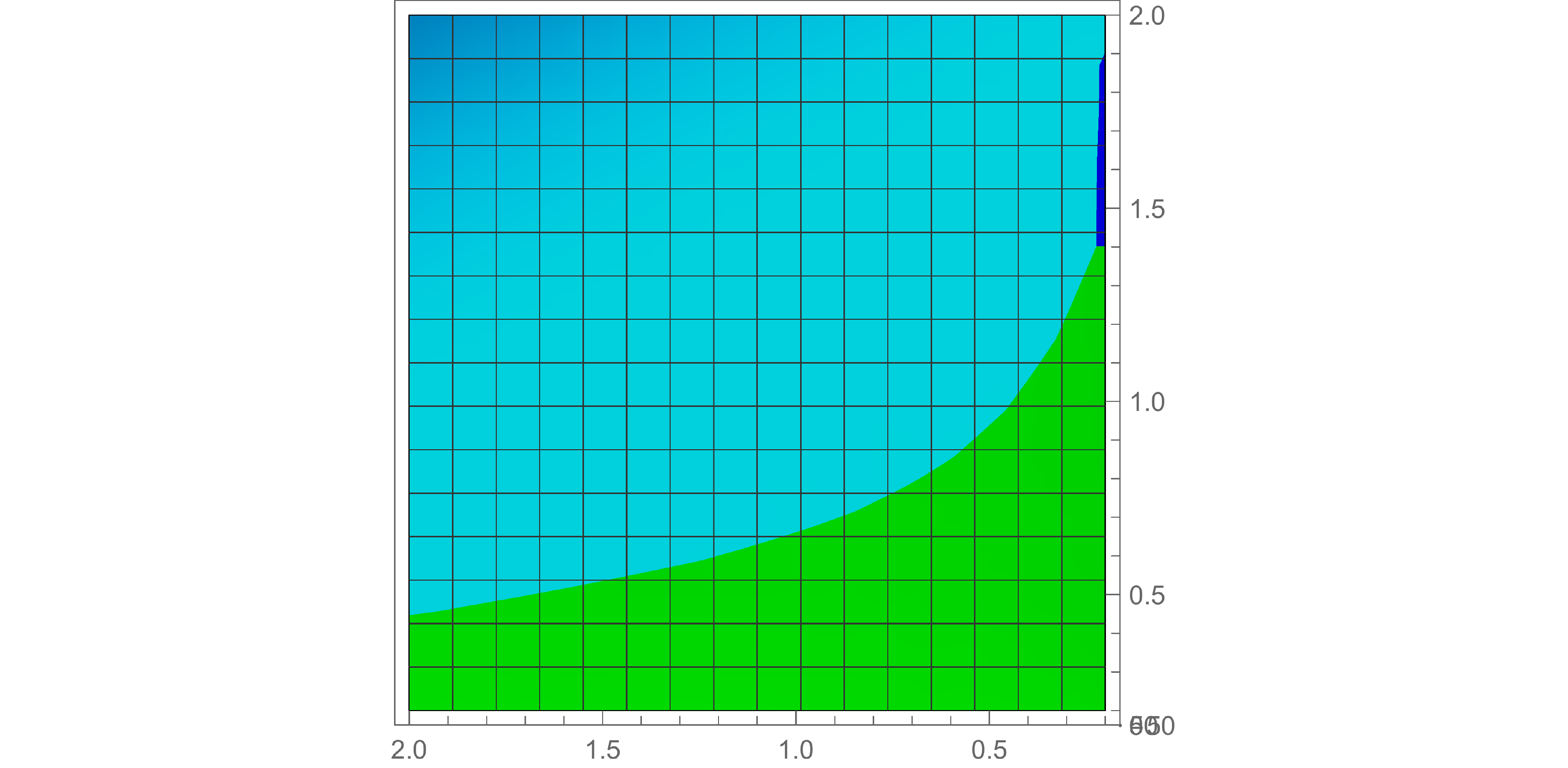}}
\put(200,80){\includegraphics[scale=0.15]{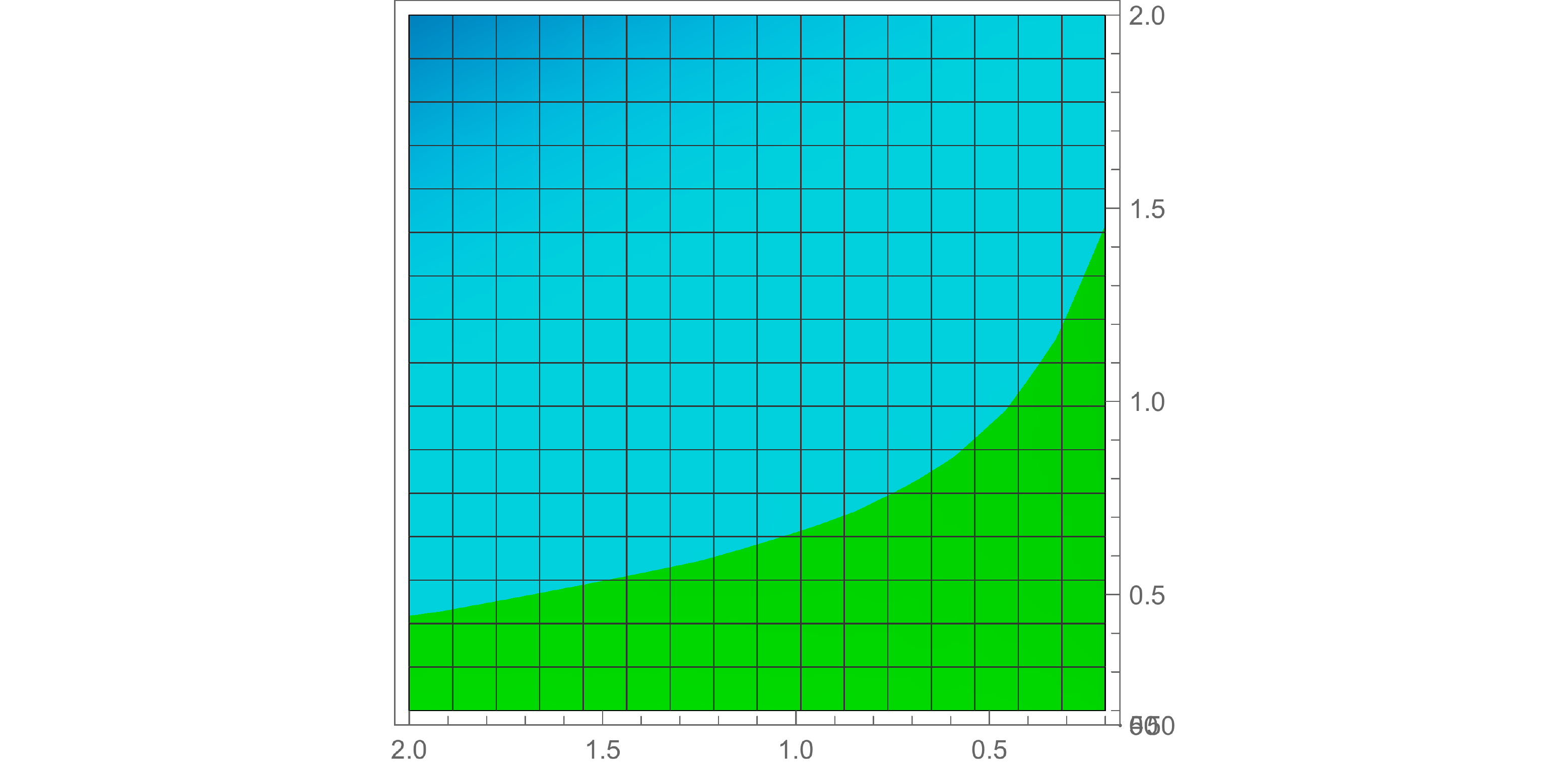}}
\put(-100,0){\includegraphics[scale=0.15]{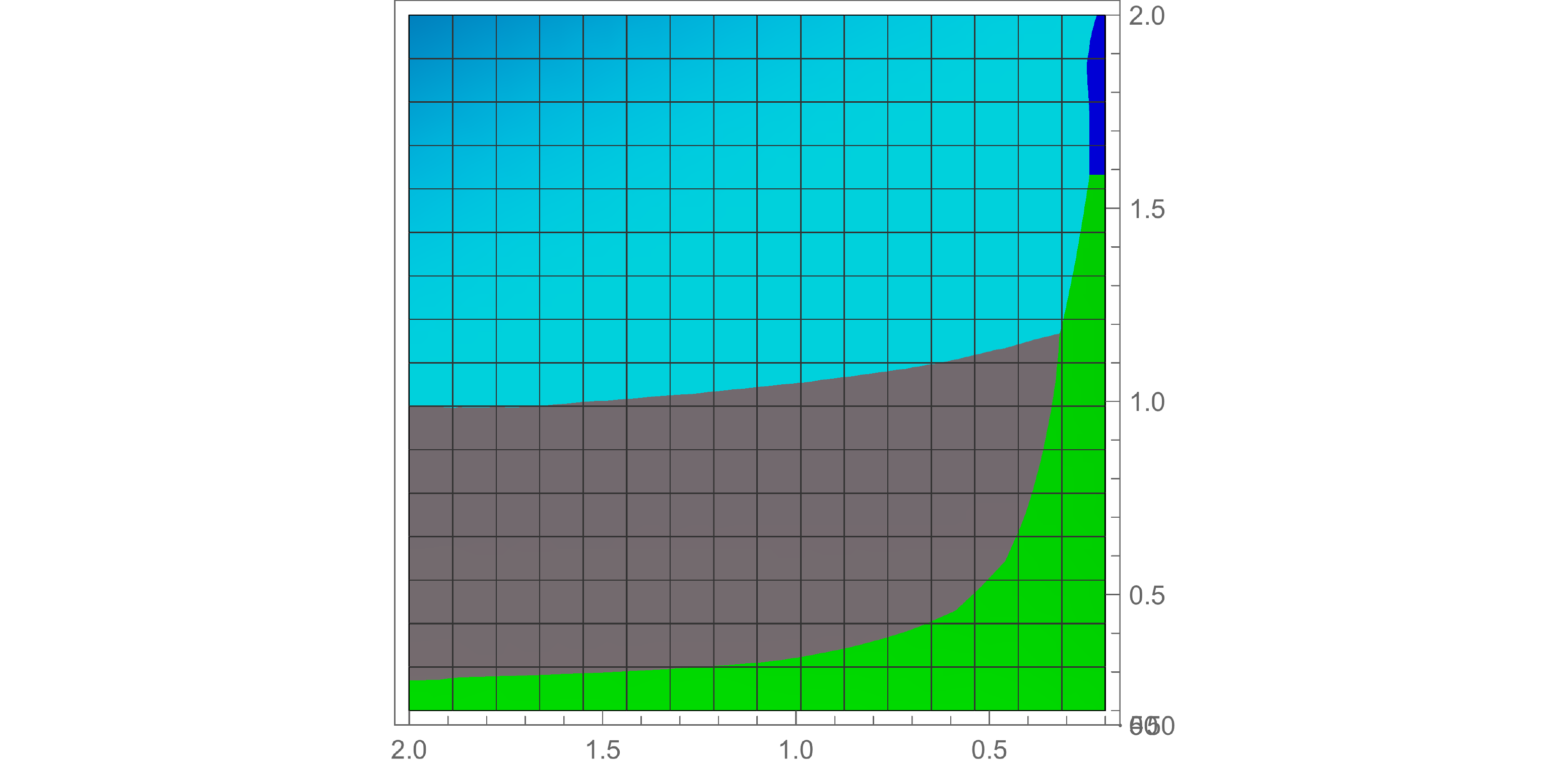}}
\put(0,0){\includegraphics[scale=0.15]{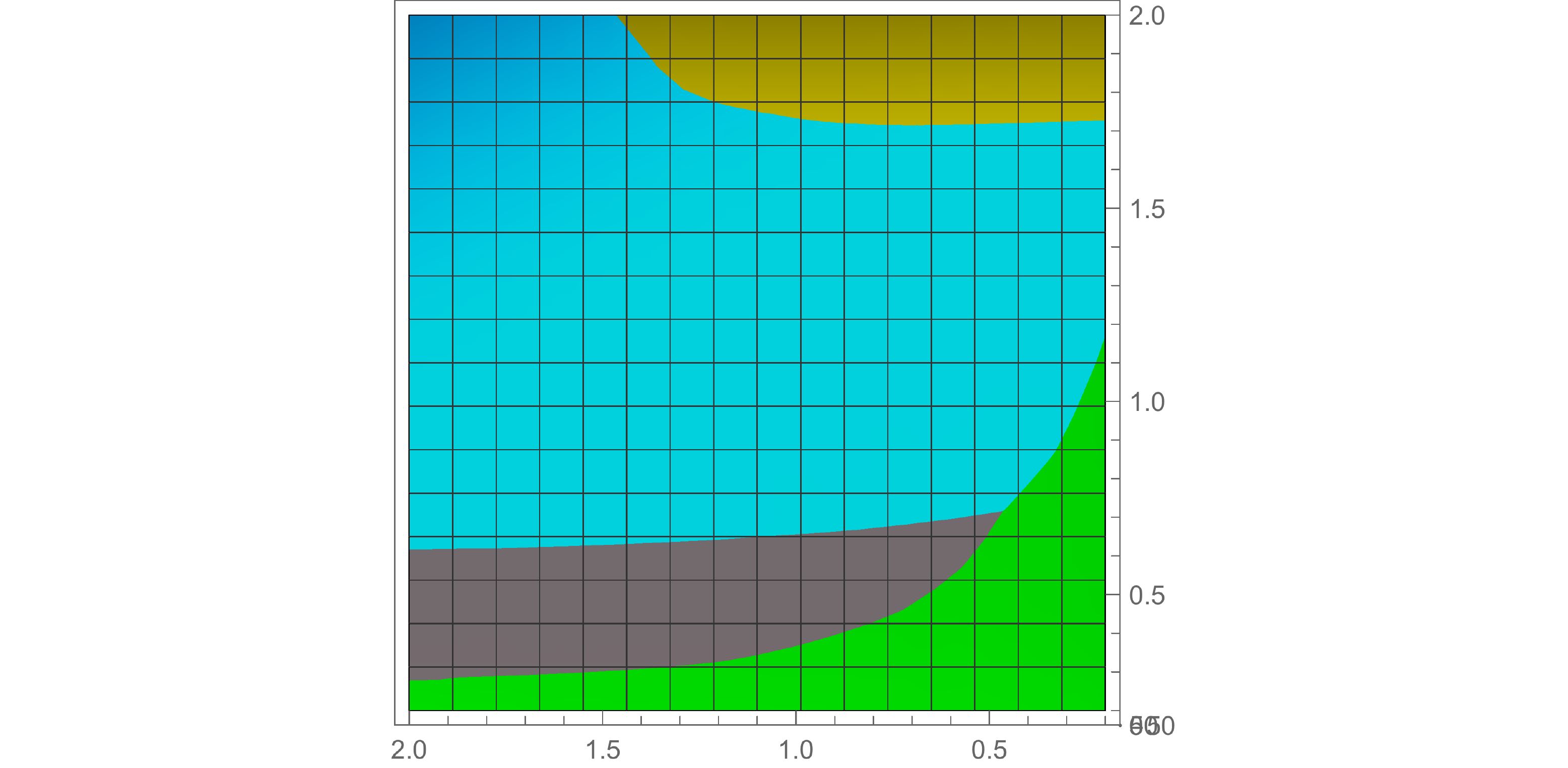}}
\put(100,0){\includegraphics[scale=0.15]{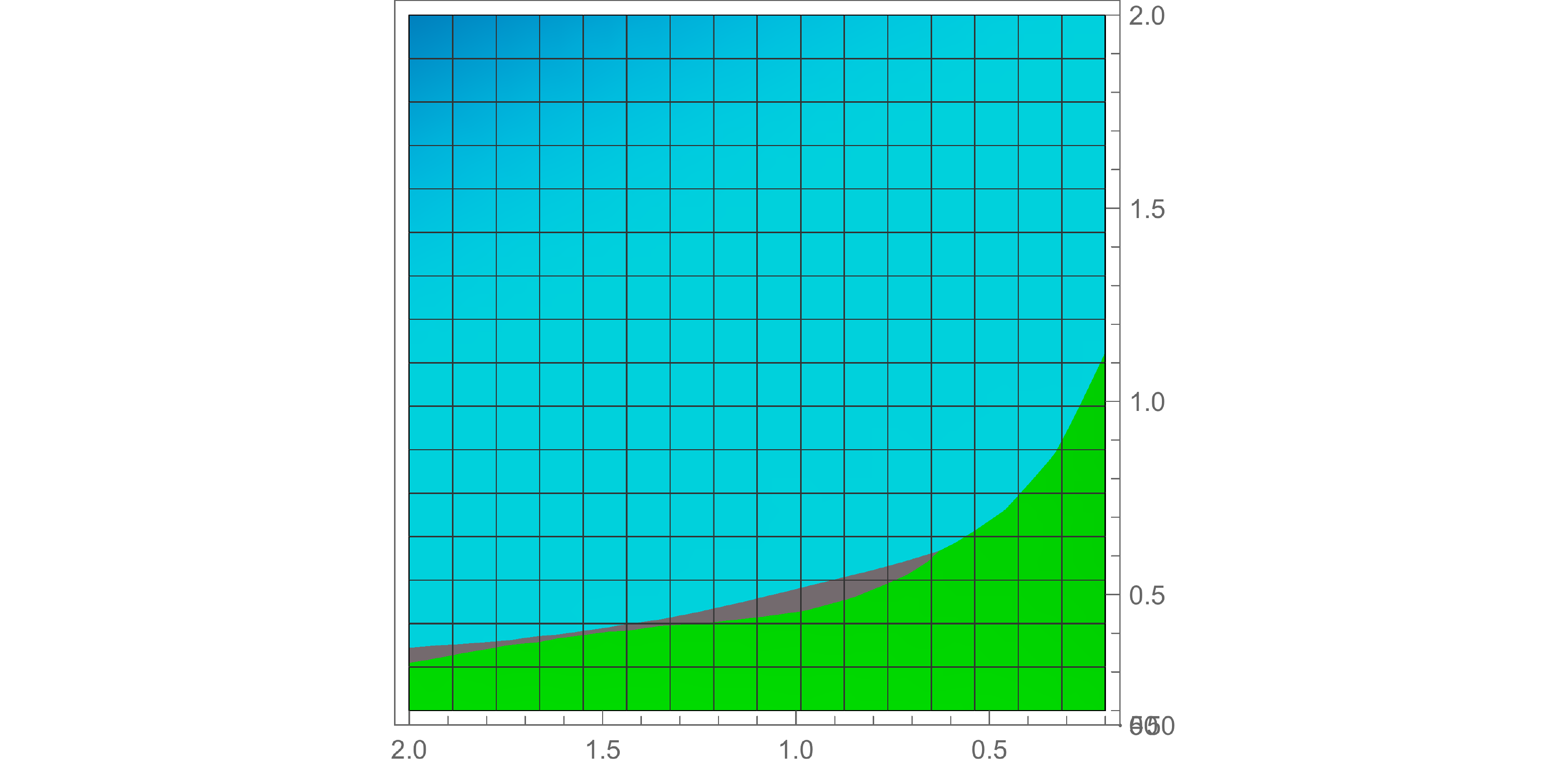}}
\put(200,0){\includegraphics[scale=0.15]{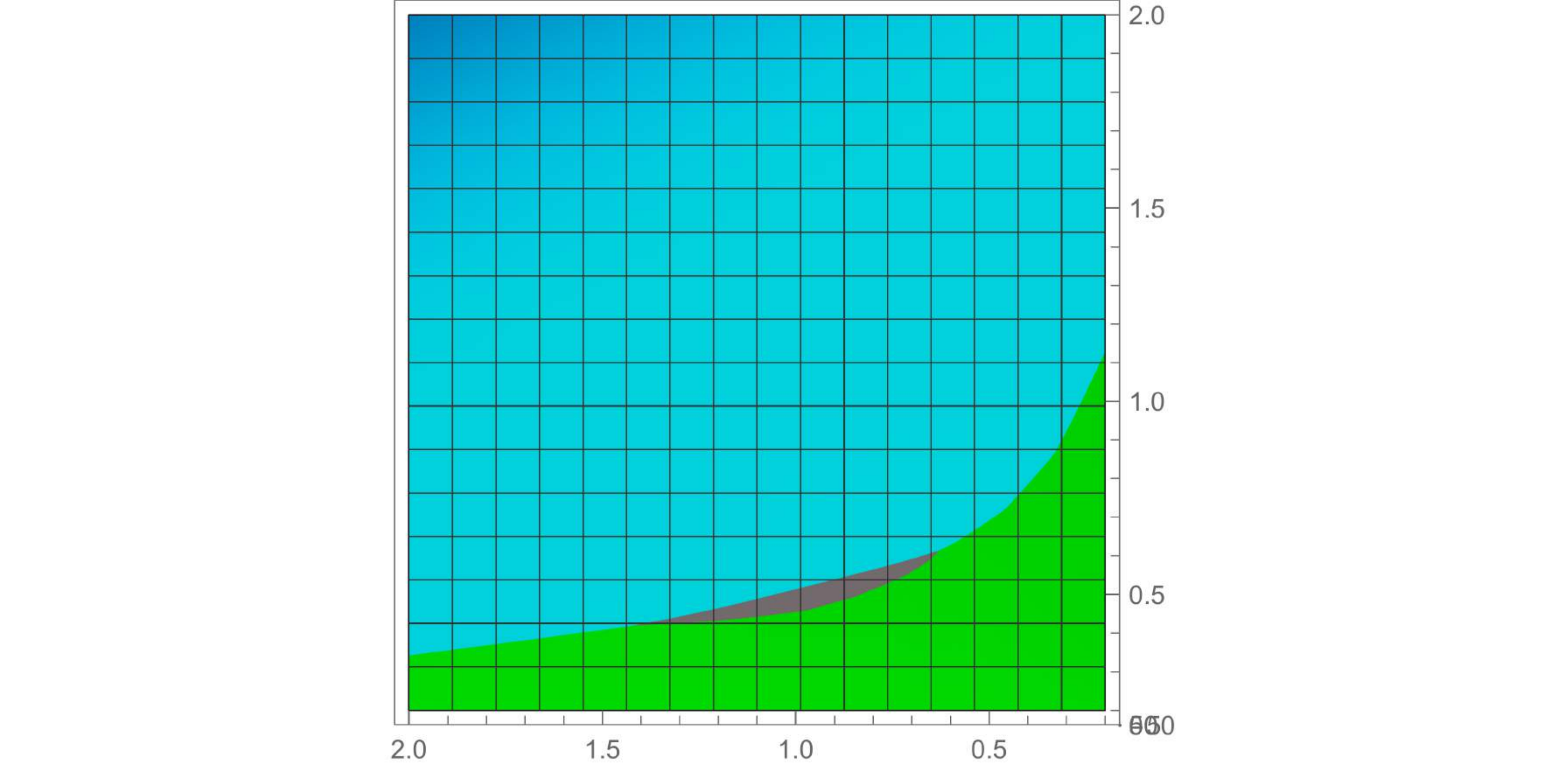}}
\put(-100,-80){\includegraphics[scale=0.15]{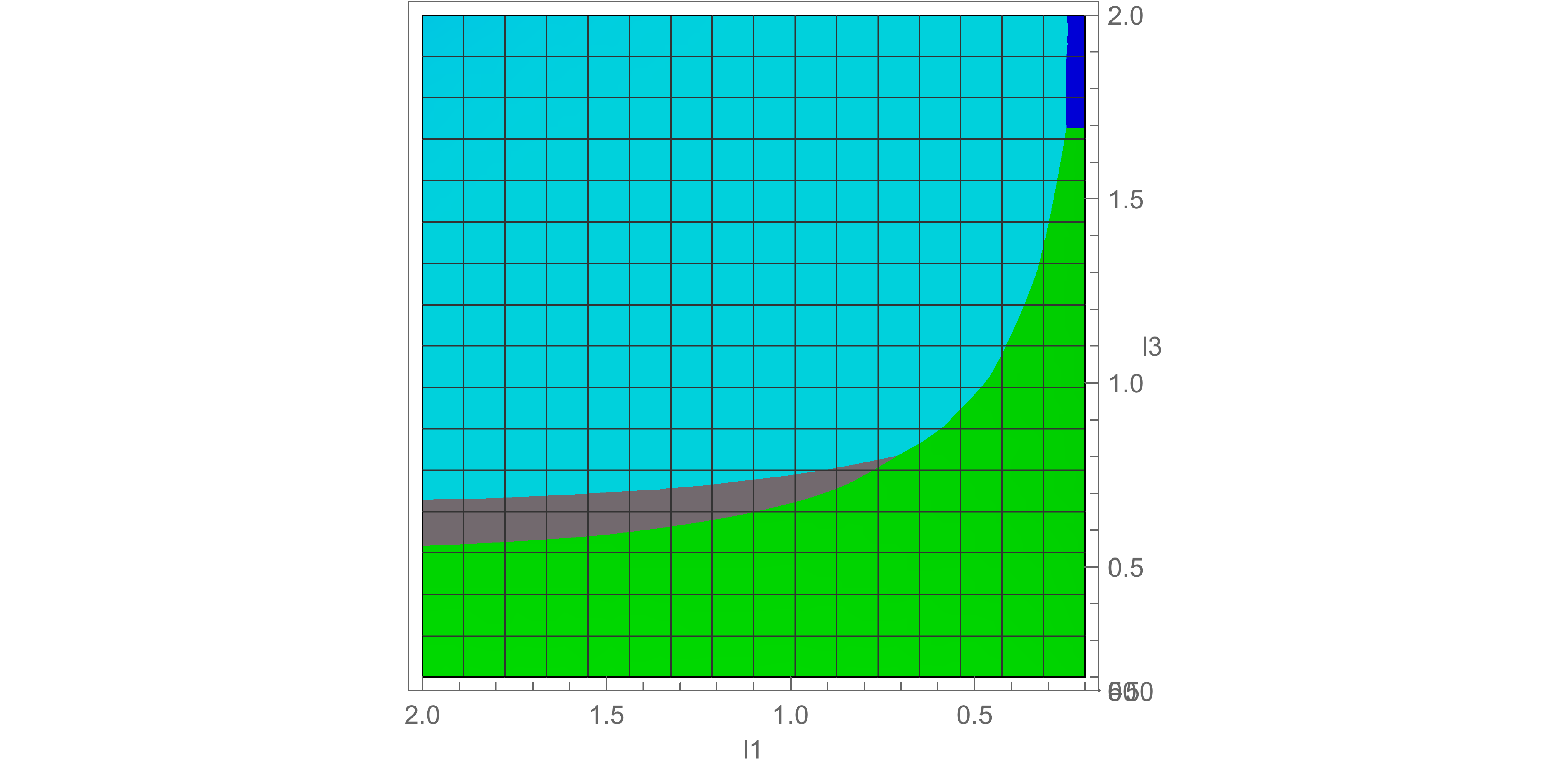}}
\put(0,-80){\includegraphics[scale=0.15]{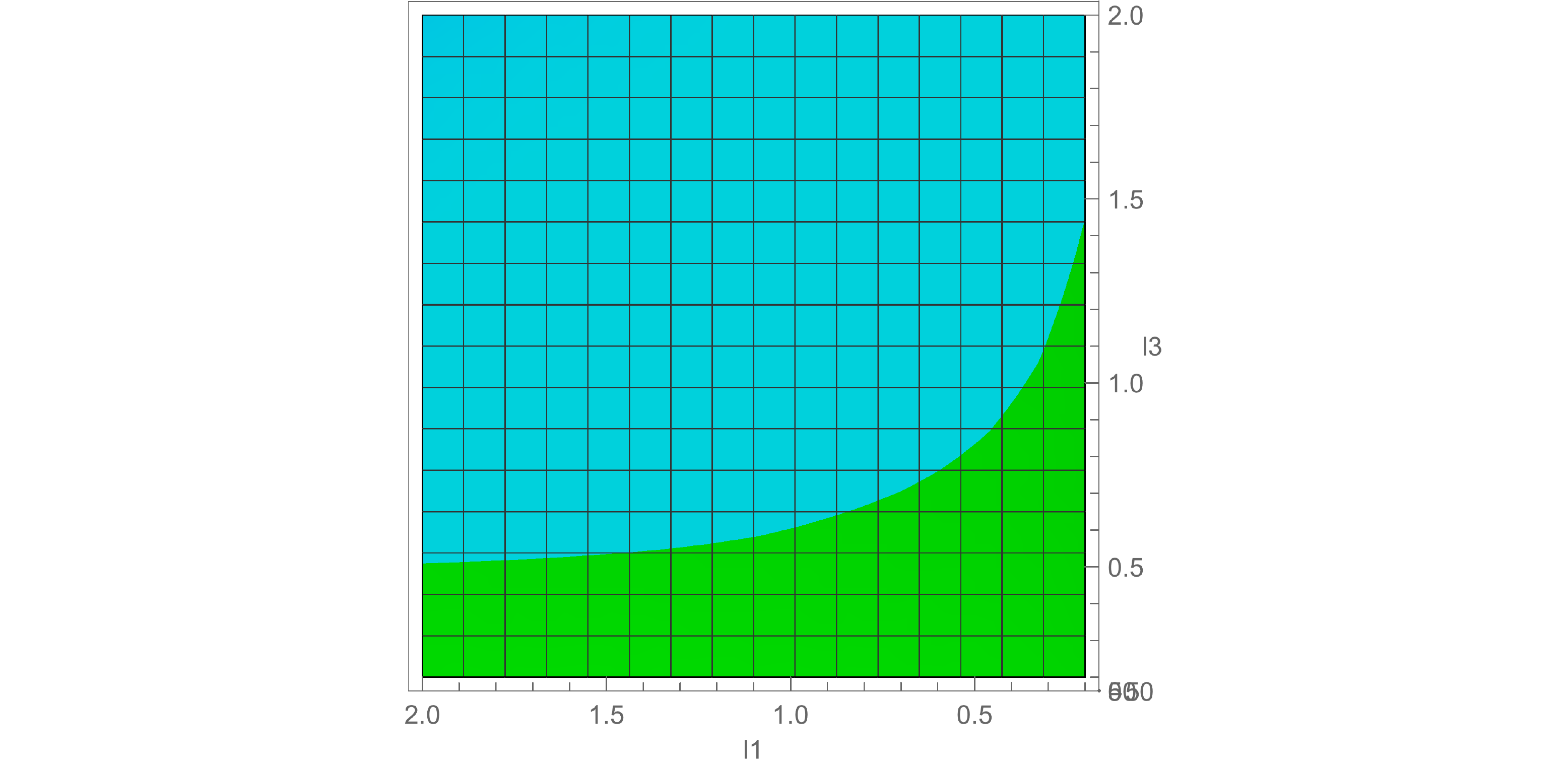}}
\put(100,-80){\includegraphics[scale=0.15]{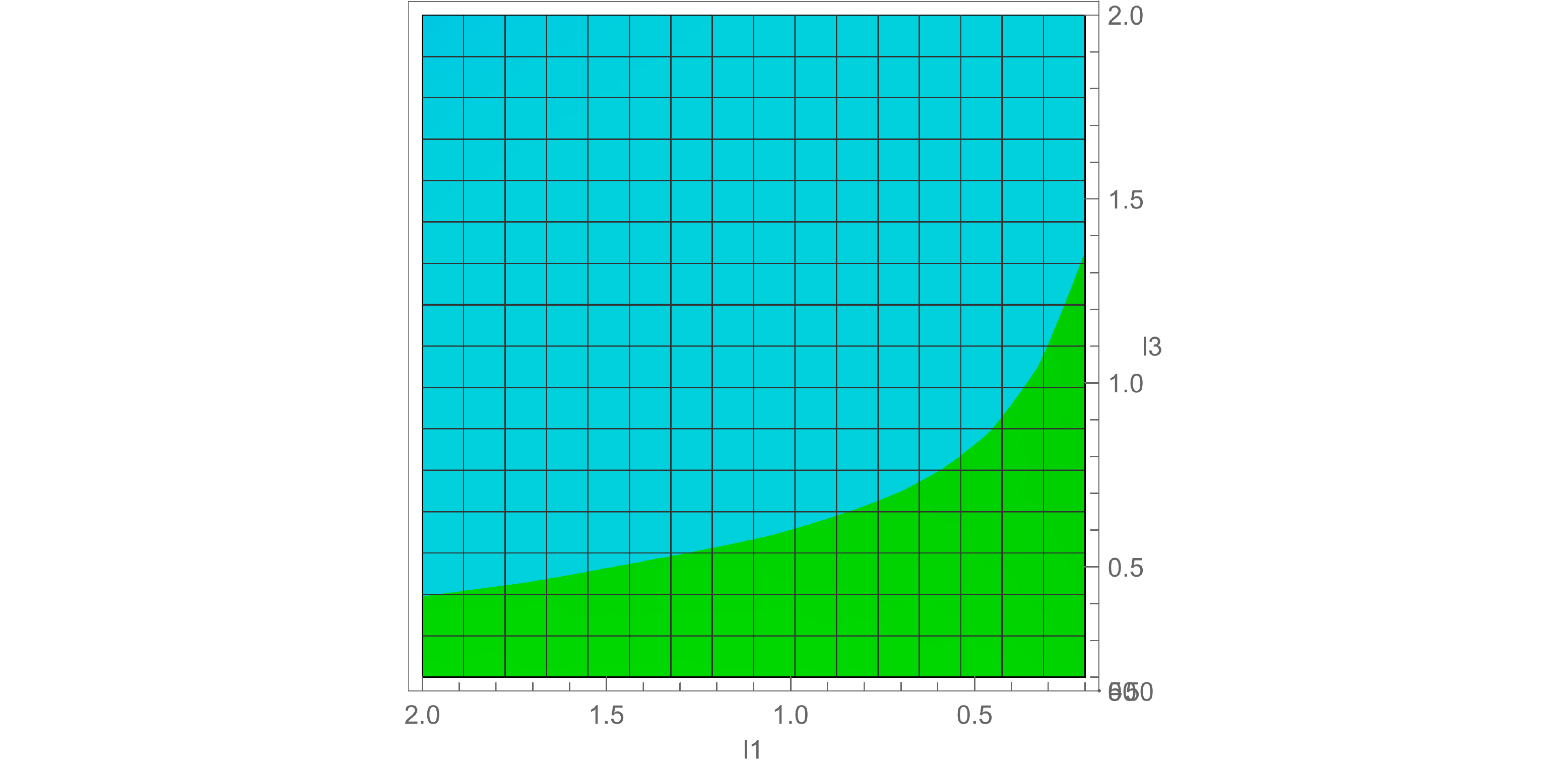}}
\put(200,-80){\includegraphics[scale=0.15]{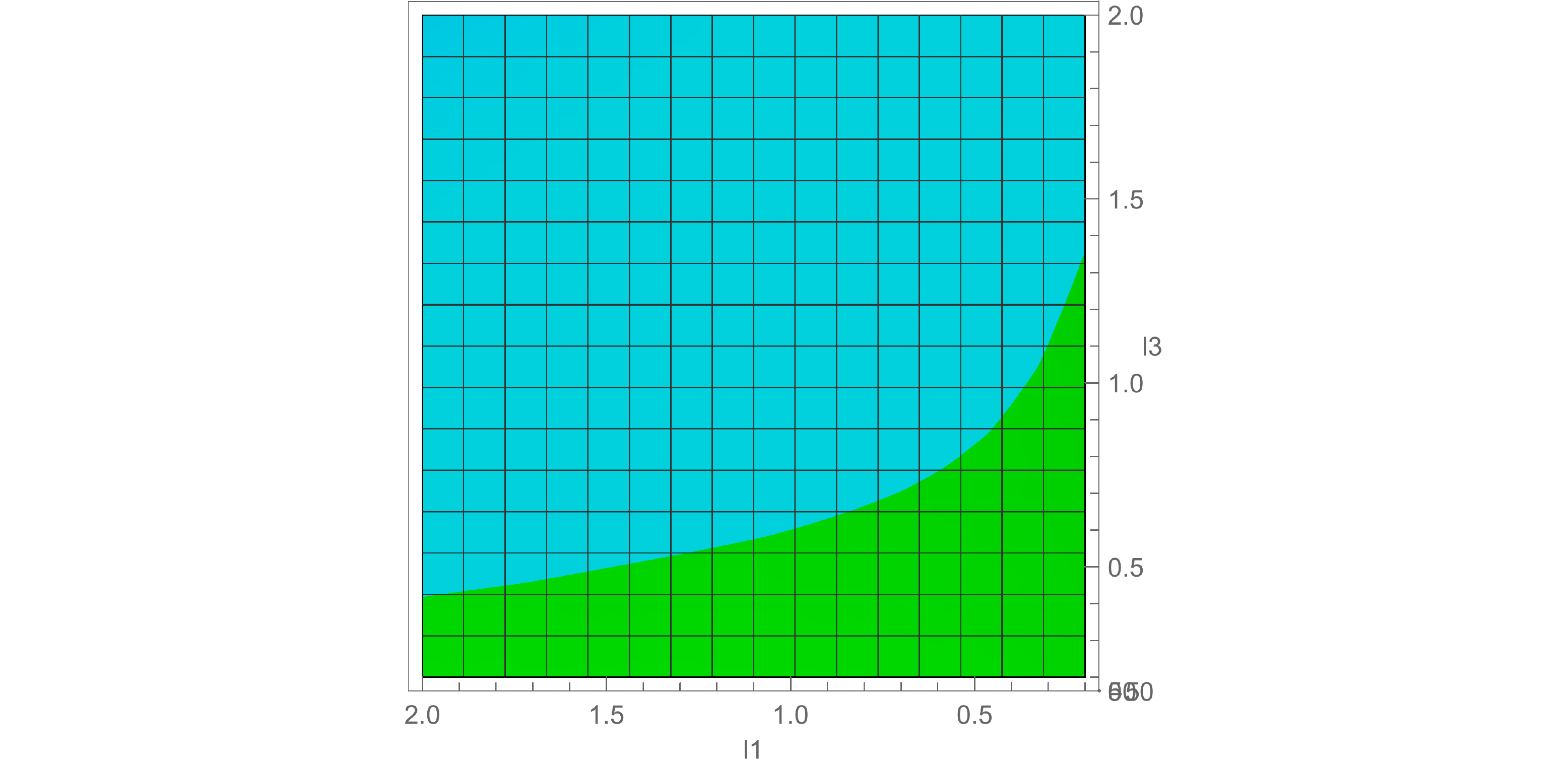}}
\put(-50,-90){$l_1$}
\put(310,-50){$l_3$}
\put(-35,150){$t=0$}
\put(50,150){$t=1$}
\put(150,150){$t=2$}
\put(250,150){$t=3$}\end{picture}
$$\,$$
$$\,$$
$$\,$$
$$\,$$
 \caption{Cartoon diagrams that show the contributions of various diagrams to the entanglement entropy for 3 strips, $A,B,C,$ with unequal lengths $l_1,l_2,l_3$ at different times during the transition  from the  black brane configuration with the small mass $m_0$ to the large mass $m$ in AdS$_3$. The green color regions correspond to the bulk surface
  $(A)||(B)||(C)$, the blue color regions correspond to the bulk surface $(A)||(B\,C)_c$, the gray color regions correspond to the bulk surface $(A\,B)_c||(C)$, the cyan color corresponds to the bulk surface  $(A,B,C)_{c,non-cr}$ and the yellow regions correspond to the bulk surface
  $(A_{_{\Huge {B}}}C)_c$. Different colors regions are separated by the curves that  are the transition lines. In the top plots $l_2=l_3, x=0.4,y=0.4$, in the middle plots $l_2=l_3, x=0.2,y=0.5$ and  in the bottom plots $l_2=0.5\,l_1$, $x=0.4,y=0.5$
 and we vary $l_1$  and $l_3$ keeping the distances between segments fixed. All plots correspond to $\alpha=0.2$.
 }
  \label{Fig:Cartoon-d2}
\end{figure}

In this section we  consider the phase structure of  holographic entanglement entropy  $S(A\cup B\cup C)$ for the Vaidya shell in the three dimensional AdS black brane background with $f=f(z,v)$
 given by \eqref{fvz} and \eqref{m-v-BB},
and 3 segments shown in Fig.\ref{Fig:Dec2}, at different times during the quench.  This quantity depends on many variables, $l_1,l_2,l_3,x,y$, $m,m_0$ and $\alpha$, and  we can perform  our analysis only numerically, see  Fig.\ref{Fig:Cartoon-d2}. In these plots  contributions of various diagrams to the entanglement entropy for 3 strips, $A,B,C,$ with unequal lengths $l_1,l_2,l_3$ at different times during the transition  from the  black brane configuration in AdS$_3$ with the small mass $m_0$ to the large mass $m$ are shown.
We see that there are plots there the contribution to the holographic entanglement entropy is given by the "engulfed" diagram.
These regions are colored yellow in images in Fig.\ref{Fig:Cartoon-d2}. The  images that contain yellow regions are located at the intersections of the  second column and top and middle rows. Hence
we have to take into account these diagrams in the analysis of the mutual information. 
To this purpose we use the diagram  technique
described in the end of Sect.\ref{Decom}.

\begin{figure}[h!]
\centering
$\,\,\,\,\,\,\,\,\,\,\,\,\,\,\,\,\,\,$
\begin{picture}(250,300)%Vaidya-BH-BHd3-05-05-1-0
\put(-100,80){\includegraphics[scale=0.15]{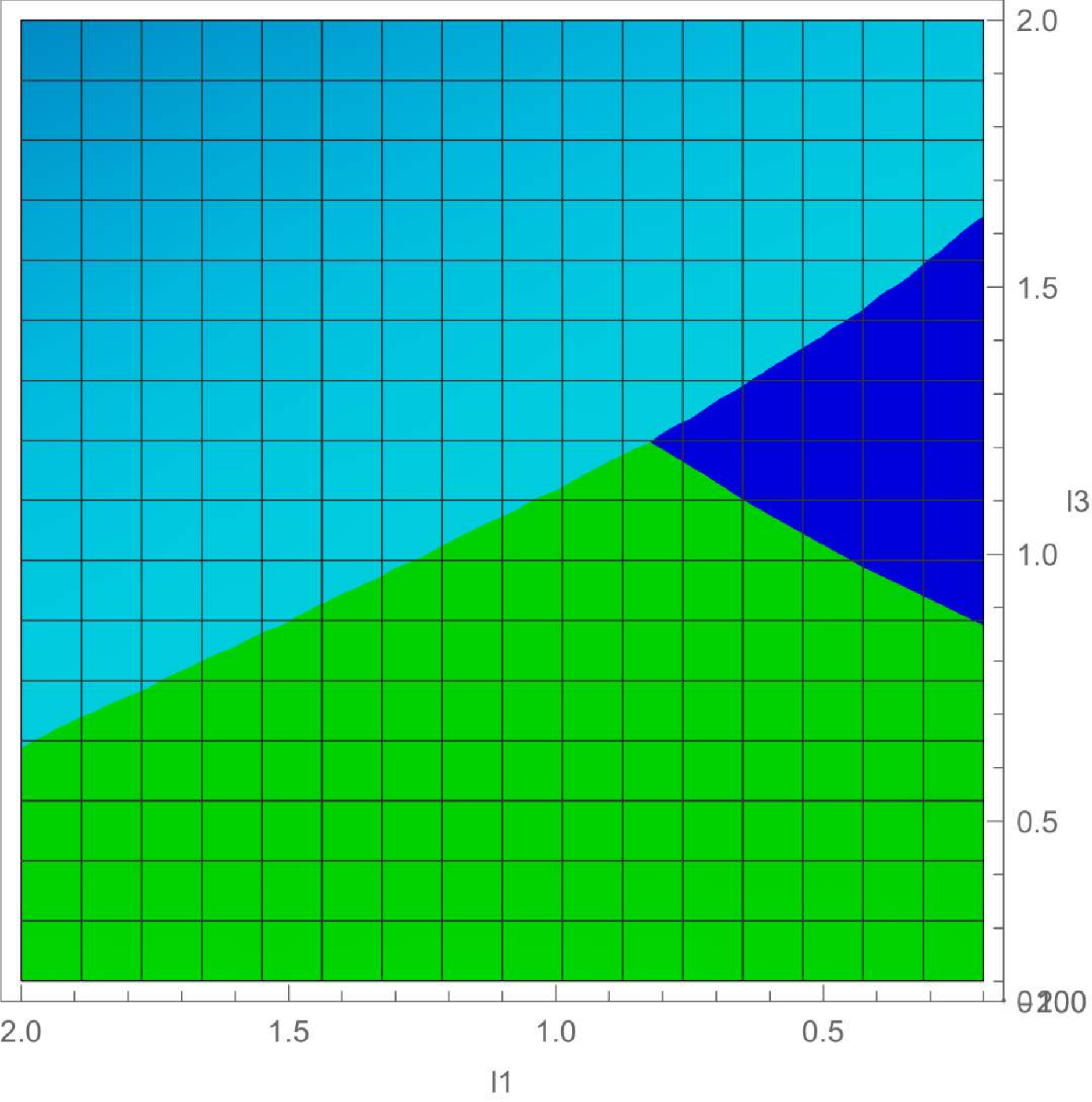}}
\put(0,80){\includegraphics[scale=0.15]{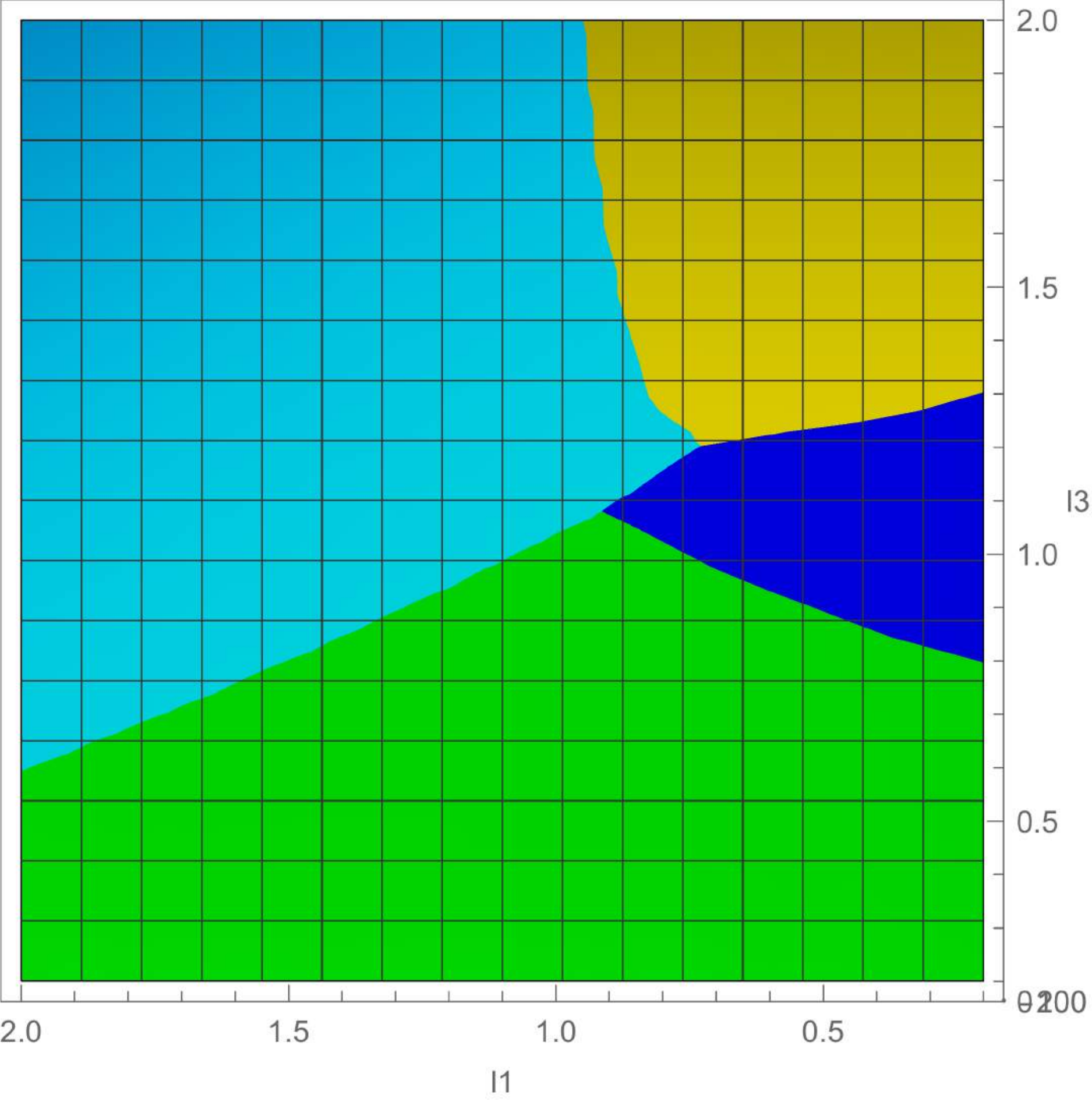}}
\put(100,80){\includegraphics[scale=0.15]{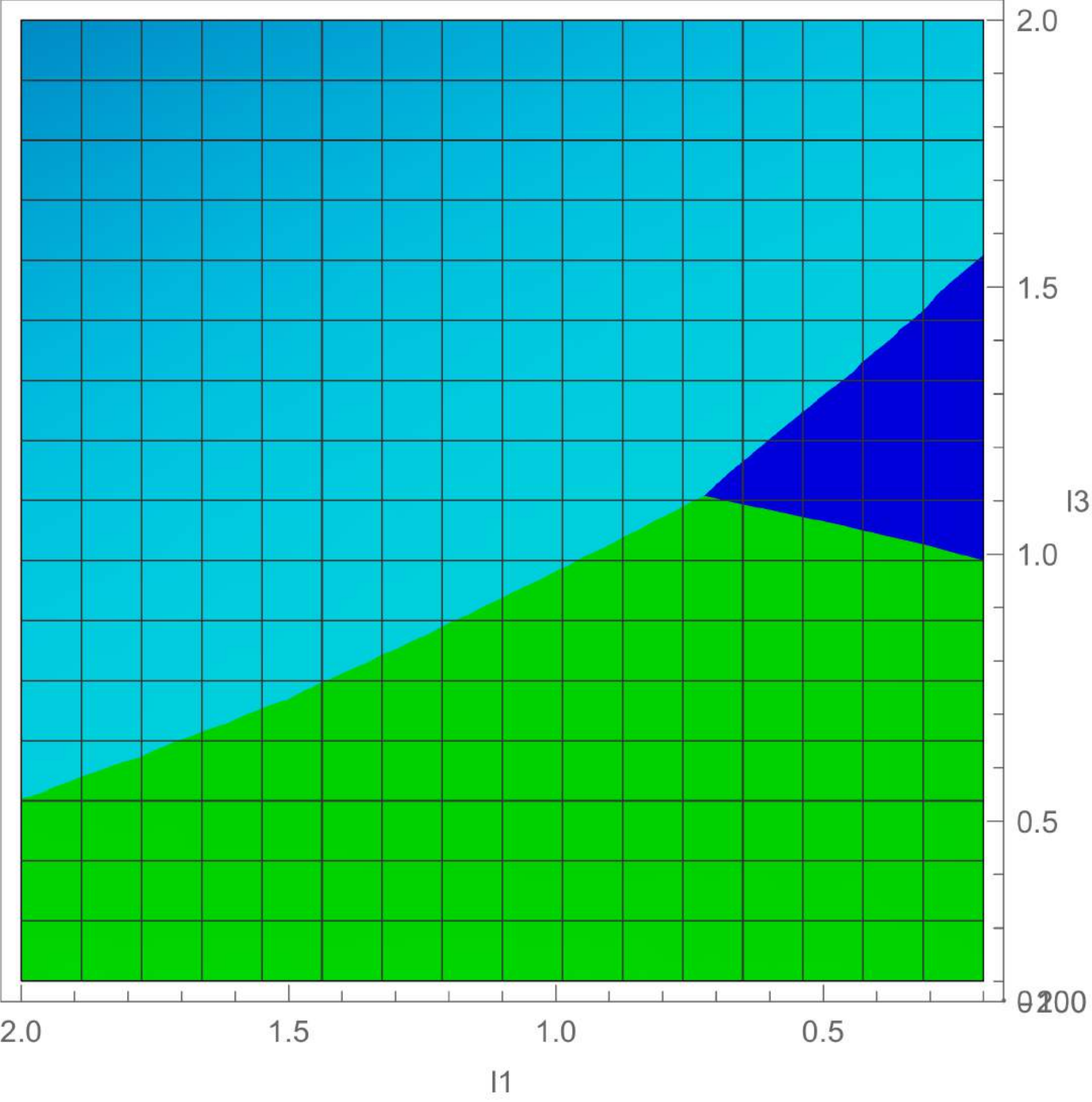}}
\put(200,80){\includegraphics[scale=0.15]{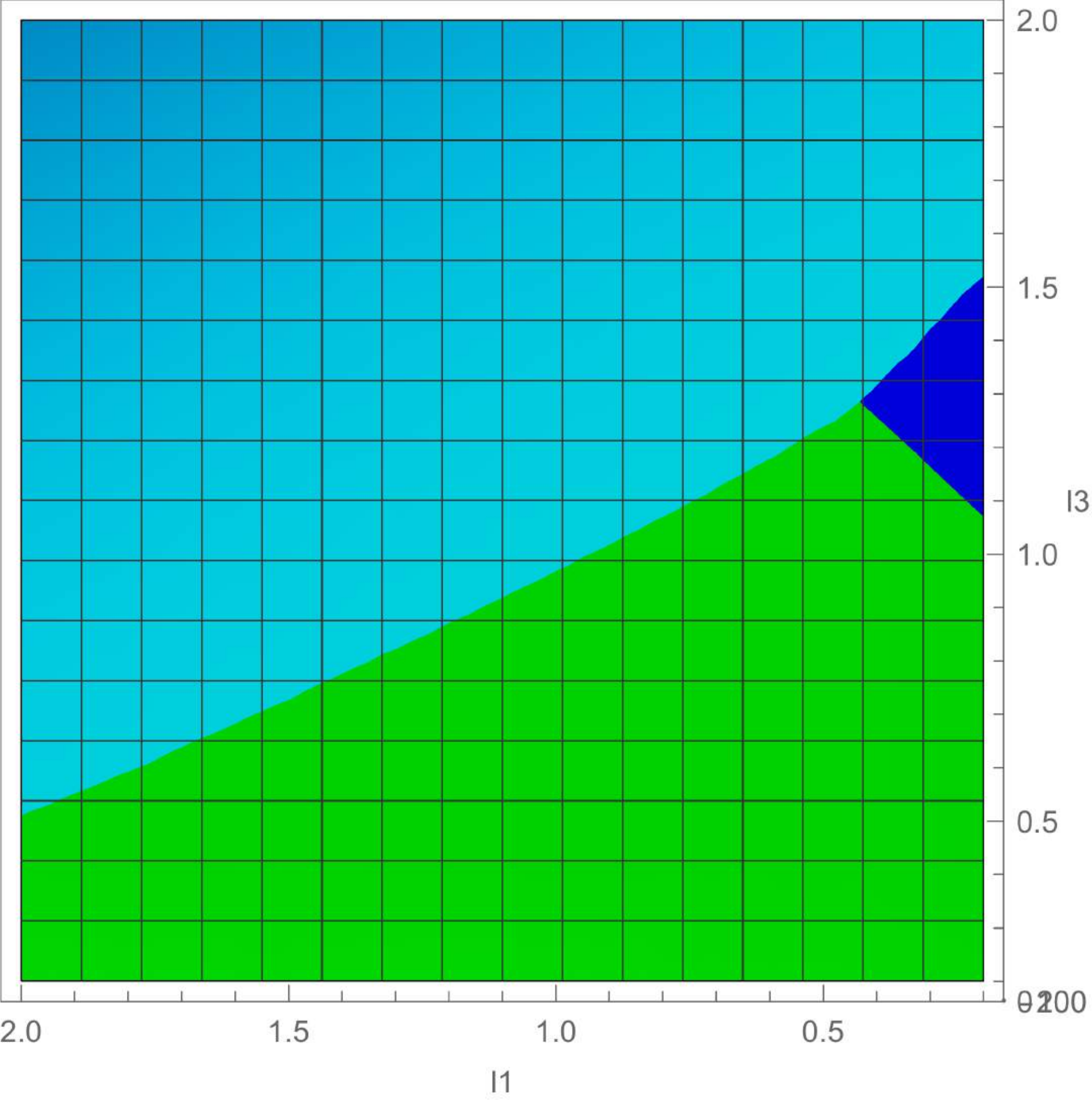}}
\put(-100,0){\includegraphics[scale=0.15]{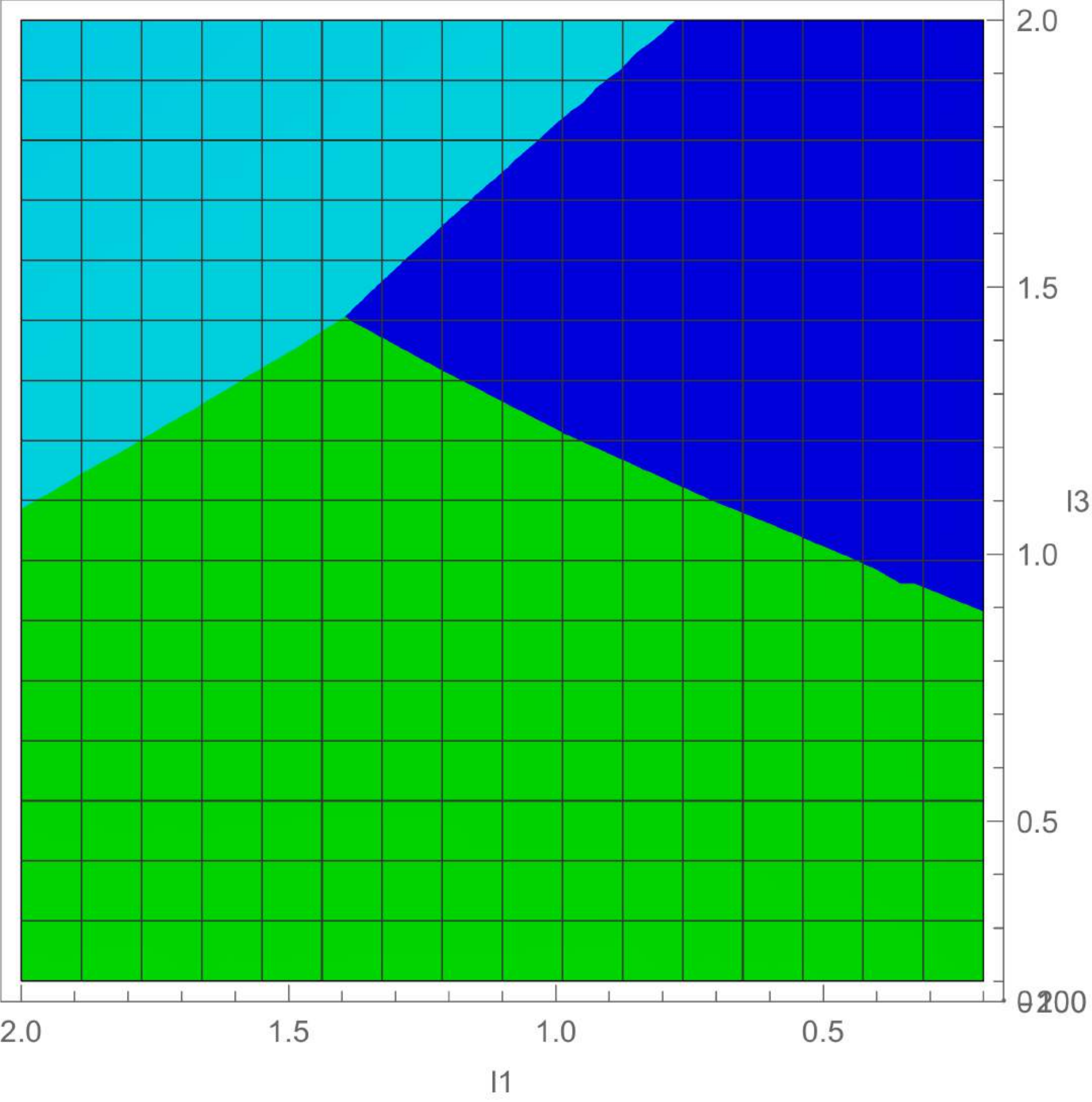}}
\put(0,0){\includegraphics[scale=0.15]{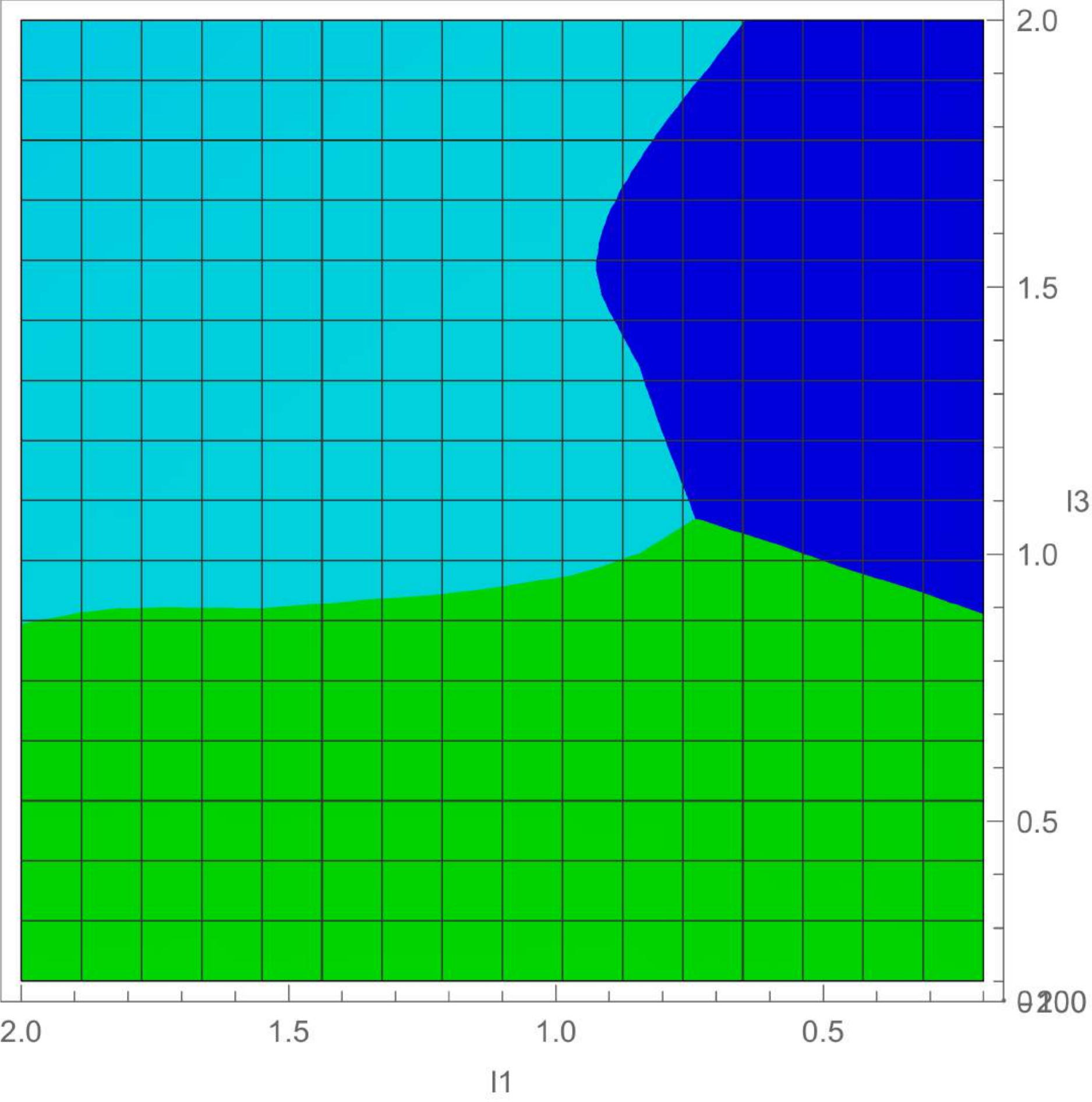}}
\put(100,0){\includegraphics[scale=0.15]{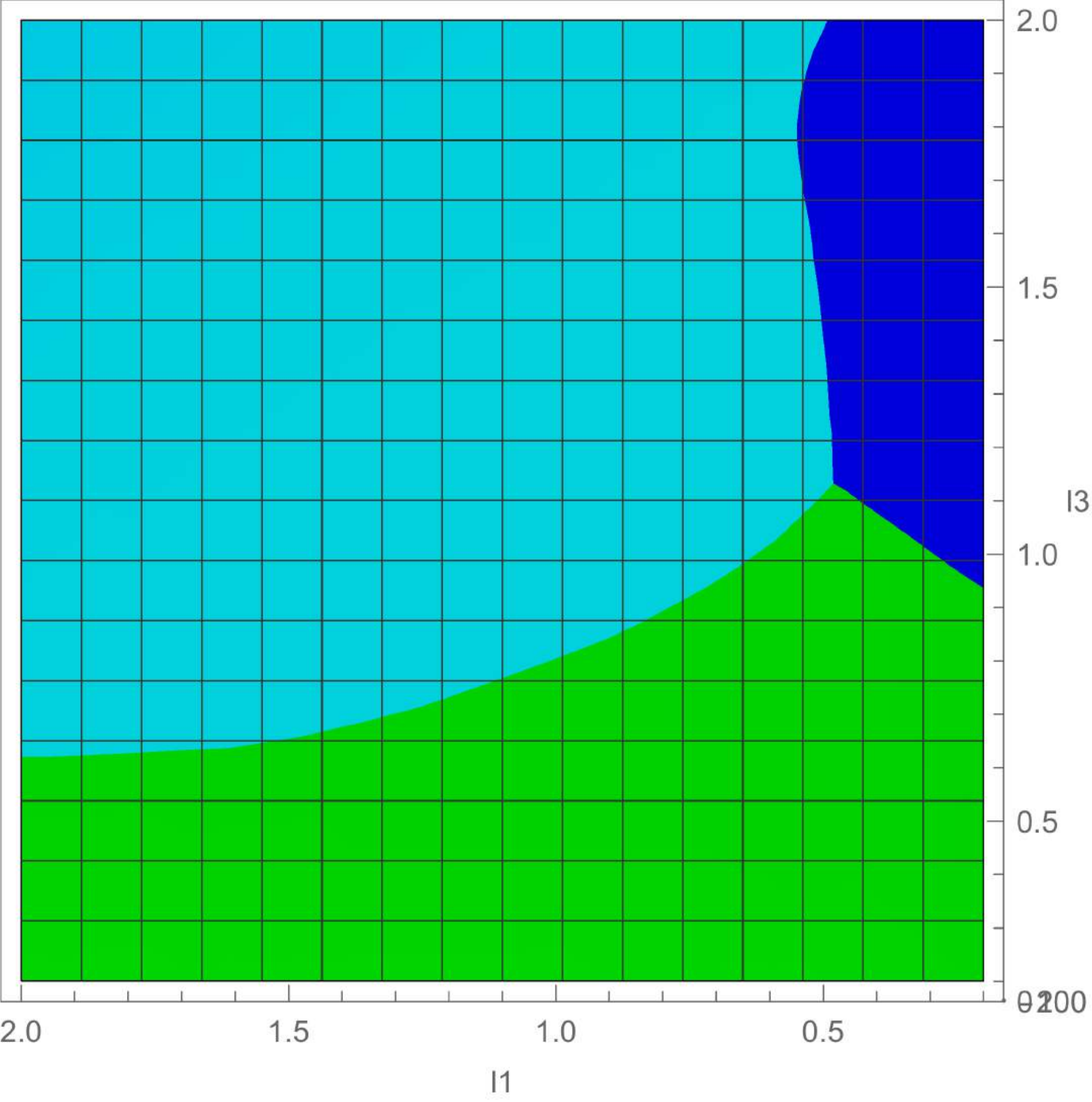}}
\put(200,0){\includegraphics[scale=0.15]{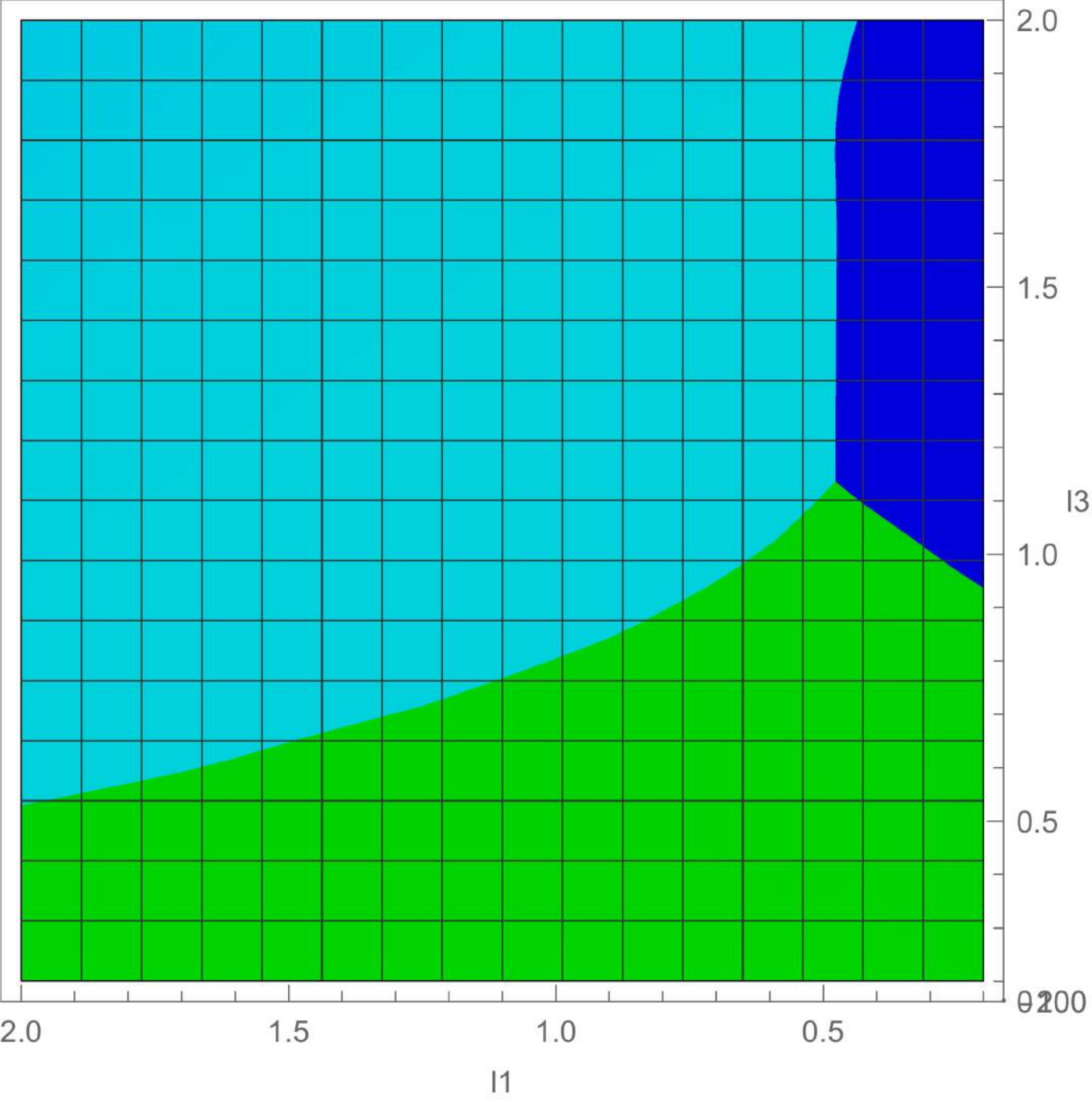}}
\put(-100,-80){\includegraphics[scale=0.15]{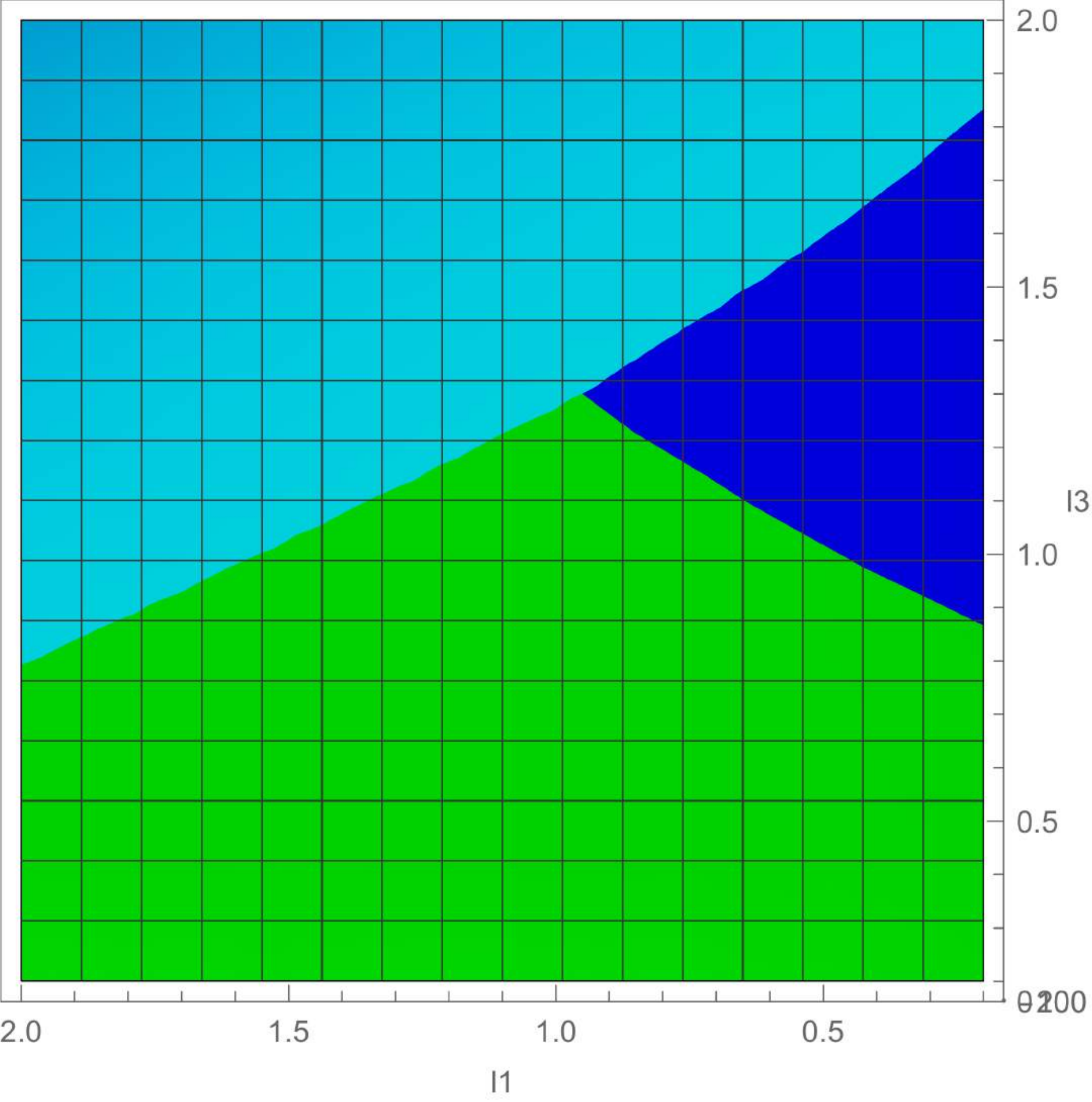}}
     \put(0,-80){\includegraphics[scale=0.15]{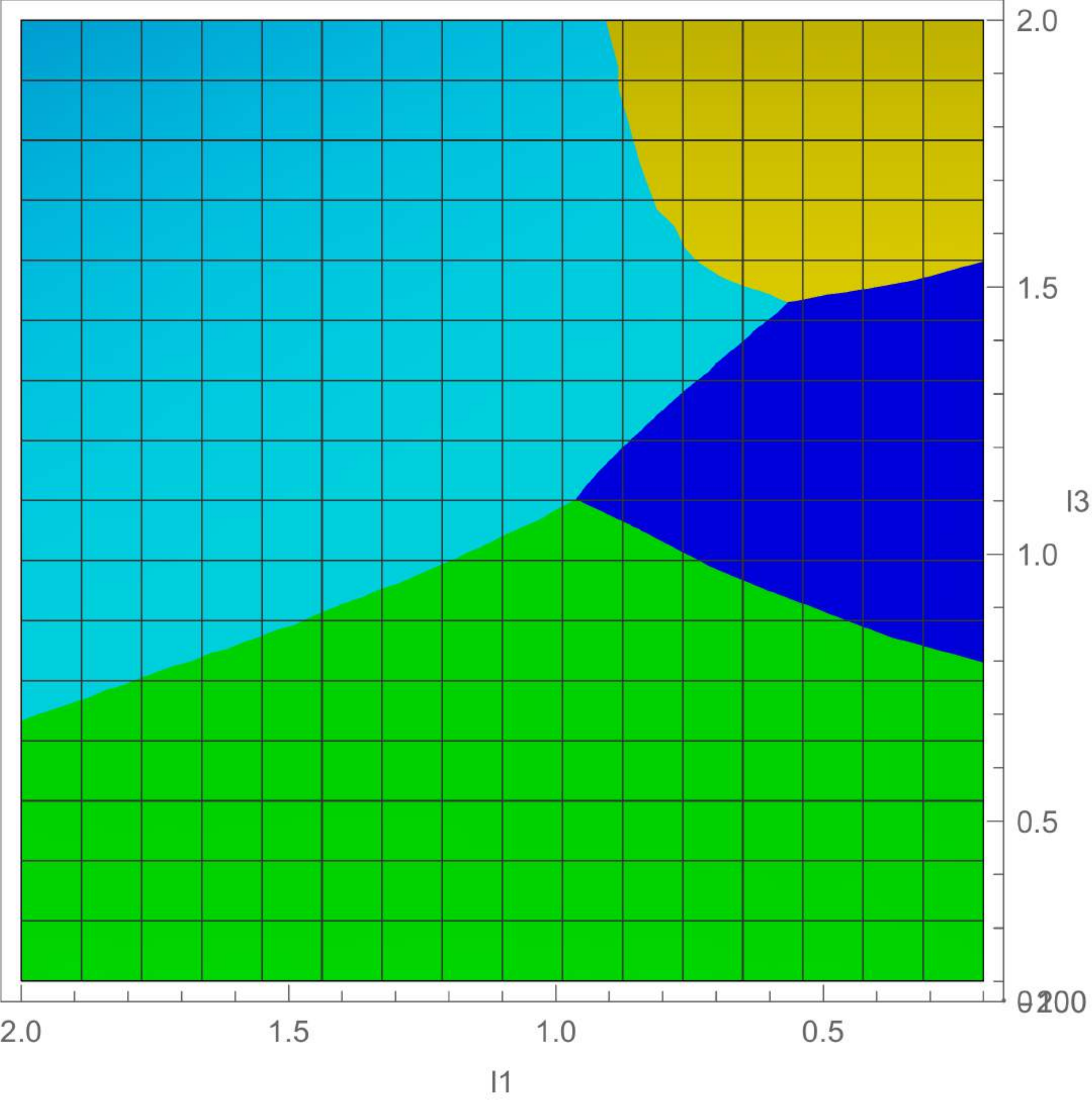}}
\put(100,-80){\includegraphics[scale=0.15]{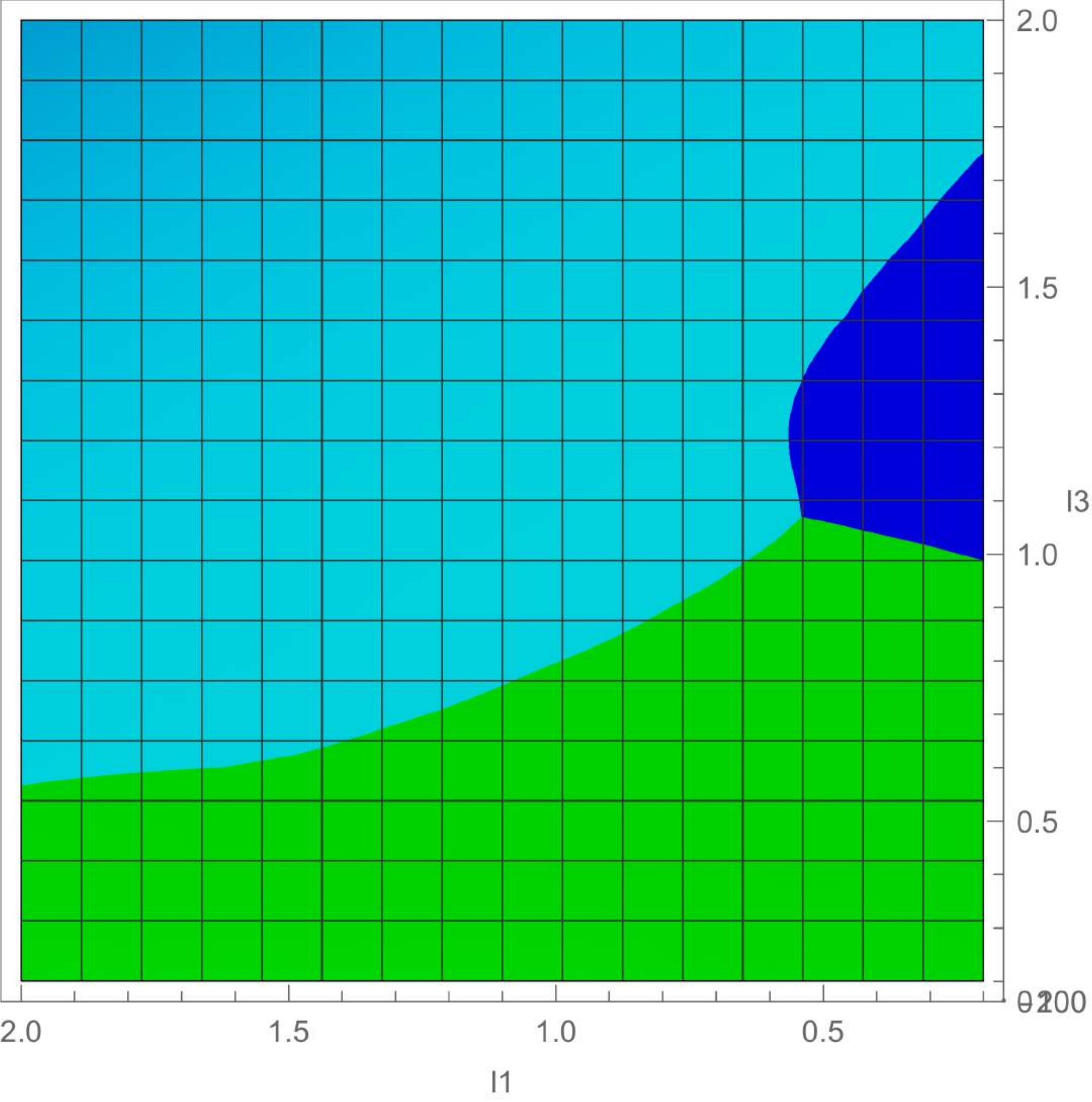}}
\put(200,-80){\includegraphics[scale=0.15]{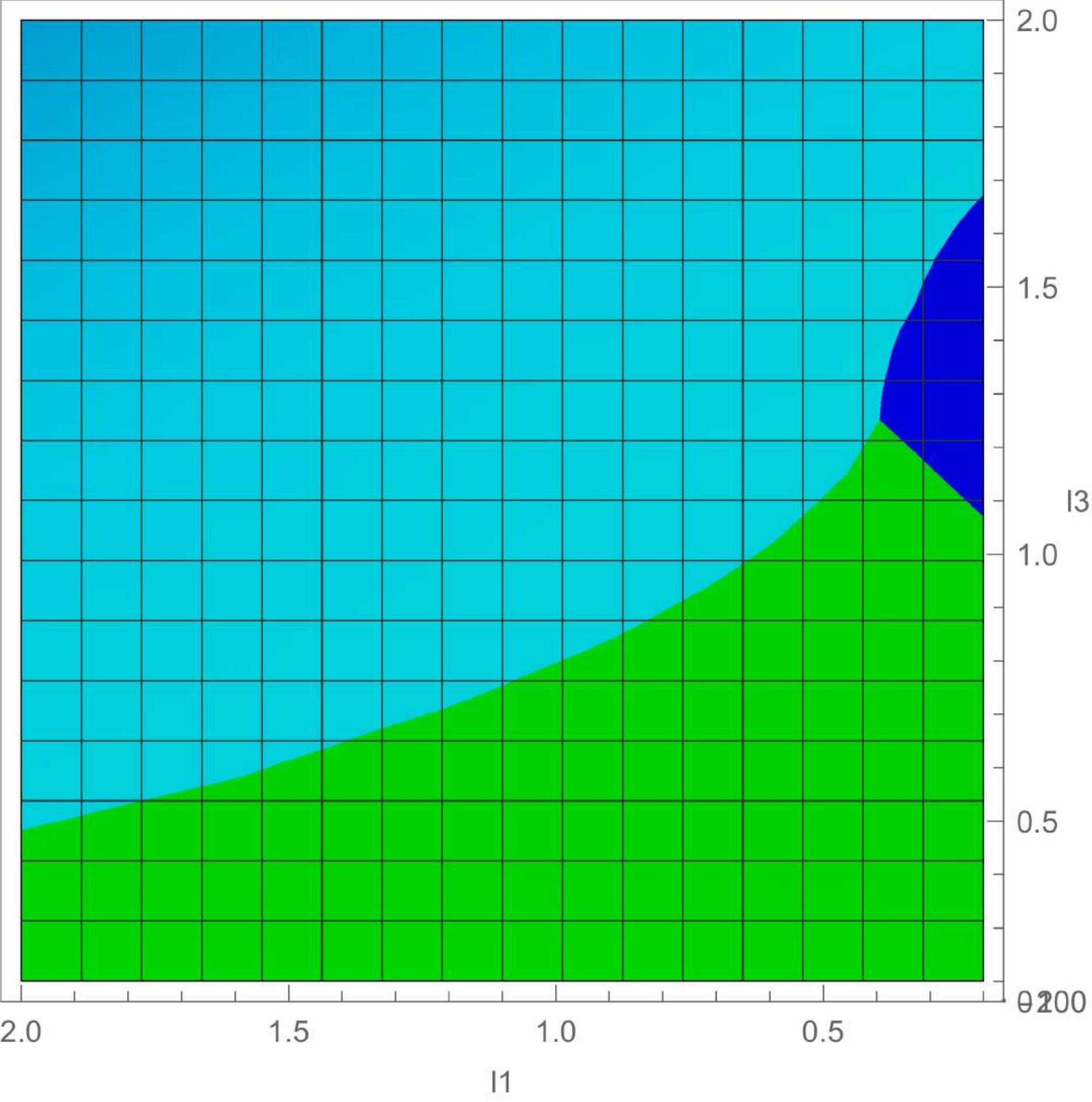}}
\put(-50,-90){$l_1$}
\put(280,-50){$l_3$}
\put(-75,170){$t=0$}
\put(50,170){$t=1$}
\put(150,170){$t=2$}
\put(250,170){$t=3$}\end{picture}
$$\,$$
$$\,$$
$$\,$$
$$\,$$
 \caption{Cartoon diagrams that show the contributions of various diagrams to the entanglement entropy for 3 strips, $A,B,C,$ with unequal lengths $l_1,l_2,l_3$ at different times during the transition  from the  black brane configuration with the small mass $m_0$ to the large mass $m$ in AdS$_4$. The green color regions correspond to the bulk surface
  $(A)||(B)||(C)$, the blue color regions correspond to the bulk surface $(A)||(B\,C)_c$, the gray color regions correspond to the bulk surface $(A\,B)_c||(C)$, the cyan color corresponds to the bulk surface  $(A,B,C)_{c,non-cr}$ and the yellow regions correspond to the bulk surface
  $(A_{_{\Huge {B}}}C)_c$. Different colors regions are separated by the curves that  are the transition lines. In the top plots $l_2=l_3, x=0.5,y=1$, in the middle plots $l_2=0.4l_3, x=0.4,y=0.5$ and  in the bottom plots $l_2=0.5\,l_1$, $x=0.5,y=1$
 and we vary $l_1$  and $l_3$ keeping the distances between segments fixed. All plots correspond to $\alpha=0.2$.
 }
  \label{Fig:Cartoon-d3}
\end{figure}
\newpage
$$\,$$
$$\,$$
\newpage

\subsection{Mutual information for  $(1,6|3)$-reduction of the  FMO complex and scrambling time}\label{MI-3segm}

	In the section we will consider also the  mutual information for  a system one part of which consists on two
	disjoint parts. Just this system corresponds to  one of simplest reduce FMO complexes  presented in Fig.\ref{Fig:fmo-1-3-6}.
	 Applying to this system the general definition  \eqref{HMI} we get
	\be\label{MI-AB-C}
		I(A\cup B;C)=S(A\cup B)+S(C)-S(A\cup B\cup C),
	\ee
	where $S(A\cup B\cup C)$ is the entanglement entropy for the
	union of three subsystems. $I(A\cup B,C)$ is related with
		the tripartite information which is  defined for a
	system consisting of three disjoint parts as
	follows
	\bea\label{3par}
		I_3(A;B;C) &=& S(A) + S(B) + S(C) - S(A \cup B) - S(A \cup C) \nn\\
		&-& S(B \cup C) + S(A \cup B \cup C).
	\eea
The tripartite information  can be positive, negative or zero, however
the holographic tripartite information is always negative, i.e. $ I_3(A;B;C)<0$.  The validity of this  inequality  means that the holographic mutual information is monogamous \cite{Hayden:2011ag}.
The tripartite information can be written in terms of mutual information as follows
	\bea\label{3par1}
		I_3(A,B,C) &=& I(A;B) + I(A;C) - I(A;B \cup C).
	\eea

\begin{figure}[h!]
$$\,$$
\centering
 \includegraphics[scale=0.75]{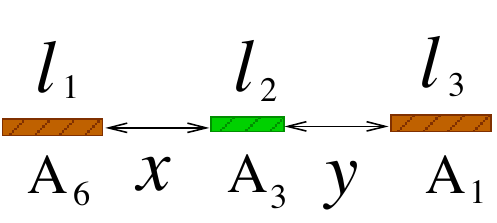}
\caption{Brown segments show  6 and 1 sites location and the green one shows  3 site location  for $(1,6|3)$-reduction   of the FMO complex,
depicted in Fig.\ref{Fig:fmo-1-3-6}. }
\label{Fig:MI136}
\end{figure}

We take $A=A_6,\,B=A_1,\,C=A_3$, where $A_i$, $i=1,3,6$ are 1,3 and 6 sites of the  $(1,6|3)$-reduction   of the FMO complex.
 We assume that $A_6,A_3,A_1$ are  three strips of  length $l_1$, $l_2$  and $l_3$ , and separated by  distances $x$ and $y$, Fig.\ref{Fig:MI136}.    

In the notations introduced in Sect.\ref{Decom} and the geometry presented in Fig.\ref{Fig:MI136}
we can write 
 \bea\label{S2-joint}
{\cal S}((A_6,A_1)_{joint-ncr})&=&{\cal S}(l_1+l_2+l_3+x+y,t)+{\cal S}(l_2+x+y,t)\\
\label{S2-dis}
{\cal S}((A_6,A_1)_{dis})&=&{\cal S}(l_1,t)+{\cal S}(l_3,t)\\
\label{S2-joint-cr}
{\cal S}((A_6,A_1)_{joint-cr})&=&{\cal S}(l_1+l_2+l_3+x+y,t)+{\cal S}(l_2+l_3+x+y,t)
\eea
where ${\cal S}(l,t)$ is the holographic entropy given by (\ref{ren-action})  or (\ref{ren-action-d2}).
\be
S(A_1\cup A_6)=\min\{{\cal S}((A_6,A_1)_{joint-ncr}),((A_6,A_1)_{dis}),((A,B)_{joint-cr})\}\ee

 To get $S(A_6 \cup A_3 \cup A_1)$ for  three segments we have to take into account the following competing contributions
\bea\label{S3-joint}
{\cal S}((A_6||A_3||A_1))&=&{\cal S}(l_1,t)+{\cal S}(l_2,t)+{\cal S}(l_3,t)\\
\label{S3-Adis}
{\cal S}(A_6||(A_3,A_1)_{ncr})&=&{\cal S}(l_1)+{\cal S}(l_2+l_3+y,t)+{\cal S}(y,t)\\
\label{S3-Cdis}
{\cal S}((A_6A_3)_{ncr}||A_1)&=&{\cal S}(l_1+l_2+x,t)+{\cal S}(x,t)+{\cal S}(l_3,t)\\
\label{S3-Bdis}
{\cal S}(A_6\underbrace{\bf A_3}A_1)_{ncr})&=&{\cal S}(l_1+l_2+l_3+x+y,t)+{\cal S}(l_2+x+y,t)+{\cal S}(l_2,t)\\
\label{S3-Bdiscr}
{\cal S}(A_6\underbrace{\bf A_3}A_1)_{cr})&=&{\cal S}(l_1+l_2+x+y,t)+{\cal S}(l_2+l_3+x+y,t)+{\cal S}(l_2,t)\\
\label{S3-Ncr}
{\cal S}((A_6A_3A_1)_{ncr})&=&{\cal S}(l_1+l_2+l_3+x+y,t)+{\cal S}(x,t)+{\cal S}(y,t)\\
\label{S3-cr1}
{\cal S}((A_6A_3A_1)_{cr,1})&=&{\cal S}(l_1+l_2+l_3+x+y,t)+{\cal S}(l_2+x,t)+{\cal S}(l_2+y,t)\\
\label{S3-cr2}
{\cal S}((A_6A_3A_1)_{cr,2})&=&{\cal S}(l_1+l_2+x+y,t)+{\cal S}(x,t)+{\cal S}(l_3+y,t)\\
\label{S3-cr3}
{\cal S}((A_6A_3A_1)_{cr,3})&=&{\cal S}(l_1+x,t)+{\cal S}(l_2+l_3+x+y,t)+{\cal S}(y,t)
\eea
and
\bea
&\,&S(A _6\cup A_3 \cup A_1)\\
&=&\min\Big\{{\cal S}((A_6||A_3||A_1)),{\cal S}(A_6||(A_3,A_1)_{ncr}),{\cal S}((A_6A_3)_{ncr}||A_1),{\cal S}(A_6\underbrace{\bf A_3}A_1)_{ncr})\nn\\
&\,&{\cal S}(A_6\underbrace{\bf A_3}A_1)_{cr}),{\cal S}((A_6A_3A_1)_{ncr}),{\cal S}((A_6A_3A_1)_{cr,2}),{\cal S}((A_6A_3A_1)_{cr,3})\Big\}
\nn\eea
Let us note, that for some particular background \cite{Ben-Ami:2014gsa} and refs therein, we can ignore contribution of (\ref{S2-joint-cr}), (\ref{S3-Bdis}), (\ref{S3-cr1}),
(\ref{S3-cr2}) and (\ref{S3-cr3}).

\begin{figure}[h!]
\centering
\begin{picture}(250,250)
\put(-80,0){\includegraphics[scale=0.7]{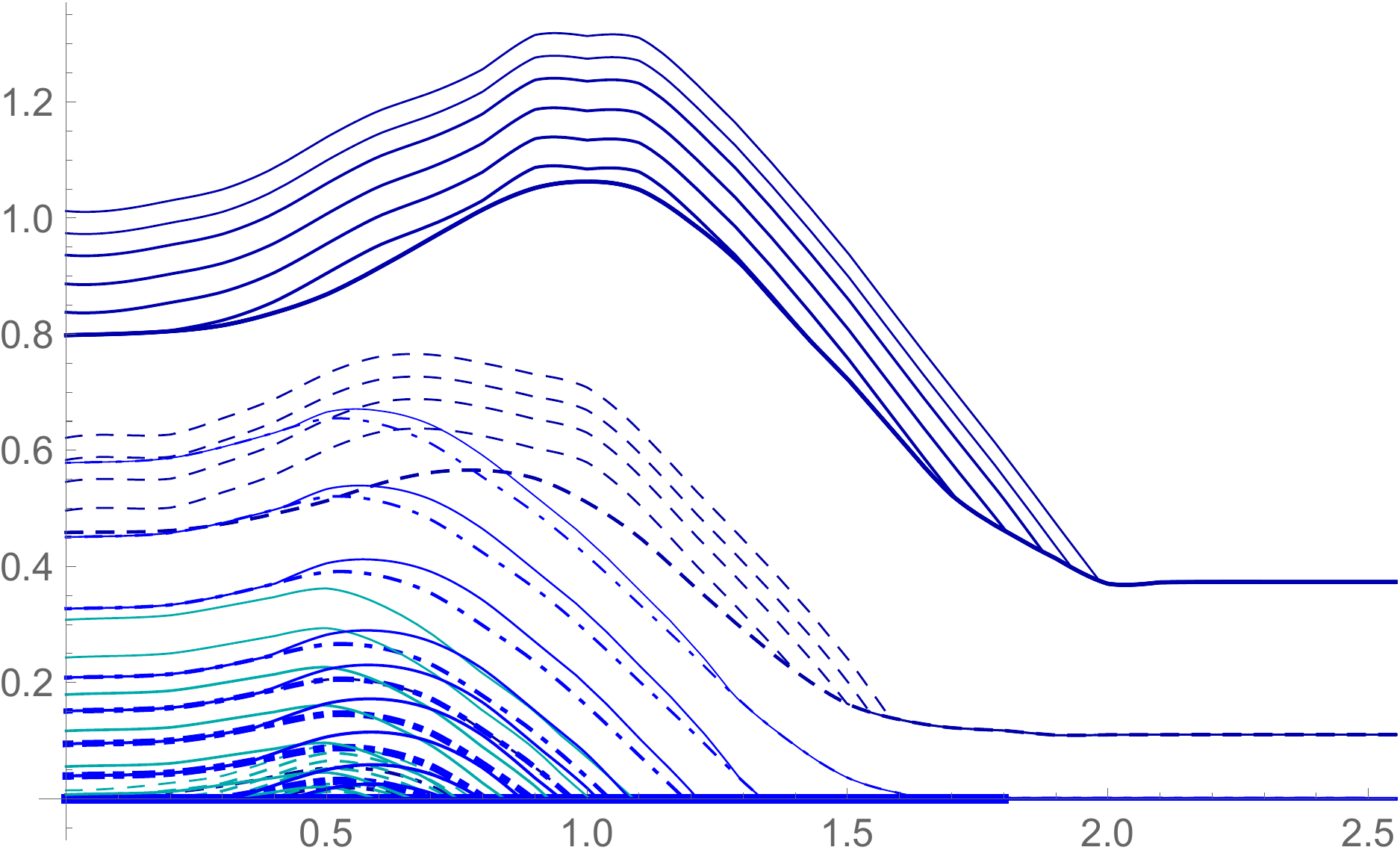}}
\put(310,8){$t$}
\put(-70,230){$I$}
\end{picture}
\caption{
%File:Vaidya-AD-seek-13-march0d3-BH-BH.
Holographic mutual information $I(A_6\cup A_1;A_3)$, up to the normalizing factor, for the Vaidya metric in the four dimensional black brane background with $f=f(z,v)$
 given by \eqref{fvz} and \eqref{m-v-BB} ($m_0=0.25,m=1$)
and the $x_1$-projection of the 3 belts configuration  shown in Fig.\ref{Fig:MI136},
as function of the boundary time t at fixed $l_1,l_2,l_3,x,y$. The  lines  of the same style correspond  
to the same $l_1,l_2,l_3$ and $y$, but different values of $x$, the distance between the sites "3" and "1" of the FMO complex. For more details see Fig.\ref{Fig:MI-3s-x-details1} and Fig.\ref{Fig:MI-3s-x-details2}. }\label{Fig:MI-3s-x-total}
\end{figure}

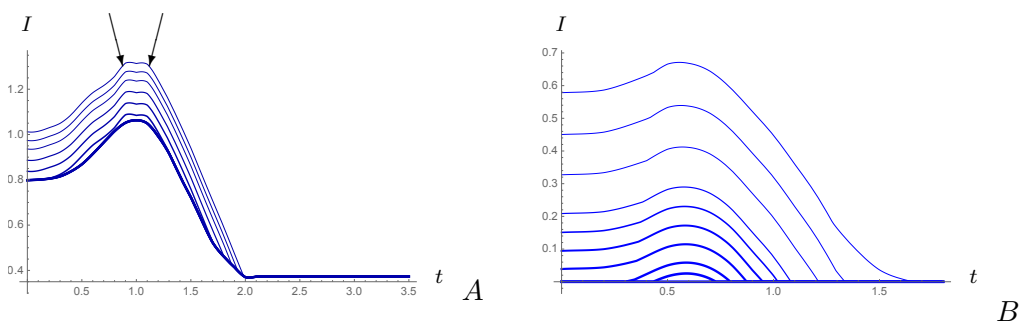
\begin{figure}[h!]
\centering
\begin{picture}(250,150)
\put(-60,0){\includegraphics[scale=0.25]{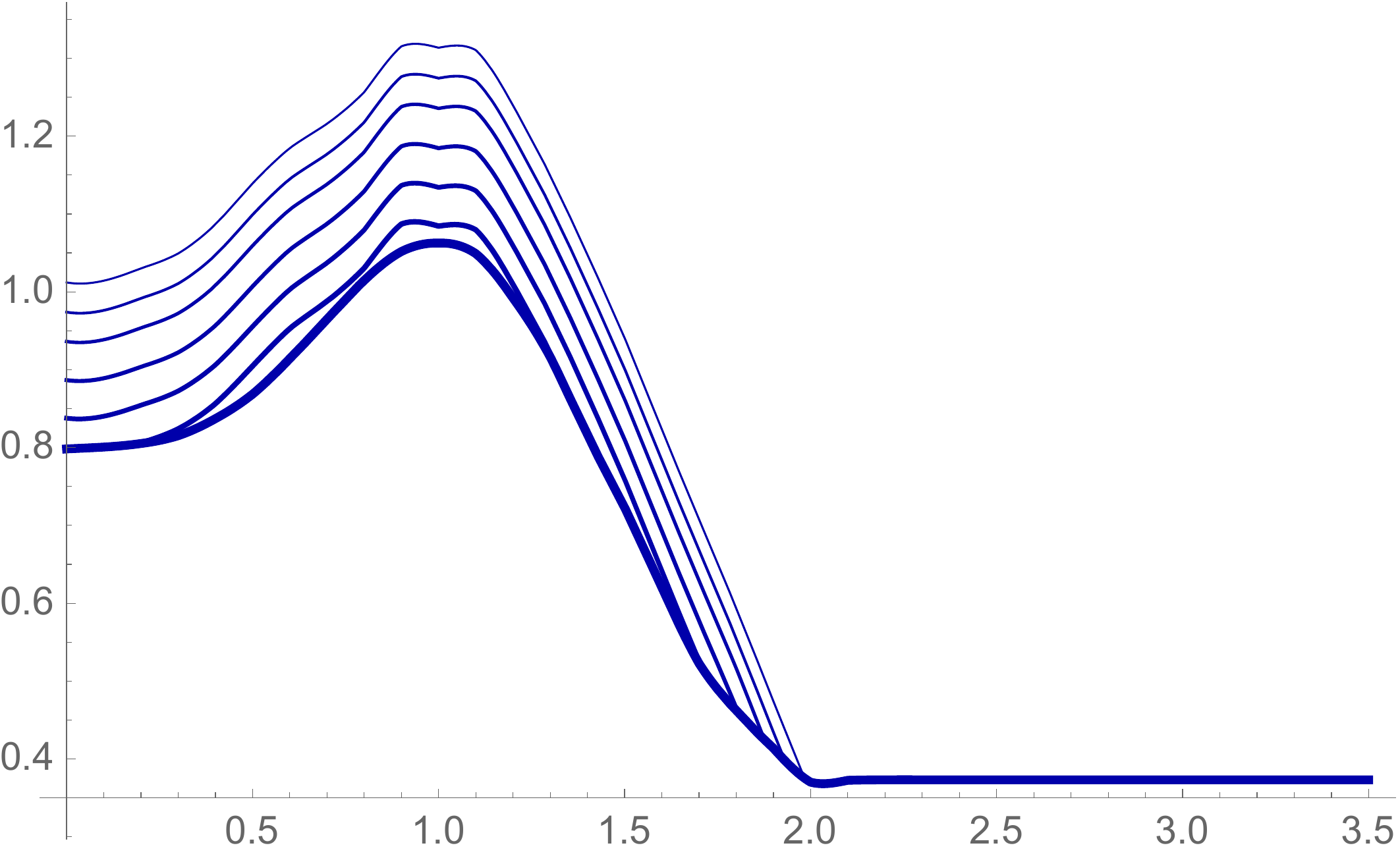}}
\put(100,8){$_t$}\put(110,0){$A$}
\put(-2,108){\vector(-1,-4){5}}
\put(-22,108){\vector(1,-4){5}}
\put(-55,103){$_I$}
\put(-60,11){\line(1,2){2}}
\put(140,0){\includegraphics[scale=0.25]{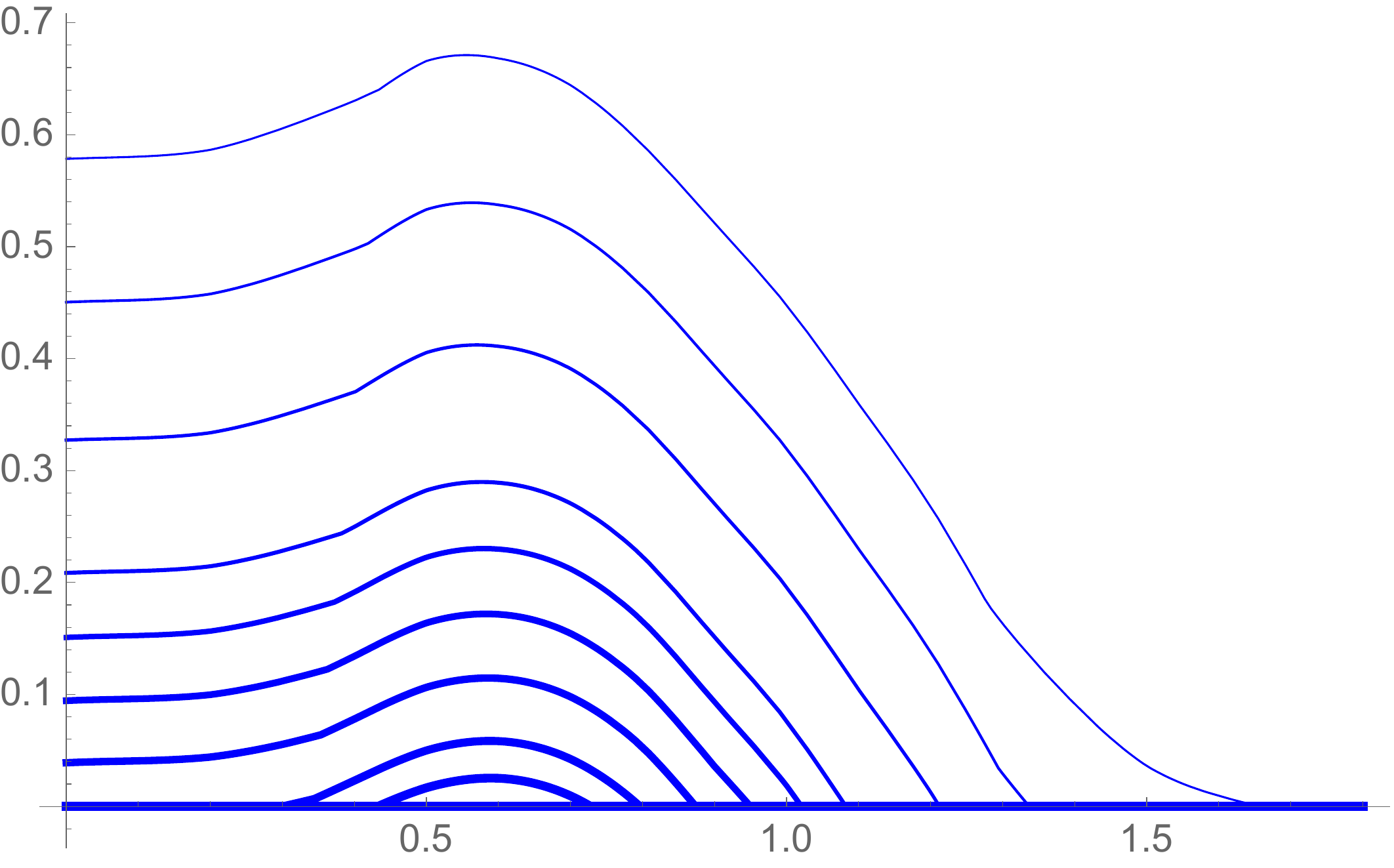}}
\put(300,8){$_t$}
\put(310,-8){$B$}
\put(145,103){$_I$}
\end{picture}
\caption{
%File:Vaidya-AD-seek-13-march0d3-BH-BH.
Detailed  plots  for Fig.\ref{Fig:MI-3s-x-total}. In plots {\bf A}: $l_1=0.3,l_2= 0.8,l_3=0.7,y=0.3$
and  $0.2<x<0.24$; in  plots {\bf B}: $l_1=0.4,l_2= 0.5,l_3=0.4,y=0.3$ and  $0.2<x<0.26$.  Values of $x$ increase going from the top curve to the bottom one. Arrows indicate the phase transitions.
}\label{Fig:MI-3s-x-details1}
\end{figure}

\begin{figure}[h!]
\centering
\begin{picture}(250,125)
\put(-60,0){\includegraphics[scale=0.25]{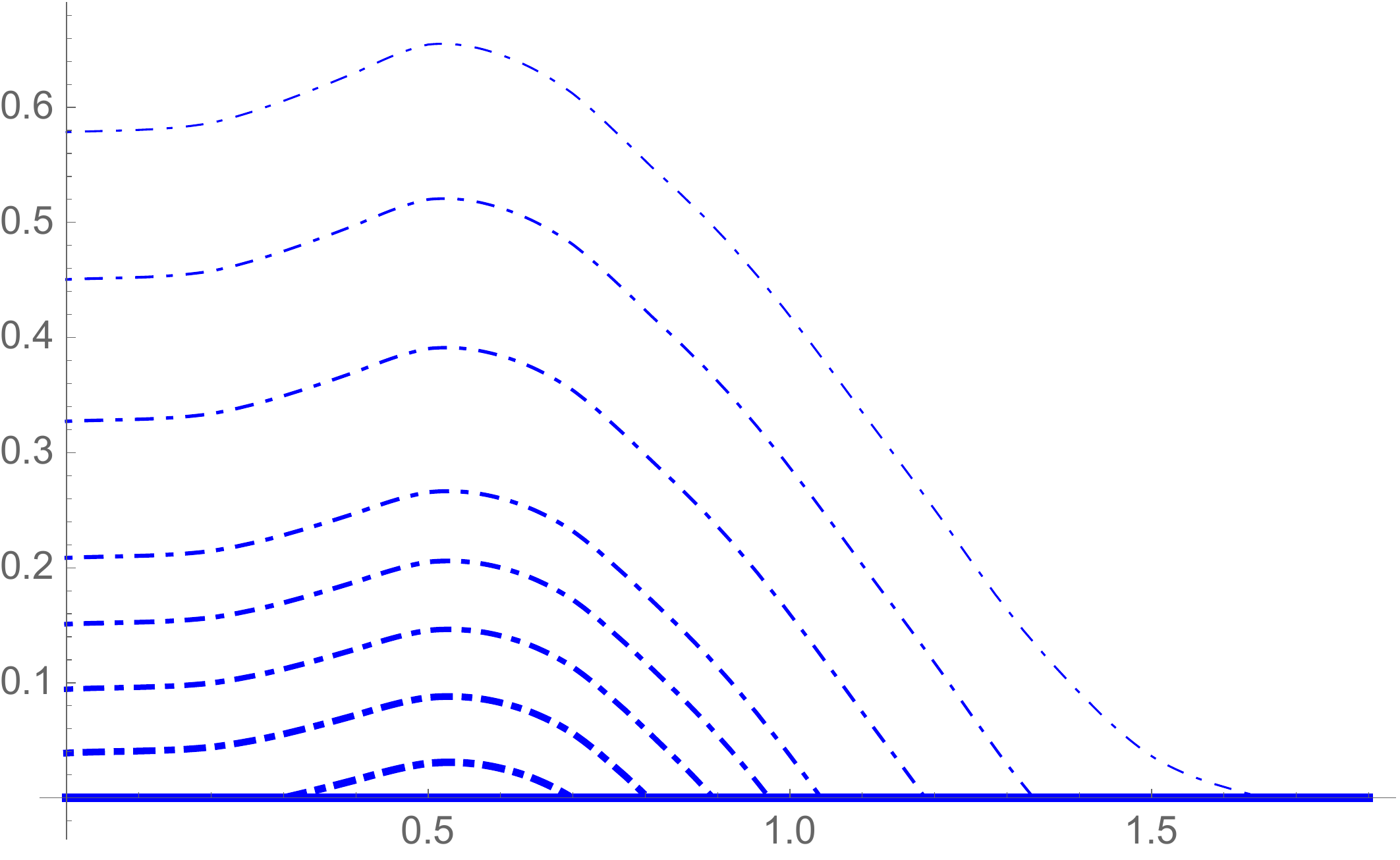}}
\put(100,8){$_t$}\put(110,8){$A1$}
\put(-55,100){$_I$}
\put(140,0){\includegraphics[scale=0.25]{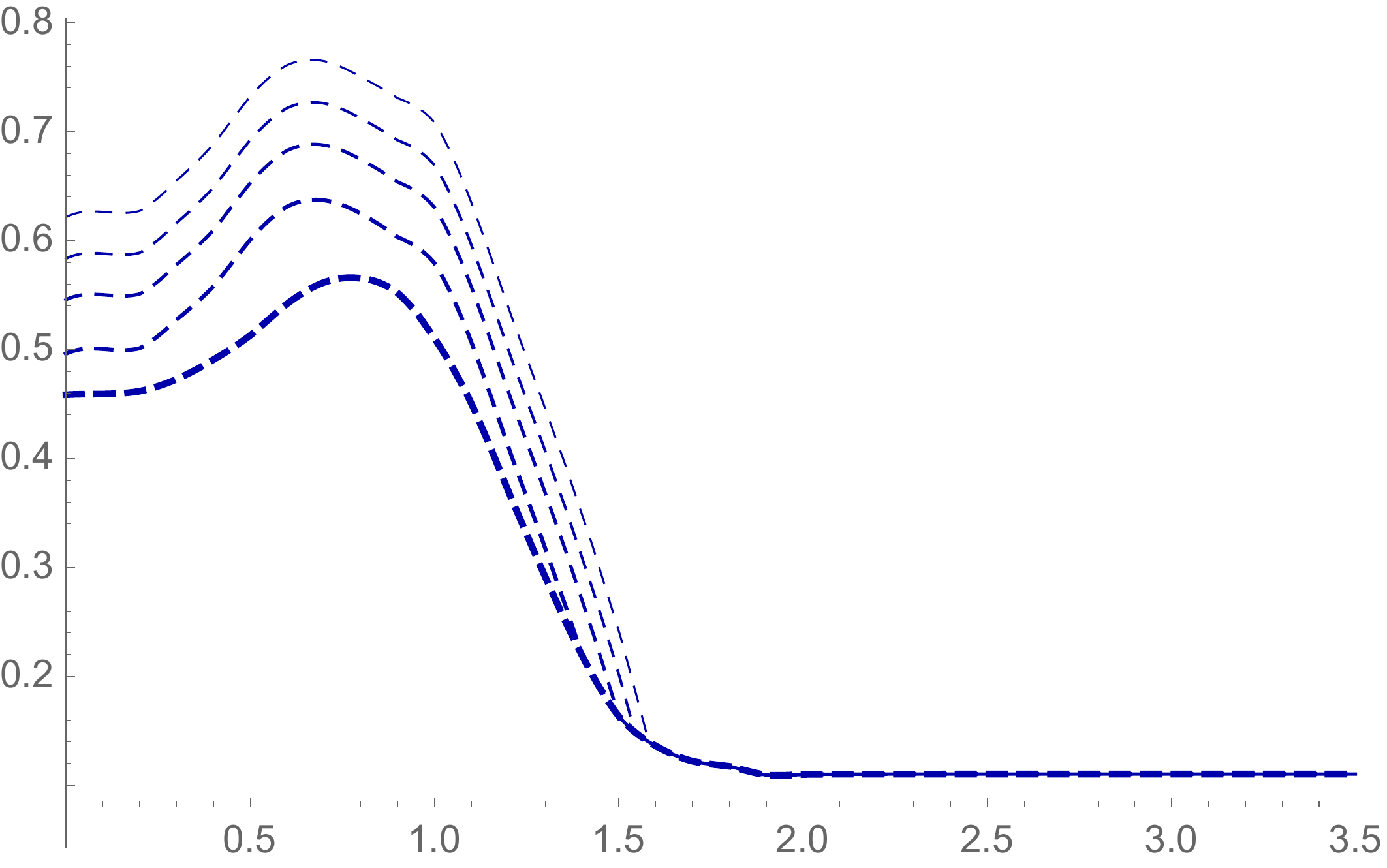}}
\put(300,8){$_t$}
\put(310,8){$B1$}
\put(145,100){$_I$}
\end{picture}
\begin{picture}(250,250)
\put(-60,120){\includegraphics[scale=0.25]{Fig3/MI-3-x-040504.pdf}}
\put(100,128){$_t$}\put(110,120){$C$}
\put(-55,223){$_I$}
\put(140,120){\includegraphics[scale=0.25]{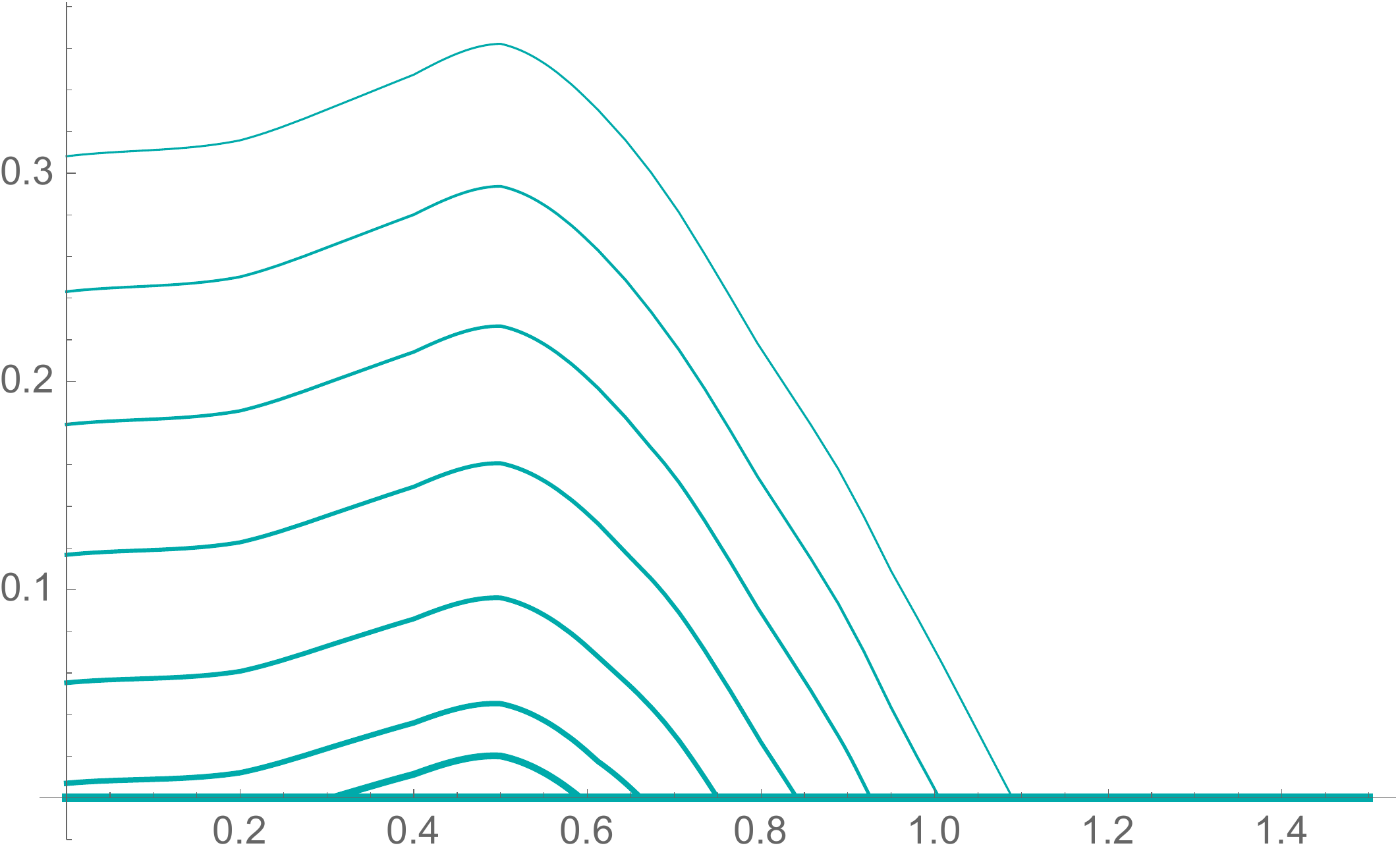}}
\put(300,128){$_t$}
\put(310,128){$D$}
\put(145,223){$_I$}
\put(-60,0){\includegraphics[scale=0.25]{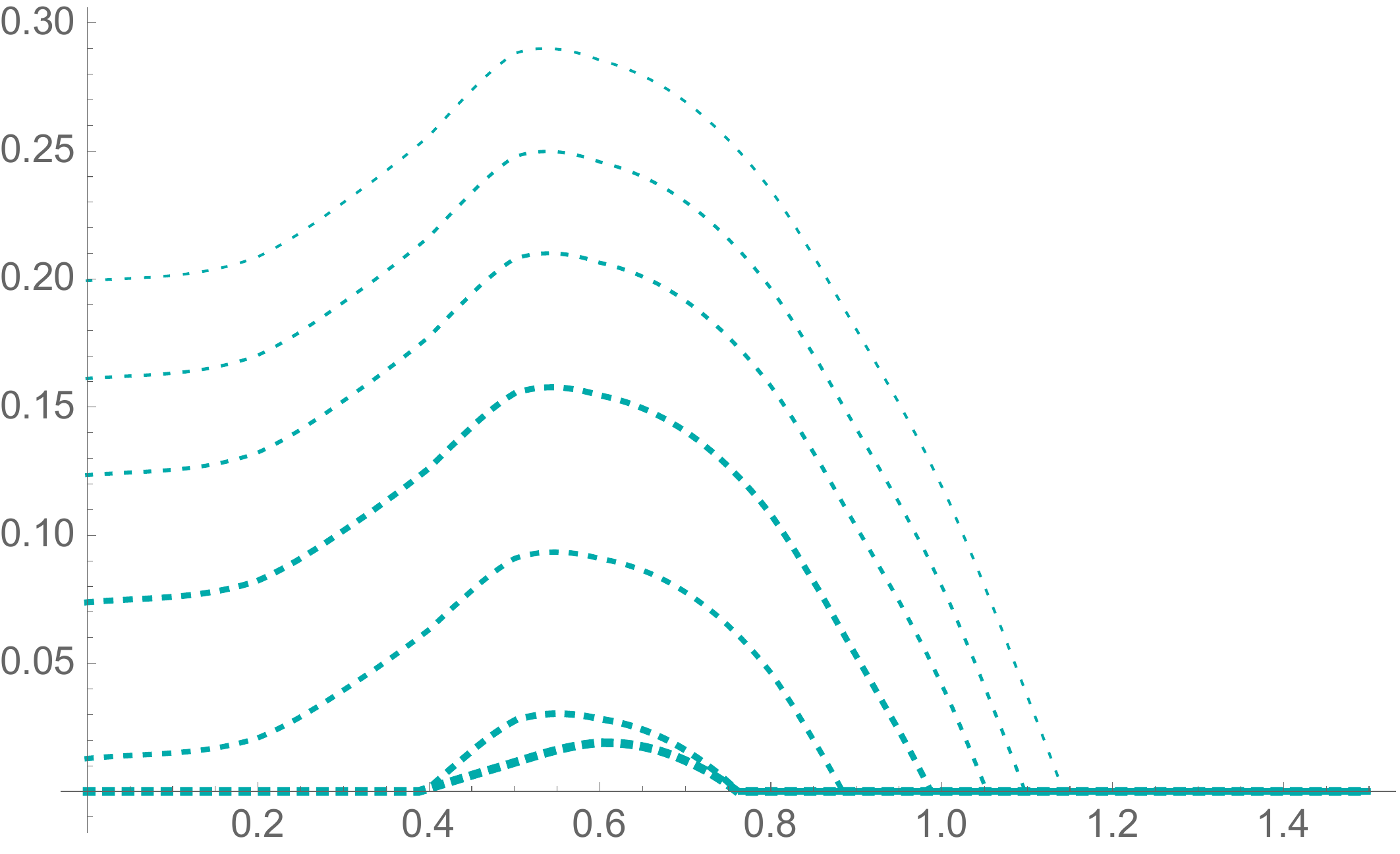}}
\put(100,8){$_t$}\put(110,8){$E$}
\put(-10,50){\vector(1,-4){5}}
\put(-11,34){\vector(1,-4){5}}
\put(-9,68){\vector(1,-4){5}}
\put(-55,100){$_I$}
\put(140,0){\includegraphics[scale=0.25]{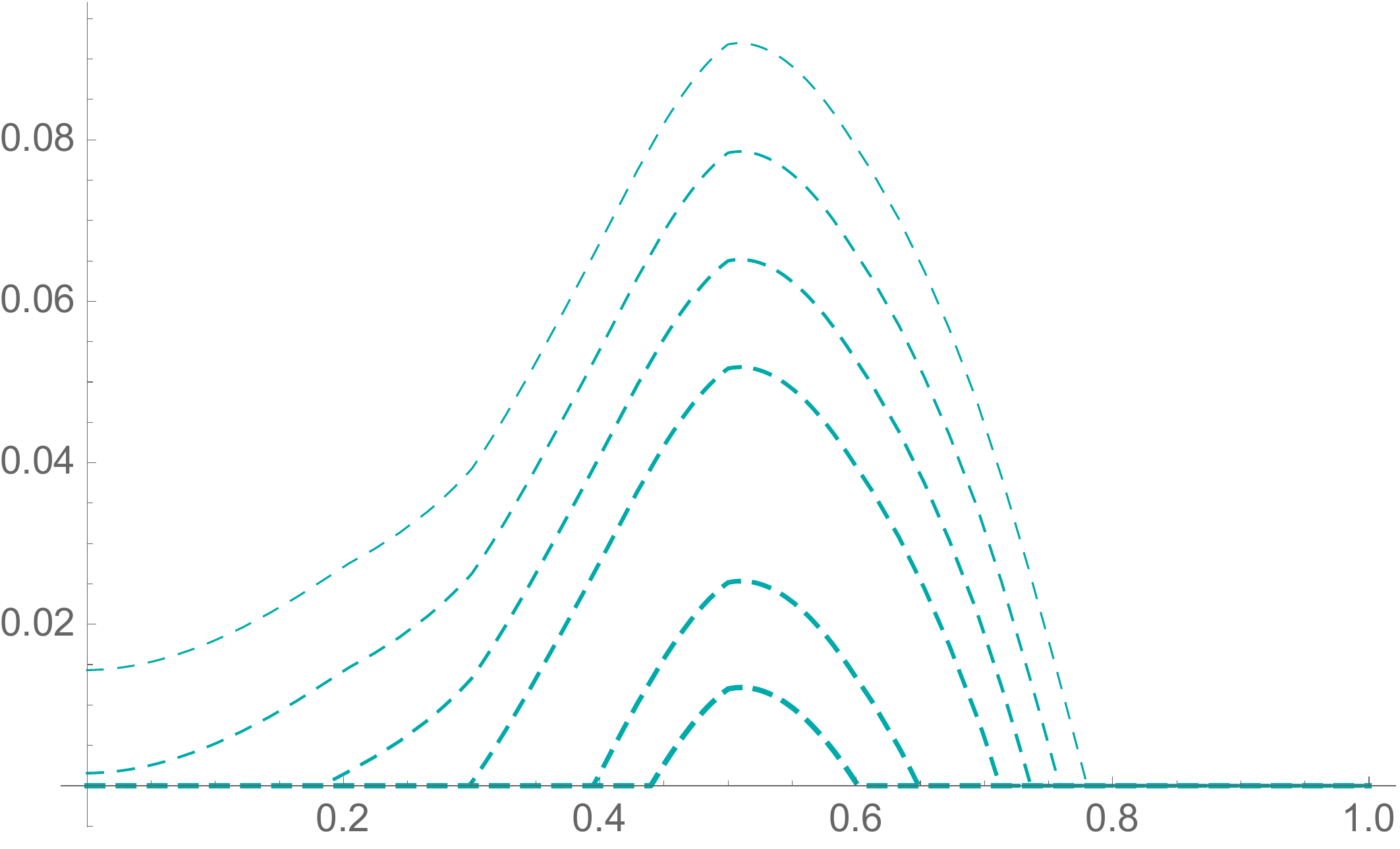}}
\put(300,8){$_t$}
\put(310,8){$F$}
\put(145,100){$_I$}
\end{picture}
$$\,$$\\
\caption{
Detailed  plots  for Fig.\ref{Fig:MI-3s-x-total}. In  plots {\bf A}: $l_1=0.4,l_2= 0.5,l_3=0.3,y=0.3$ and  $0.2<x<0.26$; {\bf B}:
$l_1=0.4,l_2= 0.7,l_3=0.3,y=0.3$ and  $0.2<x<0.235$.  Values of $x$ increase going from the top curve to the bottom one.
 {\bf C}: $l_1=0.4$, $l_2= 0.4$, $l_3=0.4$, $y=0.3$ and  $0.2<x<0.26$; {\bf D}: $l_1=0.3,l_2= 0.4,l_3=0.5$,  $y=0.3$ and $0.2<x<0.24$;  {\bf E}: $l_1=0.3$,
$l_2= 0.4$, $l_3=0.5$, $y=0.3$ and and $0.2<x<0.24$; {\bf F}: $l_1=0.3$, $l_2= 0.4$, $l_3=0.5$,$y=0.3$
and $0.2<x<0.206$.
 Values of $x$ increase going from the top curve to the bottom one. Arrow indicate the phase transitions.}
 \label{Fig:MI-3s-x-details2}
\end{figure}

The time dependence of the holographic mutual information $I(A_6\cup A_1;A_3)$ for the Vaidya metric in the four dimensional black brane background with $f=f(z,v)$
 given by \eqref{fvz} and \eqref{m-v-BB} ($m_0=0.25,m=1$)
and the $x_1$-projection of the 3 belts configuration  shown in Fig.\ref{Fig:MI136},
 at fixed $l_1,l_2,l_3,x,y$ is presented in Fig.\ref{Fig:MI-3s-x-total}.
 The  lines  of the same style (by the style we mean the color and the type of the line) correspond  
to the same $l_1,l_2,l_3$ and $y$, but different values of $x$, the distance between the sites "3" and "1" of the FMO 
complex presented in  Fig.\ref{Fig:fmo-1-3-6}.   Fig.\ref{Fig:MI-3s-x-details1} and Fig. \ref{Fig:MI-3s-x-details2} show different plots collected in Fig.\ref{Fig:MI-3s-x-total}.

We observe that the evolution of the holographic mutual information  for the case when one part of the system 
is  consisting  from two disjoints subsystems, globally  resembles the  case of two simple parts, but locally there are some changes.
 More specifically,  as in the case of two simple parts,  see Sec.\ref{Sect:HMI}, globally there are  4 different behaviors:
 the mutual information is $0$ at all times;
 the mutual information starts from a positive value and ends at another positive value;
the mutual information starts from a positive value and ends at $0$;
the mutual information starts from $0$, becomes positive for some time and ends at $0$ (the bell form of the time dependence). In the context of the FMO complex study, the last case  presents a special interest.
 Locally, we see phase transitions for some particular configurations.  
These cases are indicated by arrow in Fig.\ref{Fig:MI-3s-x-details1}  and  Fig.\ref{Fig:MI-3s-x-details2}.

$$\,$$
$$\,$$
\newpage
$$\,$$$$\,$$

\newpage

\subsection{Matching holographic calculations to the simulation results.}\label{Sect-Comparision-comp}

In this subsection we fit some results of numerical simulations for he FMO complex presented in \cite{0912.5112} by the holographic description.
The  mutual information at the physiological temperature ($300^\circ K$) calculated in \cite{0912.5112}, Fig.3c, when the  first system is the site three  and the second system  is a mixed state of sites one and six is  presented in Fig.\ref{Fig:MI-3s-x-details2} by the red dashed line. The time dependence  of the holographic mutual information $I(A_6\cup A_1;A_3)$ under the global quench by the Vaidya shell in AdS$_4$ is shown by the dark cyan line also in Fig.\ref{Fig:MI-3s-x-details2}. Note that we make rescaling for this line. We see a rather good fit of these two curves.

\begin{figure}[h!]
\centering
\includegraphics[scale=0.3]{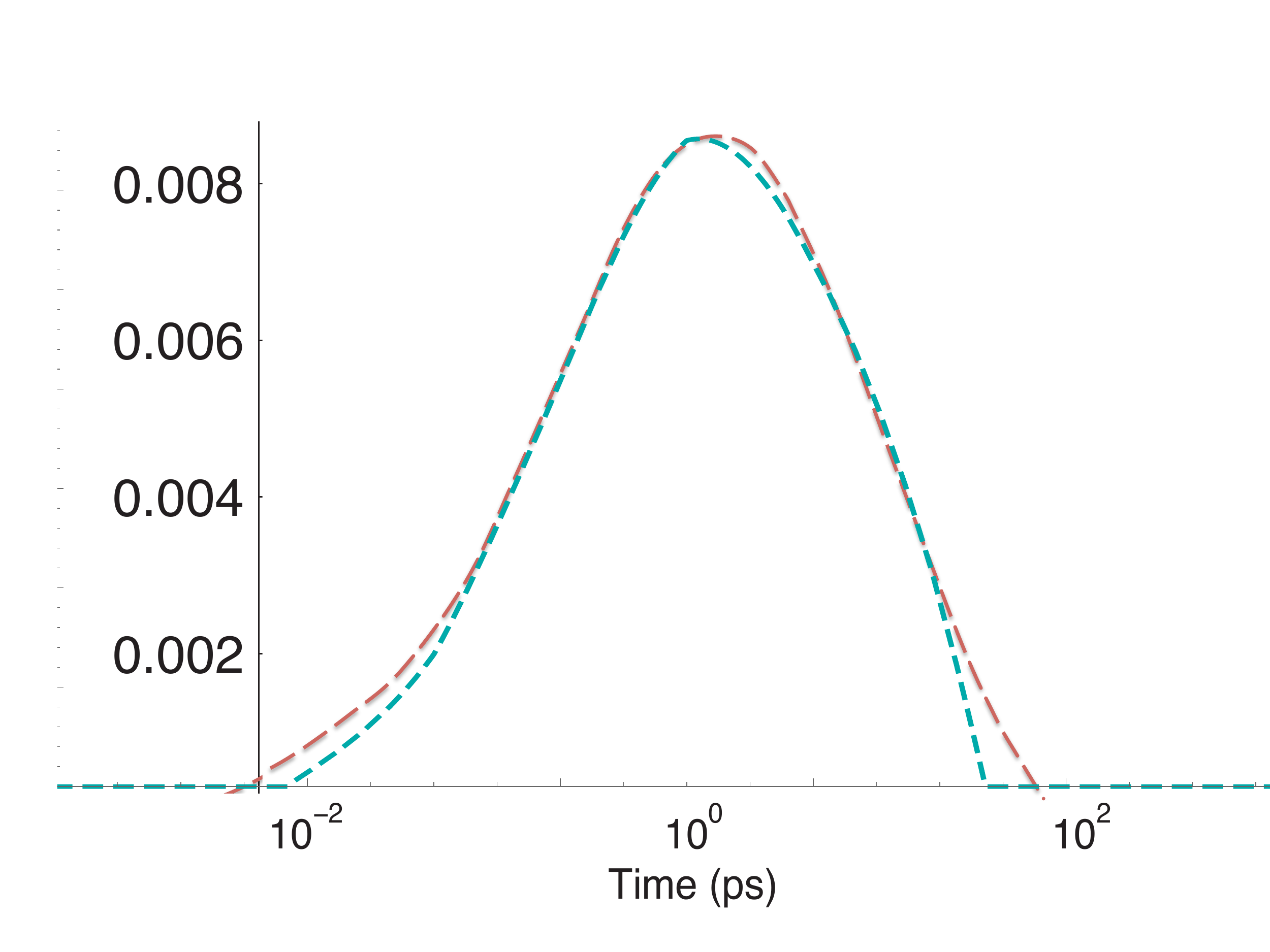}
\caption{
The dashed red curve shows  the time dependence (in the logarithmic scale) of the  mutual information at the physiological temperature ($300^\circ K$) calculated in \cite{0912.5112}, Fig. 3c, when the  fist system is the site ones and the second system  is a mixed state of sites one and six. The time dependence  of the holographic mutual information 
$I(A_6\cup A_1;A_3)$ under the global quench by the Vaidya shell in AdS$_4$ is shown by the dark cyan line.}\label{Fig:Comparison2}
\end{figure}

\section {Conclusions}
We have applied
the holographic  approach to evaluate the time dependence of entanglement entropy and
quantum mutual information in the Fenna-Matthews-Olson  protein-pigment complex in green sulfur bacteria during
the transfer of an excitation from a chlorosome antenna to a reaction center. It is shown that the time evolution of the mutual information
simulating the Lindblad master equation in some cases can be obtained by means
of a dual gravity describing black hole formation in the  Vaidya AdS spacetime or the  Vaidya AdS black brane spacetime.
The wake up and scrambling times for various partitions of the FMO complex are discussed.
We have demonstrated that some results of numerical simulations \cite{0912.5112}
for the FMO complex can be fitted by reduced holographic models containing just 2 or 3 segments, Fig.\ref{Fig:Compa},
Fig.\ref{Fig:Bell-reg}.B
and Fig.\ref{Fig:Comparison2}.

To describe another results known for the FMO complex it would be interesting to study more complex
holographic models containing  7 or 8  segments. It would be also suitable in this context to change the geometry of 
disjoint regions and consider, for example, 
the disk regions. Considering the global AdS$_{d+1}$ one can also study the compact systems. In the case of the global 
AdS$_3$ one can consider the time dependence of the mutual information not only during the global quench,
described by the Vaidya metric  \cite{1507.00306}, but also during the local quench provided by ultrarelativictic
 particle in AdS$_3$ \cite{1512.03363}.
Studying   Lifshitz  backgrounds could be useful for describing not only the FMO complex but also  other light-harvesting complexes. We suspect that the holographic approach could provide
a useful description of certain appropriate quantities not only for quantum photosynthesis but also for  other
life science phenomena studied in quantum biology such as  the process of vision,
the olfactory sense and
quantum tunneling in biomolecules.

\section{Acknowledgments} I.A. would like to thank Dmitry Ageev for  help with numerical calculations.  This work was supported by the Russian
Science Foundation (grant No. 14-11-00687).

\end{document}